\documentclass[a4paper,fleqn,usenatbib]{mnras}

\usepackage{newtxtext,newtxmath}
 
\usepackage[T1]{fontenc}
\usepackage{ae,aecompl}

\usepackage{graphicx}	% Including figure files
\usepackage{amsmath}	% Advanced maths commands
\usepackage{amssymb}	% Extra maths symbols
\usepackage{pdflscape} % To produce landscape figure or table
\usepackage{adjustbox} % To scale the table to a specific width

\title[The accretion rates and mechanisms of Herbig Ae/Be stars]{The accretion rates and mechanisms of Herbig Ae/Be stars}

\author[C. Wichittanakom et al.]{C. Wichittanakom,$^{1,2}$\thanks{E-mail: pycw@leeds.ac.uk (CW)}
R. D. Oudmaijer,$^{2}$
J. R. Fairlamb,$^{3}$
I. Mendigut\'{i}a,$^{4}$
\newauthor
M. Vioque,$^{2}$ and K. M. Ababakr$^{5}$
\\
$^{1}$Department of Physics, Faculty of Science and Technology, Thammasat University, Rangsit Campus, Pathum Thani 12120, Thailand\\
$^{2}$School of Physics and Astronomy, EC Stoner Building, University of Leeds, Leeds LS2 9JT, UK\\
$^{3}$Institute for Astronomy, University of Hawaii, 2680 Woodlawn Drive, Honolulu, HI, 96822, USA\\
$^{4}$Centro de Astrobiolog\'{i}a (CSIC-INTA), Departamento de Astrof\'{i}sica, ESA-ESAC Campus, PO Box 78,\\
28691 Villanueva de la Ca\~{n}ada, Madrid, Spain\\
$^{5}$Erbil Polytechnic University, Kirkuk Road, Erbil, Iraq
}

\date{Accepted for publication in MNRAS 493, 234--249 (2020). Accepted 2020 January 07. Received 2020 January 01; in original form 2019 October 09}

\pubyear{2020}

\begin{document}
\label{firstpage}
\pagerange{\pageref{firstpage}--\pageref{lastpage}}
\maketitle

% Abstract of the paper < 250 words for Main Journal or < 200 words for Letters
\begin{abstract}
This work presents a spectroscopic study of 163 Herbig Ae/Be stars. Amongst
these, we present new data for 30 objects. Stellar parameters such as temperature, reddening, mass,
luminosity and age are homogeneously determined. Mass accretion rates are
determined from $\rm H\alpha$ emission line measurements.  
Our data is complemented with the X-Shooter sample from previous studies and we 
update results using {\em Gaia} DR2 parallaxes giving a total of 78 objects with homogeneously
determined stellar parameters and mass accretion rates. In addition, 
mass accretion rates of an additional 85 HAeBes are determined. 
We confirm previous findings that the mass
accretion rate increases as a function of stellar mass, and the
existence of a different slope for lower and higher mass stars
respectively.  The mass
where the slope changes is determined to be
$3.98^{+1.37}_{-0.94}\,\rm M_{\sun}$. We discuss this break in the
context of different modes of disk accretion for low- and high mass
stars. Because of their similarities with T Tauri stars, 
we identify the accretion mechanism for the late-type Herbig
stars with the Magnetospheric Accretion. The possibilities for the
earlier-type stars are still open, we suggest the Boundary Layer
accretion model may be a viable alternative. Finally, we
  investigated the mass accretion - age relationship. Even using the
  superior {\em Gaia} based data, it proved hard
  to select a large enough sub-sample to remove the mass dependency in this
  relationship.  Yet, it would appear that the mass accretion does
  decline with age as expected from basic theoretical considerations.
\end{abstract}


%%%%%%%%%%%%%%%%%%%%%%%%%%%%%%%%%%%%%%%%%%%%%%%%%%

%%%%%%%%%%%%%%%%% BODY OF PAPER %%%%%%%%%%%%%%%%%%

\section{Introduction}

Herbig Ae/Be stars (HAeBes) are optically visible intermediate
pre-main sequence (PMS) stars whose masses range from about 2 to
10\,$\rm M_{\sun}$.  These PMS stars were first identified by
\cite{Herbig1960}, using three criteria: ``Stars with spectral type A
and B with emission lines; lie in an obscured region; and illuminate
fairly bright nebulosity in its immediate vicinity''.  HAeBes play an
important role in understanding massive star formation, because they
bridge the gap between low-mass stars whose formation is relatively
well understood, and high-mass stars whose formation is still
  unclear.  The study of their formation is not well understood as
  massive stars are very rare and as a consequence on average far
  away, and often optically invisible \citep{Lumsden2013}. A long
standing problem has also been that their brightness is very high,
such that radiation pressure can in principle stop accretion onto the
stellar surface \citep{Kahn1974}.  Moreover, they form very quickly
and reach the main sequence before their surrounding cloud disperses
\citep{Palla1993}.  This suggests that their evolutionary processes
are very different from T Tauri stars.

The accretion onto classical T Tauri low-mass stars is magnetically
controlled.  The magnetic field from the star truncates material in
the disc and from this point, material falls onto the star along the
field line.  After matter hits the photosphere, it produces X-ray
radiation that is absorbed by surrounding particles.  Then, these
particles heat up and re-radiate at ultraviolet wavelengths, producing
an UV-excess that can be observed and from which the accretion luminosity can be calculated   \citep{Bouvier2007, Hartmann2016}.
It was found later, that a correlation between the line
luminosity and the accretion luminosity exists, allowing accretion
rates of classical T Tauri stars to be determined from the UV-excess
and absorption line veiling \citep{Calvet2004, Ingleby2013}.

HAeBes have similar properties as classical T Tauri stars, for
instance having emission lines, UV-excess, a lower surface gravity
than main-sequence stars \citep{Hamann1992, Vink2005} and they are
usually identified by an infrared (IR) excess from circumstellar discs
\citep{van_den_Ancker2000, Meeus2001}. Because their envelopes are
radiative, no magnetic field is expected to be generated in HAeBe
stars as this usually happens in stars by convection. Indeed, magnetic
fields have rarely been detected towards them
\citep{Catala2007, Alecian2013}. As a result magnetically controlled
accretion is not necessarily expected to apply in the case of the most
massive objects, requiring another accretion mechanism. However, how
the material arrives at the stellar surface in the absence of magnetic
fields is still unclear.

Several lines of evidence have indicated a difference between the
  lower mass Herbig Ae and higher mass Herbig Be
  stars. Spectropolarimetric studies suggest that Herbig Ae stars and
T Tauri stars may form in the same process \citep{Vink2005,
  Ababakr2017}.  Accretion rates of HAeBes are determined from the
measurement of UV-excess \citep{Mendigutia2011b} who have found that
the relationship between the line luminosity and the accretion
luminosity is similar to the classical T Tauri stars.  A spectroscopic
variability study suggests that a Herbig Ae star is undergoing
magnetospheric accretion in the same manner as classical T Tauri stars
\citep{Scholler2016} while \citet{Mendigutia2011a} find the Herbig Be
stars to have different H$\alpha$ variability properties than the
Herbig Ae stars.  Therefore, magnetically influenced accretion in
HAeBes would still be possible. A study employing X-shooter spectra
for 91 HAeBes was carried out by \citet[hereafter F15 and F17
  respectively]{Fairlamb2015,Fairlamb2017}.  They determined the
stellar parameters in a homogenous fashion, derived mass accretion
rates from the UV-excess and found the relationship between accretion
luminosity and line luminosity for 32 emission lines in the range
  0.4--2.4 $\mu$m.   47 objects in
their sample were taken from \citet[table 1]{The1994}, which to this
date contains the strongest (candidate) members of the group.
The Fairlamb papers were published before the {\em Gaia} parallaxes
became available, and the distances used may need revision as
these can affect parameters such as the radius, luminosity, stellar
mass and mass accretion rate.

A {\em Gaia} astrometric study of 252 HAeBes was carried out by
\cite{Vioque2018} hereafter V18. They presented parallaxes for all known HAeBes
from the {\em Gaia} Data Release 2 ({\em Gaia} DR2) Catalogue
\citep{Gaia2016, Gaia2018}.  Also, they collected effective
temperatures, optical and infrared photometry, visual extinctions,
$\rm H\alpha$ equivalent widths, emission line profiles and binarities
from the literature.  They derived distances, luminosities, masses,
ages, infrared excesses, photometric variabilities for most of their
sample.  This is the largest astrometric study of HAeBes to date.
These {\em Gaia} DR2 parallaxes are useful for improving the stellar
parameters of the previous studies.

The main aim of this paper is to determine stellar parameters and
  accretion rates of 30 Northern HAeBes using a homogeneous approach,
  extending the mostly Southern F15 sample. In parallel, the results
of F15 will be updated using the distances determined from the {\em
  Gaia} DR2 parallaxes and a redetermined extinction towards the
objects.
As such, this becomes the largest homogeneous spectroscopic analysis
of HAeBes to date.  In addition, the accretion rates of HAeBes in the
V18 study are determined for the 104 objects for
which H$\alpha$ emission line equivalent widths are collected from the
literature or determined from archival spectra.

The structure of this paper is as follows.  Section 2 presents the
spectroscopic data observation of all targets.  Details of
observations, instrumental setups and reduction procedures are given
in this section.  Section 3 and 4 detail the determination of stellar
parameters and mass accretion rates. 
Sections 5 and 6 focus on the analysis and a discussion respectively. 
Section 7 provides a summary of the main conclusions of this paper.

\section{Observations and Data Reduction}
\subsection{IDS and sample selection}

The data were collected in June 2013 using the Intermediate Dispersion
Spectrograph (IDS) instrument and the RED+2 CCD detector with $2048
\times 4096$ pixels (pixel size 24\,\micron), which is attached to the
Cassegrain focus of the 2.54-m Isaac Newton Telescope (INT) at the
Observatorio del Roque de los Muchachos, La Palma, Spain.  The
observations spanned six nights between the $\rm 20^{th}$ and the $\rm
25^{th}$ June 2013.  Bias frames, flat field frames, object frames and
arc frames of Cu-Ar and Cu-Ne comparison lamp were taken each night to
prepare for the data reduction.

In the first two nights the spectrograph was set up with a
  1200\,lines\,mm$^{-1}$ R1200B diffraction grating and a 0.9 or 1.0
  arcsecond wide slit in order to obtain spectra across the Balmer
  discontinuity.  The wavelength coverage was 3600--4600\,\AA,
centred at 4000.5\,\AA.  In this range the hydrogen lines of $\rm
H\gamma, H\delta$, H(7-2), H(8-2), H(9-2) and H(10-2) can be analysed.
This combination provides a reciprocal dispersion of
0.53\,\AA\,pixel$^{-1}$ with a spectral resolution of $\sim$ 1\,\AA\ and a
resolving power of $R\sim4000$.

For the next two nights the R1200Y grating was used to observe spectra
in the visible.  Its spectral range covers 5700 to 6700\,\AA, centred
at 6050.4\,\AA.  This range includes the \ion{He}{i} line at
5876\,\AA, the [\ion{O}{i}$_{\lambda6300}$] line and the $\rm H\alpha$
line at 6562\,\AA.  This setup results in a reciprocal dispersion of
0.52\,\AA\,pixel$^{-1}$ and a resolving power of $R\sim6500$.  
  The final 2 nights used the R1200R grating to observe the spectral
  range of 8200--9200\,\AA, centred at 8597.2 and 8594.4\,\AA. This
range covers the \ion{O}{i} line at 8446\,\AA, all lines of the
\ion{Ca}{ii} triplet and many lines of the Paschen series.  With this
setup the reciprocal dispersion becomes 0.51\,\AA\,pixel$^{-1}$,
giving a resolving power of $R\sim9000$.

The sample consists of 45 targets, with 30 HAeBes, and 15 standard stars.
26 Herbig stars were chosen from the catalogue of \citet[table
  1]{The1994} and 4 from \cite{Vieira2003}.  About 67 per cent of
HAeBes in the final sample are in the northern hemisphere.  There are
7 HAeBes which are also in the sample of F15.  Spectra
of standard stars were observed each night for spectral comparisons.
A log of the observations of the 30 HAeBes is shown in Table~\ref{tab:log}
while a log of the observations of the standard stars is presented in
Table~\ref{tab:A1_standard_stars} (See Appendix A in the online version of this paper).

The data reduction was performed using the Image Reduction and Analysis
Facility ({\small IRAF}\footnote{{\small IRAF} is distributed by the
  National Optical Astronomy Observatory, which is operated by the
  Association of Universities for Research in Astronomy (AURA) under a
  cooperative agreement with the National Science Foundation.}).
Standard procedures were used in order to process all frames.
Generally, there are three steps of data reduction; bias subtraction,
flat field division and wavelength calibration.  First, many bias
frames were taken and averaged to reduce some noise and the bias level
was removed from the CCD data.  Next, flat field division was used to
remove the variations of CCD signal in each pixel.  All flat frames
were averaged before subtracting the bias.  Then, all of the object
frames were divided by the normalised flat-field frame in order to
correct the pixel-to-pixel variation of the detector sensitivity.  The
object frame was extracted to a one-dimensional spectrum by defining
the extraction aperture on the centre of the profile and subtracting
the sky background.  Finally, the data were wavelength calibrated.
The accuracy of the wavelength calibration is measured from an RMS of
the fit and is less than 10 per cent of the reciprocal dispersion
($<0.05$\,\AA\,pixel$^{-1}$).  For illustration, the $\rm H\alpha$
profiles of all 30 objects are displayed in Figure~\ref{fig:B1_EW} (See Appendix B in the online version of this paper).

\section{Stellar Parameters and Mass Accretion Rate Determinations}

Measurement of accretion rates requires accurate stellar parameters of
the star in question. Therefore, this section aims to determine
accretion rates by first determining the stellar parameters, such as
the effective temperature, surface gravity, distance, radius,
reddening, luminosity, mass and age. These will be determined by
combining the IDS spectra, stellar model atmosphere grids,
photometry from the literature, the {\em Gaia} DR2 parallaxes and
stellar isochrones.
 
\subsection{Effective temperature and surface gravity}
To estimate the effective temperature of the target, a comparison of the known standard star spectra with the unknown target spectrum is performed.
The list of standard stars including spectral types is shown in Table~\ref{tab:A1_standard_stars} (See Appendix A in the online version of this paper).
The conversion from the spectral type to effective temperature can be found from \cite{Straizys1981}.
Then, the range of effective temperatures was explored in more detail with spectra computed from model atmospheres.

The model atmospheres used in this work are grids of BOSZ-Kurucz model atmospheres computed by \cite{Bohlin2017}.
The BOSZ models are calculated from ATLAS-APOGEE ATLAS9 \citep{Meszaros2012} which came from the original ATLAS code version 9 \citep{Kurucz1993}.
The metallicity $[\rm M/H]=0$, carbon abundance $\rm [C/H]=0$, alpha-element abundance $[\alpha/ \rm H]=0$, microturbulent velocity $\xi=2.0$\,km\,s$^{-1}$, rotational broadening velocity $v\sin i=0.0$\,km\,s$^{-1}$, and instrumental broadening $R=5000$ are adopted.
This instrumental broadening is chosen for matching the resolution of the blue spectra.
The range of effective temperature $T_{\rm eff}$ is from 3500\,K to 30000\,K with steps of 250\,K (from 3500\,K to 12000\,K), 500\,K (from 12000\,K to 20000\,K) and 1000\,K (from 20000\,K to 30000\,K).
The range of surface gravity $\log(g)$ is from 0.0\,dex to 5.0\,dex with steps of 0.5\,dex.
By using linear interpolation, $\log(g)$ with steps of 0.1\,dex were calculated.

The procedure to obtain the effective temperature and surface gravity
in this work follows the same method by F15.  The effective
temperature and surface gravity determination is carried out by 
  primarily comparing the wings of the observed hydrogen Balmer lines
$\rm H\gamma$, $\rm H\delta$ and $\rm H\epsilon$ with synthetic
profiles produced with solar metallicity from BOSZ models.  The shapes
of the hydrogen profiles are not very sensitive to metallicity
\citep{Bohlin2017} while their widths are dominated by pressure
  broadening, while rotation hardly affects the shapes.  $\rm
H\alpha$ is not used for spectral typing because it is frequently
strongly in emission for HAeBes.  This phenomenon can affect the shape
of the wings and cause difficulty of fitting the BOSZ models to target
spectra.  $\rm H\beta$ is also not used for spectral typing because it
appears around the edge of the blue spectrum which makes a proper
characterization of $\rm H\beta$'s line profile troublesome.

In order to carry out the fitting, both the observed profiles and the
synthetic profiles are normalised based on the continuum on both the
blue and red side of the profile.  Next, the observed wavelengths of
the target profiles were corrected to the vacuum wavelength of the
synthetic profile using the {\small IRAF DOPCOR} task.  The effective
temperature and surface gravity are obtained from the fit of the
normalised synthetic spectra to the normalised observed spectra by
considering continuum features and the wings of the profile above the
normalised intensity of 0.8.  This intensity was chosen because this
part of the wings is sensitive to variation of the $\log(g)$ while the
central part of the profile of a Herbig Ae/Be star can be contaminated
by emission.

\begin{figure}
\centering
\includegraphics[width=\columnwidth]{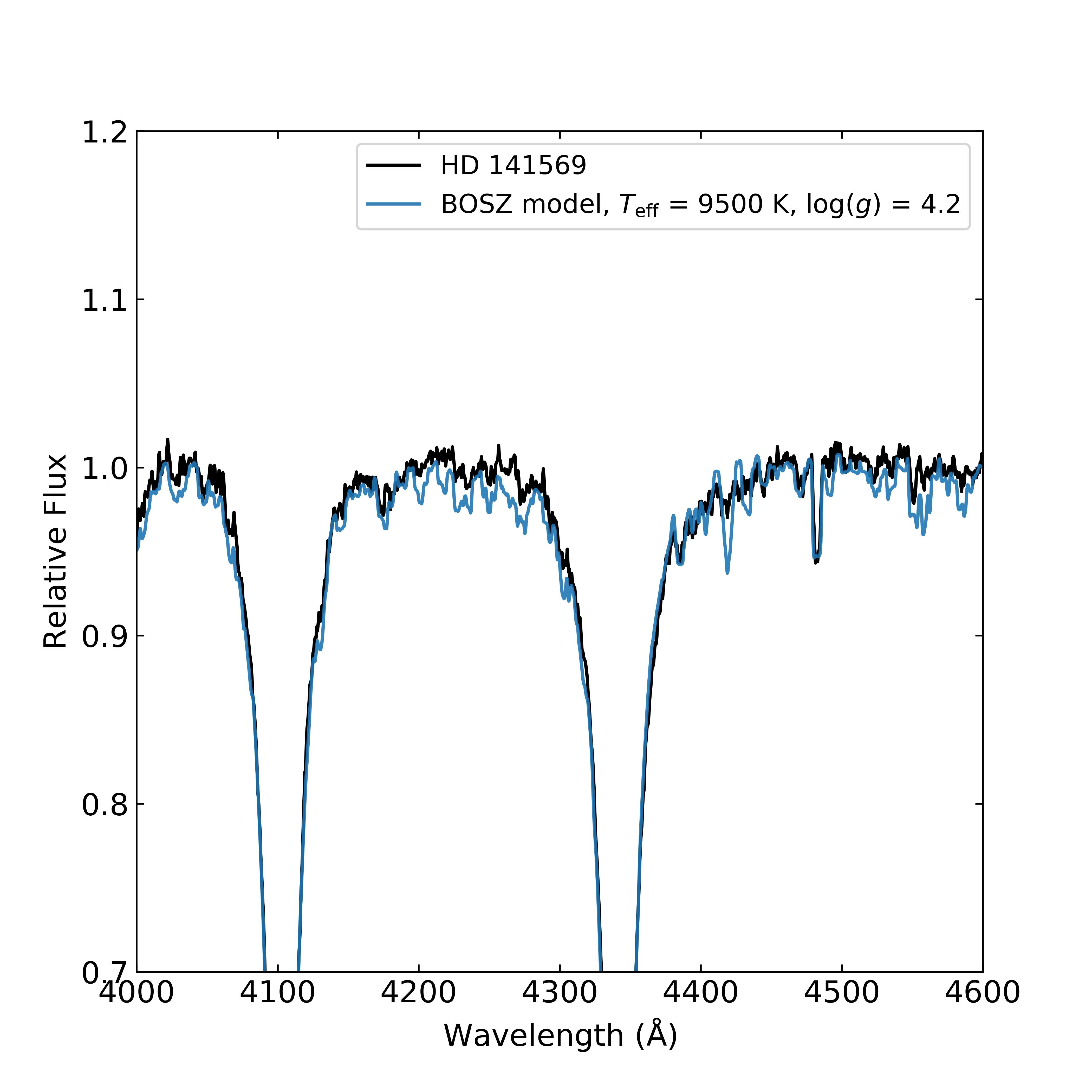}
\caption{The normalised spectrum of HD~141569 fits with BOSZ model of $T_{\rm eff}=9500$\,K and $\log(g)=4.2$.}
\label{fig:HD141569_Hgamma}
\end{figure}

The width of the Balmer lines depends upon both effective temperature
and surface gravity.  Different combinations of $T_{\rm eff}$ and
$\log(g)$, can create the same width.  This degeneracy can be solved
by visual inspection of absorption features in the wings and continuum
either side of the lines.  The final value for the effective
temperature and surface gravity are the average value of the best fit
for each profile ($\rm H\gamma$, $\rm H\delta$ and $\rm H\epsilon$).
The uncertainties of $T_{\rm eff}$ and $\log(g)$ were chosen to be the
typical difference in the values determined for the three lines
respectively.  If the standard error becomes zero or less than that,
the step size will be adopted.  An example of spectral typing by
fitting the Balmer line profile of HD~141569 (black) and the BOSZ
model (blue) is demonstrated in Figure~\ref{fig:HD141569_Hgamma}.

\begin{landscape}
\begin{table}
  \caption{Log of observations of 30 HAeBes. Column 1 gives the object
    name. Columns 2 and 3 are right ascension (RA) in the units of
    time (\fh \, \fm \, \fs) and declination (DEC) in the units of
    angle (\fdg \, \farcm \, \farcs) respectively. Column 4 lists the
    observation dates. Columns 5--7 present the exposure times for
    each grating. The signal-to-noise ratios for each grating are
    given in columns 8--10. These were determined using a
    20\,\AA\ wide line free spectral regions centred at the wavelength
    4200, 6050 and 8800\,\AA\ for R1200B, R1200Y and R1200R grating
    respectively.  Spectral types and photometry are listed along with
    references in columns 11--16.}
  \label{tab:log}
\begin{tabular}{lcclcccccccccccc}
\hline
Name&				RA& 				DEC& 				Obs Date& 	\multicolumn{3}{c}{Exposure Time (s)}&	\multicolumn{3}{c}{SNR}&	Spectral Type&	\multicolumn{5}{c}{Photometry (mag)}\\[3pt]
 & 					(J2000)& 		(J2000)& 			(June 2013)&		R1200B& 	R1200Y& 	R1200R&	B& 			Y& 				R&		&			{\em B}&	{\em V}&	{\em R$_c$}&	{\em R}&	{\em I$_c$}\\[3pt]
\hline
V594~Cas& 	 	00:43:18.3&	+61:54:40.1& 	21; 23; 24& 		600& 		600&			600&			152&			160&				50&		B8$^{1}$&		11.08&		10.51&		10.08&				-&				9.55$^{13}$\\[3pt]
PDS~144&		15:49:15.3&	-26:00:54.8&		22; 24&				-&				900&			900&			-&				97&				83&		A5 V$^{2}$&		13.28&		12.79&		12.49&				-&				12.16$^{2}$\\[3pt]
HD~141569&		15:49:57.8&	-03:55:16.3&		20; 22; 24&		60&			60&			60&			302&			385&				194&		A0 Ve$^{3}$&		7.23&		7.13&		7.08&				-&				7.02$^{2}$\\[3pt]
HD~142666&		15:56:40.0&	-22:01:40.0&		20; 22; 24&		90; 180&	90; 150&	150&			19&			147&				47&		A8 Ve$^{3}$&		9.17&		8.67&		8.35&				-&				8.01$^{2}$\\[3pt]
HD~145718&	 	16:13:11.6&	-22:29:06.7&		21; 23; 25&		90&			180&			180&			30&			160&				74&		A8 IV$^{2}$&		9.62&		9.10&		8.79&				-&				8.45$^{2}$\\[3pt]
HD~150193&		16:40:17.9&	-23:53:45.2&		21; 23; 24; 25&	90&			300&			120; 300&	54&			258&				96&		A2 IVe$^{3}$&		9.33&		8.80&		8.41&				-&				7.93$^{14}$\\[3pt]
PDS~469&		17:50:58.1&	-14:16:11.8&		23; 25&				-&				1200&		1200&		-&				127&				62&		A0$^{2}$&		13.33&		12.77&		12.39&				-&				11.95$^{2}$\\[3pt]
HD~163296&		17:56:21.3&	-21:57:21.9&		20; 23; 24; 25&	60&			60&			60&			60&			270&				125&		A1 Vep$^{3}$&	6.94&		6.83&		6.77&				-&				6.73$^{14}$\\[3pt]
MWC~297&		18:27:39.5&	-03:49:52.1&		21; 22; 24; 25&	900&			10; 60&		1200; 600&	22&		16&				102&		B0$^{2}$&		14.28&		12.03&		10.19&				-&				8.80$^{14}$\\[3pt]
VV~Ser&			18:28:47.9&	+00:08:39.8&		21; 23; 25&		900&			600&			600&			91&			180&				113&		B6$^{4}$&		12.82&		11.81&		11.10&				-&				10.31$^{13}$\\[3pt]
MWC~300&		18:29:25.7&	-06:04:37.3&		21; 23; 25&		900&			600; 60&	600&			48&			29&				19&		B1 Ia+[e]$^{5}$&		12.81&		11.78&		11.09&				-&				10.58$^{14}$\\[3pt]
AS~310&			18:33:21.3&	-04:58:04.8&		20; 22; 24&		40; 900&	1200&		1200&		35&			144&				128&		B1e$^{4}$&		13.56&		12.59&		11.93&				-&				11.20$^{14}$\\[3pt]
PDS~543&		18:48:00.7&	+02:54:17.1&		21; 22; 24&		900&			1200&		1200&		39&			153&				149&		B1$^{2}$&		14.56&		12.52&		11.25&						-&		9.96$^{2}$\\[3pt]
HD~179218&		19:11:11.3&	+15:47:15.6&		21; 23; 25&		60&			120&			120&			140&			311&				166&		A0 IVe$^{3}$&		7.47&		7.39&		7.34&				-&				7.29$^{2}$\\[3pt]
HD~190073&		20:03:02.5&	+05:44:16.7&		21; 23; 25&		60&			120&			120; 300&	65&			205&				90&		A0 IVp+sh$^{2}$&		7.86&		7.73&		-&						7.66$^{15}$&		-\\[3pt]
V1685~Cyg&		20:20:28.2&	+41:21:51.6&		20; 22; 24; 25&	600&			600; 60&	600; 60&	120&			114&				82&		B2 Ve$^{3}$&		11.48&		10.69&		-&						9.63$^{16}$&		-\\[3pt]
LkHA~134&		20:48:04.8&	+43:47:25.8&		20; 22; 24&		600&			600&			600&			79&			192&				95&		B8$^{6}$&		12.02&		11.35&		-&						10.56$^{16}$&		-\\[3pt]
HD~200775&		21:01:36.9&	+68:09:47.8&		20; 22; 24; 25&	60&			60; 30&		30; 60&		161&			192&				50&		B3$^{4}$&		7.77&		7.37&		-&						6.84$^{16}$&		-\\[3pt]
LkHA~324&		21:03:54.2&	+50:15:10.0&		21; 23; 25&		900&			1200&		1200&		52&			166&				123&		B8$^{4}$&		13.71&		12.56&		11.85&				-&				11.09$^{13}$\\[3pt]
HD~203024&		21:16:03.0&	+68:54:52.1&		21; 23; 24; 25&	120&			180&			180&			56&			187&				65&		A5 V$^{3}$&		9.24&		9.01&		9.02&						-&		8.62$^{17}$\\[3pt]
V645~Cyg&		21:39:58.3&	+50:14:20.9&		21; 23; 25&		900&			1200&		1200&		20&			86&				152&		O8.5$^{7}$&		14.55&		13.47&		-&						12.28$^{16}$&		-\\[3pt]
V361~Cep&		21:42:50.2&	+66:06:35.1&		20; 22; 24; 25&	300&			300&			300&			146&			219&				84&		B4$^{4}$&		10.65&		10.21&		9.89&				-&				9.50$^{13}$\\[3pt]
V373~Cep&		21:43:06.8&	+66:06:54.2&		20; 22; 24&		900&			1200&		1200&		96&			143&				103&		B8$^{8}$&		13.22&		12.33&		11.7&				-&				11.01$^{13}$\\[3pt]
V1578~Cyg&		21:52:34.1&	+47:13:43.6&		21; 23; 24; 25&	300&			300; 600&	300&			86&			225&				191&		A0$^{4}$&		10.53&		10.13&		-&						9.67$^{16}$&		-\\[3pt]
LkHA~257&		21:54:18.8&	+47:12:09.7&		20; 22; 23; 24&	900&			1200&		1200&		54&			179&				129&		A2e$^{9}$&		14.00&		13.29&		12.86&				-&				12.39$^{17}$\\[3pt]
SV~Cep&			22:21:33.2&	+73:40:27.1&		20; 22; 24&		240&			300&			300&			87&			188&				112&		A2 IVe$^{3}$&		11.37&		10.98&		-&				10.58$^{16}$&				-\\[3pt]
V375~Lac&		22:34:41.0&	+40:40:04.5&		23; 25&				-&				1200&		1200&		-&				109&				71&		A7 Ve$^{10}$&		14.22&		13.38&		12.84&				-&				12.25$^{13}$\\[3pt]
HD~216629&		22:53:15.6&	+62:08:45.0&		20; 22; 24&		90&			180&			180&			91&			251&				110&		B3 IVe+A3$^{11}$&	10.01&		9.28&		-&						8.53$^{16}$&		-\\[3pt]
V374~Cep&		23:05:07.5&	+62:15:36.5&		20; 22; 24&		600&			450&			450&			121&			173&				138&		B5 Vep$^{12}$&		11.14&		10.28&		9.71&				-&				9.12$^{17}$\\[3pt]
V628~Cas&		23:17:25.6&	+60:50:43.4&		21; 23; 24; 25&	900&			600; 60&	600&			26&			130&				149&		B0eq$^{1}$&		12.63&		11.37&		10.38&				-&				9.37$^{13}$\\[3pt]
\hline
\end{tabular}
\\[3pt]
{\bf References.} $^{(1)}$ \citet{Hillenbrand1992}; $^{(2)}$ \citet{Vieira2003}; $^{(3)}$ \citet{Mora2001}; $^{(4)}$ \citet{Hernandez2004}; $^{(5)}$ \citet{Wolf1985}; $^{(6)}$ \citet{Herbig1958}; $^{(7)}$ \citet{Clarke2006}; $^{(8)}$ \citet{Dahm2015}; $^{(9)}$ \citet{Walker1959}; $^{(10)}$ \citet{Calvet1978}; $^{(11)}$ \citet{Skiff2014}; $^{(12)}$ \citet{Garrison1970}; $^{(13)}$ \citet{Fernandez1995}; $^{(14)}$ \citet{deWinter2001}; $^{(15)}$ \citet{Oudmaijer2001}; $^{(16)}$ \citet{Herbst1999}; $^{(17)}$ \citet{Zacharias2012}.
\end{table}
\end{landscape}

For a given $T_{\rm eff}$, normally, the higher the $\log(g)$, the
broader the Balmer profile. Therefore, the region of the wings of the
hydrogen profile near the continuum level can be used to obtain both
effective temperature and surface gravity as was mentioned earlier.
Unfortunately, there is a non-linear relationship between the surface
gravity and width of the Balmer profile for objects that have $T_{\rm
  eff} < 8000$\,K.  For this reason, a spectroscopic $\log(g)$ can not
be determined and the surface gravity is calculated instead using the
stellar mass and radius (see later).  For 3 objects (PDS~144S, PDS~469
and V375~Lac) no near UV blue spectra were taken during the
observations. Fortunately, a FEROS spectrum of PDS~469\footnote{Based
  on observations collected at the European Southern Observatory under
  ESO programmes 084.A-9016(A).} was found in the ESO Science Archive
Facility and was used to determine temperature and surface gravity.
In the case of PDS~144S and V375~Lac, visible spectra including $\rm
H\alpha$ profiles in combination with spectral types from the
literature were investigated in order to adopt their effective
temperatures.

  7 objects show extremely strong emission lines.
  These strong emissions have an influence on the wings on even the
  whole Balmer profiles.  Both $T_{\rm eff}$ and $\log(g)$ could
    therefore not be determined by this method.  Instead, for
    these objects, estimated temperatures from the literature are
    adopted. The stars for which this process is performed on are
  noted in the last column of Table~\ref{tab:all_stellar_parameters}.

\begin{figure}
\centering
\includegraphics[width=\columnwidth]{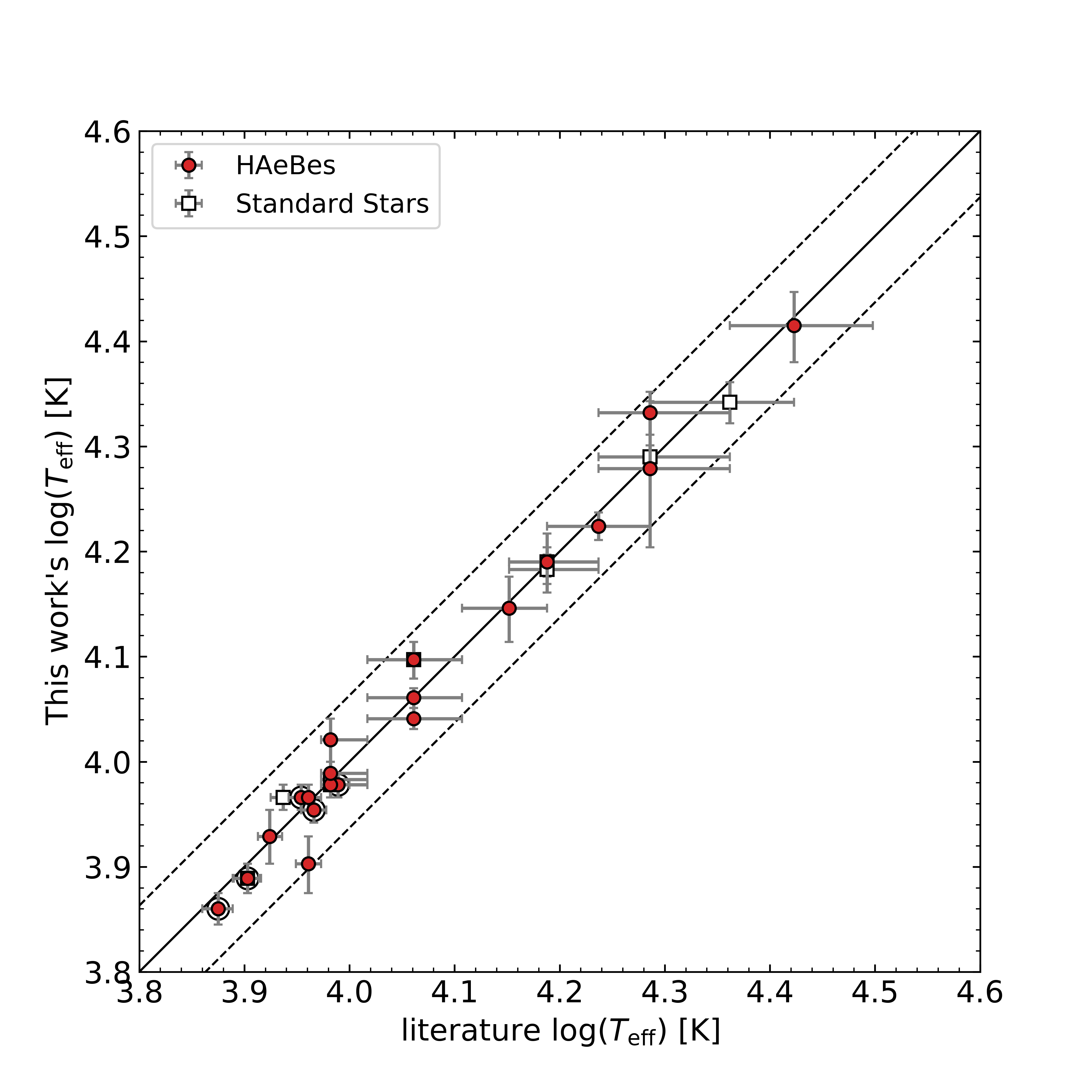}

\caption{The effective temperature derived in this work compared
  to temperatures derived from spectral types in the literature listed in
  Table~\ref{tab:log} and Table~\ref{tab:A1_standard_stars}. The
  spectral type was converted to temperature using the values provided
  in \protect\cite{Straizys1981}, and an uncertainty of a subclass of
  spectral type was assigned for the temperature.  HAeBes and standard
  stars are denoted in circle and square respectively.  Circles with
  larger circle around them indicate the objects that temperature from
 \protect\cite{Fairlamb2015} is used.  The standard deviation ($\sigma$) between both
  $\log(T_{\rm eff})$ is only 0.02. The solid line is the expected
  line of correlation and the dashed lines are $3\,\sigma$ deviation
  from the solid line.}
\label{fig:logT}
\end{figure}

There are 7 objects in this work that overlap with F15.  The
difference in temperatures of these objects is on average 180\,K,
  which is smaller than the step-size used by both studies.
Figure~\ref{fig:logT} compares the effective temperature derived in
this work (Table~\ref{tab:all_stellar_parameters}) with estimated
values from the literature. The good correlation would suggest
  that our method is reliable, the fact that it is applied to the
  entire sample ensures a homogeneous study.

\subsection{Visual extinction, distance and radius}

The second step is to use the synthetic BOSZ spectral energy
distribution and previous photometry results to determine the visual
extinction or reddening ($A_V$) by fitting the synthetic surface flux
density to observed photometry.  In the case of zero extinction or for
extinction corrected photometry, the flux density $f$ of the BOSZ
model and the observed fluxes differ by the ratio of distance to
the star and its radius ($D/R_*$).  The spectral energy distribution
grid of the BOSZ models are set up for the effective temperatures from the
previous step with a range of scaling factors $D/R_*$.  The value of
$\log(g)$ does not have a significant effect on the spectral energy
distribution shape.  Therefore, $\log(g)=4.0$ was adopted at this
stage.

\begin{figure}
\centering
\includegraphics[width=\columnwidth]{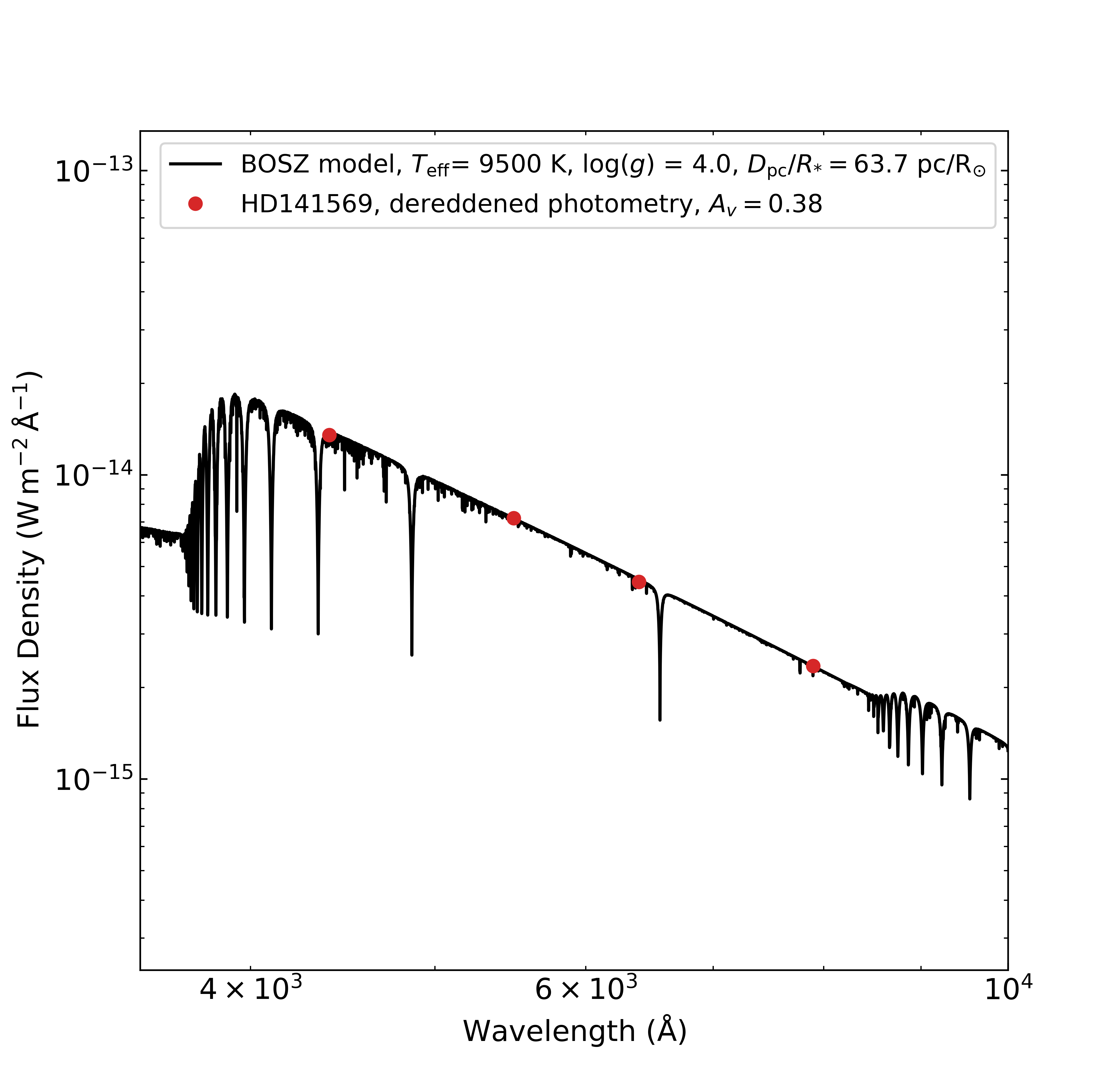}
\caption{
 The synthetic spectral energy distribution BOSZ model (black line) is
 fitted to the dereddened photometry (red point) of HD~141569. The
 observed {\em BVR$_c$I$_c$} photometry of HD~141569 were taken from
 \protect\cite{Vieira2003}.  A synthetic spectral energy distribution BOSZ
 model with $T_{\rm eff}=9500$\,K and $\log(g)=4.0$ is fitted to the
 dereddened photometry.  This provides the reddening
 $A_V=0.38^{+0.02}_{-0.03}$\,mag and the scaling factor $D/R_* =
 63.7^{+1.1}_{-0.7}$\,pc/$\rm R_{\sun}$.
}
\label{fig:HD141569_photometry}
\end{figure}

The observed photometry is dereddened using the extinction values
$A_\lambda /A_V$ from \cite{Cardelli1989} with the standard ratio of
total to selective extinction parameter $R_V=3.1$ and zero-magnitude
fluxes from \cite{Bessell1979}.  By varying $A_V$ in steps of
0.01\,magnitude, the best fit of the photometry and BOSZ models will
then yield the best fitting reddening values.  Only the
{\em BVRI} magnitudes are used, as the {\em U}-band and {\em JHKLM}
photometry are often affected by the Balmer continuum excess and
IR-excess emission respectively, i.e. they can not automatically be
used in the fitting.  All of the observed {\em BVR$_c$I$_c$} or {\em
  BVR} photometry from the literature are shown in
Table~\ref{tab:log}.  For 3 objects (HD~203024, LkHA~257 and V374~Cep) their
Sloan photometry needed to be converted to Johnson-Cousins
photometry. This was done using the transformation equations provided
by \cite{Smith2002}.  The uncertainties in the resulting $A_V$ and
$D/R_*$ are assigned to be at values that resulted twice of the minimum
chi-squared value.  Figure~\ref{fig:HD141569_photometry} demonstrates
an example of photometry fitting.

The scaling factor $D/R_*$ allows us to calculate stellar radius
$R_*$, provided the stellar distance $D$ is known. The {\em Gaia} DR2
catalogue provides astrometric parameters, such as positions, proper
motions and parallaxes for more than 1.3 billion targets including
most of the known HAeBes.  The distances to most of our targets were
determined by V18 using {\em Gaia} DR2 parallaxes. 
Re-normalised unit weight error (RUWE) is used to select sources with good astrometry. 
We adopted RUWE $<1.4$ as a criterion for good parallaxes (see {\em Gaia} Data Release 2 document).
7 stars have low quality parallaxes, and for 2 stars no parallaxes are presented in 
the {\em Gaia} archive. These are noted in the final column of
Table~\ref{tab:all_stellar_parameters}. In total, stellar radii could
be determined for 26 out of the 30 targets. 

\subsection{Stellar luminosity, mass and age}

Using the stellar radius $R_*$ and the effective temperature $T_{\rm
  eff}$, the luminosity $L_*$ can be determined from the
Stefan-Boltzmann law.  The next step is to estimate the mass and age
of the HAeBes using isochrones.
Stellar isochrones of
\cite{Marigo2017} from 0.01 to 100\,Myr are used in order to extract a
mass and an age of the target from the luminosity-temperature
Hertzsprung-Russell diagram. A metallicity $Z=0.01$ and helium mass
fraction $Y=0.267$ are chosen, because these values are close to solar
values.

After each star is placed on the HR diagram, the two closest points on
an isochrone are used to obtain the mass of the star by interpolating
between those points.  Uncertainties of mass and age are derived from
the error bars of the effective temperature $T_{\rm eff}$ and the
luminosity $L_*$ on the HR diagram.  All determined parameters of all
targets are presented in Table~\ref{tab:all_stellar_parameters}.  The
positions of the 21 HAeBes on the HR diagram are represented with red
symbols in Figure~\ref{fig:HR_all}
As mentioned above, the 9 targets that are not included in the plot have low quality
parallaxes or do not have parallaxes at all.
Since the 3 objects (HD~142666, HD~145718 and SV~Cep) have had $T_{\rm
  eff}<8000$\,K, their surface gravity cannot be determined from
fitting the spectra with stellar atmospheric models.  Instead, their
surface gravities were calculated from the stellar mass and radius
derived from the parallexes instead.  These are also noted in the
$\log(g)$ column of Table~\ref{tab:all_stellar_parameters}.

\begin{table*}
\caption{Determined stellar parameters. Columns 2--9 are effective temperature, surface gravity, visual extinction, distance, radius, luminosity, mass and age respectively. Distance $D$ in column 5 is obtained from \protect\cite{Vioque2018}. This table is available in electronic form at \url{https://cdsarc.unistra.fr/viz-bin/cat/J/MNRAS/493/234}.}
\label{tab:all_stellar_parameters}
\begin{tabular}{lcccccccc}
\hline
Name&	$T_{\rm eff}$&	$\log(g)$&			$A_V$&		$D$&		$R_*$&					$\log(L_*)$&			$M_*$&					Age\\[3pt]
&			(K)&					[cm\,s$^{-2}$]&	(mag)&		(pc)&		($\rm R_{\sun}$)&	[$\rm L_{\sun}$]&	($\rm M_{\sun}$)&	(Myr)\\
\hline
V594~Cas& $11500^{+250}_{-250}$& $3.70^{+0.10}_{-0.10}$& $2.27^{+0.18}_{-0.23}$& $569^{+16}_{-14}$& $3.86^{+0.52}_{-0.53}$& $2.37^{+0.15}_{-0.17}$& $3.51^{+0.45}_{-0.42}$& $1.29^{+0.53}_{-0.38}$\\[3pt]
PDS~144S& $7750^{+500}_{-500}$ $^{a}$& $4.00^{+0.30}_{-0.30}$& $1.02^{+0.07}_{-0.07}$& -$^{d}$& -& -& -& -\\[3pt]
HD~141569& $9500^{+250}_{-250}$& $4.20^{+0.10}_{-0.10}$& $0.38^{+0.02}_{-0.03}$& $110.63^{+0.91}_{-0.88}$& $1.74^{+0.03}_{-0.04}$& $1.34^{+0.06}_{-0.07}$& $2.06^{+0.02}_{-0.15}$& $5.89^{+1.87}_{-0.64}$\\[3pt]
HD~142666& $7250^{+250}_{-250}$& $4.00^{+0.10}_{-0.10}$ $^{c}$& $0.82^{+0.07}_{-0.08}$& $148.3^{+2}_{-1.9}$& $2.21^{+0.11}_{-0.12}$& $1.08^{+0.10}_{-0.11}$& $1.64^{+0.12}_{-0.11}$& $7.76^{+1.79}_{-1.30}$\\[3pt]
HD~145718& $7750^{+250}_{-250}$& $4.20^{+0.10}_{-0.10}$ $^{c}$& $1.10^{+0.06}_{-0.06}$& $152.5^{+3.2}_{-3}$& $1.85^{+0.10}_{-0.10}$& $1.05^{+0.10}_{-0.10}$& $1.62^{+0.07}_{-0.03}$& $8.71^{+0.84}_{-1.12}$\\[3pt]
HD~150193& $9250^{+250}_{-250}$& $4.10^{+0.10}_{-0.10}$& $1.88^{+0.16}_{-0.20}$& $150.8^{+2.7}_{-2.5}$& $2.34^{+0.26}_{-0.27}$& $1.56^{+0.14}_{-0.15}$& $2.12^{+0.21}_{-0.12}$& $4.57^{+0.93}_{-1.02}$\\[3pt]
PDS~469& $9500^{+750}_{-750}$ $^{a}$& $3.80^{+0.30}_{-0.30}$& $1.94^{+0.12}_{-0.13}$& -$^{d}$& -& -& -& -\\[3pt]
HD~163296& $9000^{+250}_{-250}$& $4.10^{+0.10}_{-0.10}$& $0.29^{+0.01}_{-0.02}$& $101.5^{+2}_{-1.9}$& $1.87^{+0.05}_{-0.05}$& $1.31^{+0.07}_{-0.07}$& $1.95^{+0.07}_{-0.07}$& $6.03^{+0.28}_{-0.27}$\\[3pt]
MWC~297& $24000^{+2000}_{-2000}$ $^{b}$& $4.00^{+0.10}_{-0.10}$& $7.87^{+0.41}_{-0.64}$& $375^{+22}_{-18}$& $9.28^{+3.12}_{-3.04}$& $4.41^{+0.39}_{-0.50}$& $14.53^{+6.11}_{-4.84}$& $0.04^{+0.07}_{-0.02}$\\[3pt]
VV~Ser& $14000^{+1000}_{-1000}$& $4.30^{+0.30}_{-0.30}$& $3.74^{+0.22}_{-0.27}$& -$^{d}$& -& -& -& -\\[3pt]
MWC~300& $23000^{+2000}_{-2000}$ $^{b}$& $3.00^{+0.20}_{-0.20}$& $3.85^{+0.21}_{-0.28}$& $1400^{+250}_{-160}$& $6.02^{+1.96}_{-1.44}$& $3.96^{+0.39}_{-0.39}$& $10.09^{+3.76}_{-1.89}$& $0.09^{+0.09}_{-0.05}$\\[3pt]
AS~310& $26000^{+2000}_{-2000}$& $4.40^{+0.35}_{-0.35}$& $3.86^{+0.20}_{-0.24}$& $2110^{+350}_{-240}$& $5.70^{+1.71}_{-1.27}$& $4.13^{+0.36}_{-0.36}$& $11.60^{+3.92}_{-2.18}$& $0.07^{+0.08}_{-0.04}$\\[3pt]
PDS~543& $28500^{+2500}_{-2500}$ $^{b}$& $4.00^{+0.10}_{-0.10}$& $7.11^{+0.28}_{-0.40}$& $1410^{+240}_{-160}$& $16.24^{+5.99}_{-4.57}$& $5.19^{+0.42}_{-0.45}$& $30.02^{+18.28}_{-10.85}$& $0.01^{+0.01}_{-0.01}$\\[3pt]
HD~179218& $9500^{+250}_{-250}$& $3.95^{+0.10}_{-0.10}$& $0.33^{+0.02}_{-0.02}$& $266^{+5.6}_{-5.2}$& $3.62^{+0.12}_{-0.11}$& $1.98^{+0.07}_{-0.07}$& $2.86^{+0.16}_{-0.20}$& $2.04^{+0.47}_{-0.26}$\\[3pt]
HD~190073& $9750^{+250}_{-250}$& $3.50^{+0.10}_{-0.10}$& $0.20^{+0.04}_{-0.04}$& $870^{+100}_{-70}$& $9.23^{+1.28}_{-0.94}$& $2.84^{+0.16}_{-0.14}$& $5.62^{+0.78}_{-0.65}$& $0.28^{+0.14}_{-0.10}$\\[3pt]
V1685~Cyg& $23000^{+4000}_{-4000}$ $^{b}$& $4.06^{+0.10}_{-0.10}$& $3.33^{+0.34}_{-0.51}$& $910^{+46}_{-39}$& $5.48^{+1.51}_{-1.51}$& $3.88^{+0.49}_{-0.61}$& $9.53^{+4.57}_{-2.55}$& $0.11^{+0.27}_{-0.07}$\\[3pt]
LkHA~134& $11000^{+250}_{-250}$& $4.00^{+0.10}_{-0.10}$& $2.44^{+0.19}_{-0.24}$& $843^{+36}_{-31}$& $4.53^{+0.71}_{-0.70}$& $2.43^{+0.17}_{-0.19}$& $3.77^{+0.56}_{-0.56}$& $1.02^{+0.60}_{-0.34}$\\[3pt]
HD~200775& $19000^{+3000}_{-3000}$& $4.27^{+0.25}_{-0.25}$& $1.85^{+0.15}_{-0.17}$& -$^{d}$& -& -& -& -\\[3pt]
LkHA~324& $12500^{+500}_{-500}$& $4.00^{+0.10}_{-0.10}$& $3.94^{+0.14}_{-0.16}$& $605^{+16}_{-14}$& $3.15^{+0.34}_{-0.33}$& $2.34^{+0.16}_{-0.17}$& $3.36^{+0.41}_{-0.30}$& $1.51^{+0.49}_{-0.41}$\\[3pt]
HD~203024& $8500^{+500}_{-500}$& $3.83^{+0.29}_{-0.29}$& $0.52^{+0.24}_{-0.31}$& -$^{e}$& -& -& -& -\\[3pt]
V645~Cyg& $30000^{+7000}_{-7000}$ $^{b}$& $3.75^{+0.35}_{-0.35}$& $4.21^{+0.29}_{-0.40}$& -$^{d}$& -& -& -& -\\[3pt]
V361~Cep& $16750^{+500}_{-500}$& $4.00^{+0.10}_{-0.10}$& $1.97^{+0.14}_{-0.15}$& $893^{+35}_{-31}$& $4.34^{+0.52}_{-0.48}$& $3.12^{+0.15}_{-0.15}$& $5.56^{+0.65}_{-0.57}$& $0.40^{+0.16}_{-0.11}$\\[3pt]
V373~Cep& $11500^{+1250}_{-1250}$ $^{b}$& $3.50^{+0.50}_{-0.50}$& $3.24^{+0.20}_{-0.23}$& -$^{d}$& -& -& -& -\\[3pt]
V1578~Cyg& $10500^{+500}_{-500}$& $3.80^{+0.20}_{-0.20}$& $1.46^{+0.08}_{-0.08}$& $773^{+30}_{-27}$& $4.77^{+0.41}_{-0.37}$& $2.39^{+0.15}_{-0.15}$& $3.74^{+0.45}_{-0.41}$& $1.02^{+0.39}_{-0.28}$\\[3pt]
LkHA~257& $9250^{+250}_{-250}$& $4.05^{+0.10}_{-0.10}$& $2.28^{+0.07}_{-0.08}$& $794^{+18}_{-16}$& $1.84^{+0.11}_{-0.11}$& $1.35^{+0.10}_{-0.10}$& $1.98^{+0.06}_{-0.04}$& $5.76^{+1.32}_{-0.51}$\\[3pt]
SV~Cep& $8000^{+500}_{-500}$& $4.37^{+0.11}_{-0.11}$ $^{c}$& $0.78^{+0.04}_{-0.05}$& $344.3^{+4}_{-3.8}$& $1.48^{+0.05}_{-0.06}$& $0.91^{+0.14}_{-0.15}$& $1.63^{+0.04}_{-0.14}$& $11.00^{+10.40}_{-2.49}$\\[3pt]
V375~Lac& $8000^{+750}_{-750}$ $^{a}$& $4.30^{+0.10}_{-0.10}$& $2.31^{+0.14}_{-0.16}$& -$^{e}$& -& -& -& -\\[3pt]
HD~216629& $21500^{+1000}_{-1000}$& $4.00^{+0.10}_{-0.10}$& $2.86^{+0.13}_{-0.15}$& $805^{+31}_{-27}$& $7.59^{+0.94}_{-0.83}$& $4.04^{+0.18}_{-0.18}$& $10.83^{+1.71}_{-1.54}$& $0.07^{+0.04}_{-0.02}$\\[3pt]
V374~Cep& $15500^{+1000}_{-1000}$& $3.50^{+0.10}_{-0.10}$& $3.20^{+0.15}_{-0.18}$& $872^{+40}_{-35}$& $7.72^{+1.05}_{-0.98}$& $3.49^{+0.22}_{-0.23}$& $7.50^{+1.46}_{-1.28}$& $0.16^{+0.12}_{-0.06}$\\[3pt]
V628~Cas& $31000^{+5000}_{-5000}$ $^{b}$& $4.00^{+0.10}_{-0.10}$& $5.05^{+0.36}_{-0.55}$& -$^{d}$& -& -& -& -\\[3pt]
\hline
\end{tabular}
\\
\begin{flushleft}
{\bf Notes.}
$^{(a)}$ Stars for which NUV-B spectra were not obtained.
$^{(b)}$ Stars which display extremely strong emission lines. 
$^{(c)}$ Stars for which parallactic $\log(g)$ is used.
$^{(d)}$ Stars which have low quality parallaxes in the {\em Gaia} DR2 Catalogue (see the text for discussion).
$^{(e)}$ Stars which do not have parallaxes in the {\em Gaia} DR2 Catalogue.
\end{flushleft}
\end{table*}

\subsection{Extending the sample with HAeBes in the southern hemisphere}

The original driver for the INT observations presented here was to
extend the spectroscopic analysis of the, mostly southern, sample of
F15 to the northern hemisphere.  However,
F15  determined spectroscopic distances using their spectral
derived values of the surface gravity or literature values for the distances to their
91 HAeBes. The availability of
{\em Gaia}-derived distances warrants a re-determination of the stellar
parameters. To this end we use the distances from the {\em Gaia} DR2
parallax (V18). In parallel, we re-assessed the
extinction values using the same photometric bands {\em BVR$_c$I$_c$} as above
to ensure consistency in the determination of the extinction.  Most of
the photometry used for the photometry fitting can be found in
\citet[table A1]{Fairlamb2015}. We collated Sloan photometry from
\citet{Zacharias2012} for 3 objects (HD~290500, HT~CMa and HD~142527) and one object, HD~95881, from {\em APASS}\footnote{https://www.aavso.org/apass-dr10-download} DR10 and  converted these to the Johnson-Cousins system.
Moreover, we also used the brightest {\em V}-band magnitude of Johnson {\em BVR} photometry for 2 objects, HD~250550 and KK~Oph, and Johnson {\em BVRI} photometry for Z~CMa from \cite{Herbst1999}.
The redetermined stellar parameters
for the 91 HAeBes from F15 are listed in
Table~\ref{tab:C1_Fairlamb_para}
(See Appendix C in the online version of this paper).
For the sample as a whole, the luminosities resulting from the revised
distances and extinctions are on average similar to the original F15
values, however the contribution to the scatter around the mean
differences is dominated by the new distances.  We thus conclude that
the improvement in distance values dominates that of the extinction
when arriving at the final luminosities.

\begin{figure*}
\centering
\includegraphics[width=12cm]{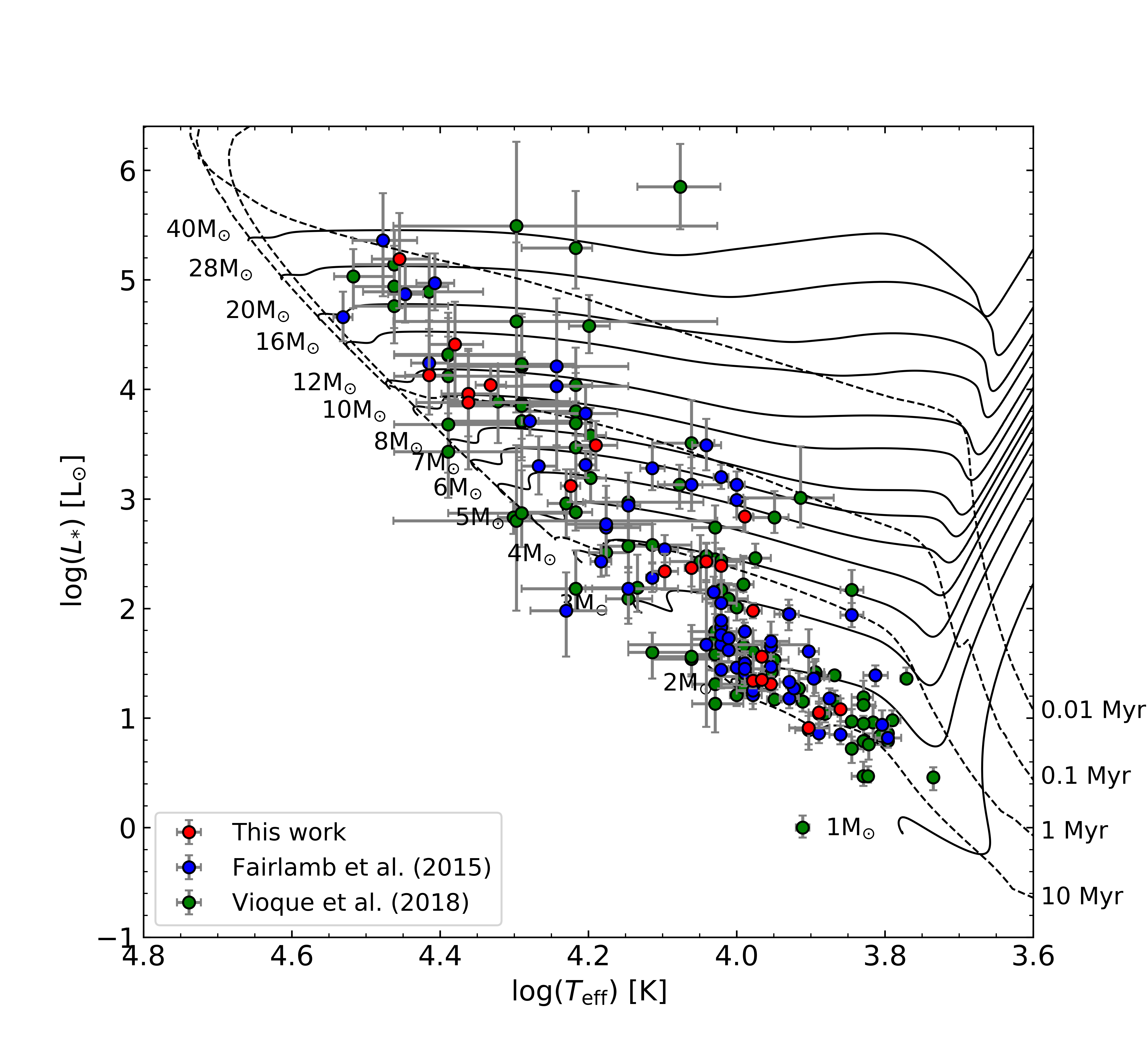}
\caption{The placement of all 184 HAeBes in the HR diagram. The red
  circles are this work sample of 21 HAeBes, whereas 62 HAeBes in the
  sample of \protect\cite{Fairlamb2015} are denoted as the blue circles and 101 HAeBes in \protect\cite{Vioque2018} are shown as the green circles. All of these objects satisfy the condition RUWE $<1.4$. The PMS tracks with initial mass
  from 1 to 40\,$\rm M_{\sun}$ \protect\citep{Bressan2012,Tang2014}
  are plotted as solid lines and isochrones of 0.01, 0.1, 1 and
  10\,Myr \protect\citep{Marigo2017} are  plotted as dashed
  lines.}
\label{fig:HR_all}
\end{figure*}

We now have the largest spectroscopic sample of HAeBes and the
homogeneously determined stellar parameters including mass accretion
rates, which will be determined in the next section.  7 targets in
F15's sample which are also in this work's sample were left out,
leaving 84 HAeBes in their sample.
Unfortunately, 22 out of the 84 objects (not including PDS~144S) have
low quality or do not have a {\em Gaia} DR2 parallax.  These objects
cannot be placed on the HR diagram at this stage. The final sample
contains 83 Herbig Ae/Be objects. Figure~\ref{fig:HR_all}
demonstrates all these HAeBes ($21+62$) placed in the HR diagram.
Many targets are gathered around $2\,\rm M_{\sun}$ and few targets are
located at high-mass tracks.  This can be understood by the initial
mass function IMF \citep{Salpeter1955}.  Low-mass stars are more
common than high-mass stars.  Moreover, low-mass stars evolve slower
across the HR diagram.

\subsection{Combining with additional HAeBes with {\em Gaia} DR2 data from V18}
           
In order to expand the sample further, the rest of 144/252 objects in
the sample of V18 were investigated. There are 101 objects which satisfy the condition RUWE $<1.4$. 
They are also included in Figure~\ref{fig:HR_all}. H$\alpha$ equivalent widths are
provided for most objects in \citet[their table 1 and table
  2]{Vioque2018} with references.  Spectra for 14 objects without
H$\alpha$ data in the V18 sample were found in the ESO Science Archive
Facility. These were downloaded and their H$\alpha$ EWs were
  measured.  The stellar radii and surface gravities were calculated
  using the temperatures, luminosities and masses provided in V18.
The intrinsic equivalent widths were measured from BOSZ models for the
temperature provided in \citet[table 1 and table 2]{Vioque2018} and
the calculated surface gravity. This will enable us to compute mass
accretion rates for a further 85 objects below.
Table~\ref{tab:D1_Vioque_Mdot} presents all of the emission lines
measurements and determined accretion rates of the V18 sample (See Appendix D in the online version of this paper).

\section{Accretion rate determination}
We now will determine the mass accretion rates of the combined
  sample of Herbig Ae/Be stars.  The underlying assumption of the
methodology is that the stars accrete material according to the MA
paradigm, which, as demonstrated by \cite{Muzerolle2004}, can explain
the observed excess fluxes in Herbig Ae/Be stars. The infalling
  material shocking the photosphere gives rise to UV-excess emission,
  whose measurement can be converted into an accretion luminosity.
Knowledge of the stellar parameters can then return a value of the
mass accretion rate.

F15 derived the accretion rates for their sample from the UV
  excess following the methodology of \cite{Muzerolle2004} and
  extending the work of \citet{Mendigutia2011b}.  Unfortunately, as
  mentioned before, our current observational set-up did not allow for
  such accurate measurements. However, as for example pointed out by
\cite{Mendigutia2011b}, the accretion luminosity  correlates
with the line strengths of various types of emission lines.  They will
therefore also correlate with the mass accretion rate (see also
\citealt{Fairlamb2017}). As such, the line strengths do provide an
  observationally cheap manner to derive the accretion rate of an
  object without having to resort to the rather delicate and
  time-consuming process of measuring the UV excess. We should note
  that despite this, it is not clear whether the observed correlation
  is intrinsically due to accretion or some other effect
  \citep{Mendigutia2015a}.

We have chosen the H$\alpha$ line for the accretion luminosity
determination, as these lines are strongest lines present in the
spectra. To arrive at a measurement of the total line emission, we
need to take account of the fact that the underlying line absorption
is filled in with emission. Therefore, the intrinsic absorption
equivalent width $EW_{\rm int}$ needs to be subtracted from the
observed equivalent width $EW_{\rm obs}$. The intrinsic EW is
  measured from the synthetic spectra corresponding to the effective temperature
  and surface gravity determined earlier.

The line luminosity $L_{\rm line}$ was calculated using the
unreddened stellar flux at the wavelength of H$\alpha$ and distance to
the star.  The relationship between accretion luminosity and line
luminosity goes as (cf. e.g. \citealt{Mendigutia2011b}).

\begin{equation} \label{eq:Lacc_Lline}
\log\left( \frac{L_{\rm 
acc}}{\rm L_{\sun}} \right)=A+B \times \log\left( \frac{L_{\rm line}}{\rm L_{\sun}} \right),
\end{equation}

where $A$ and $B$ are constants corresponding to the intercept and the
gradient of the relation between $\log(L_{\rm acc}/\rm L_{\sun})$ and
$\log(L_{\rm line}/\rm L_{\sun})$ respectively.  This relationship
has, most recently, been determined for 32 accretion diagnostic
emission lines by F17.  For the case of $\rm H\alpha$,
the constants are $A=2.09\pm0.06$ and $B=1.00\pm0.05$.  The mass accretion
rate is then determined from the accretion luminosity, stellar radius and
stellar mass by Equation \ref{eq:Macc}.

\begin{equation} \label{eq:Macc}
\dot M_{\rm acc}=\frac{L_{\rm acc}R_*}{GM_*}
\end{equation}

Table \ref{tab:EW_Mdot} summarises the EW measurements for the $\rm
H\alpha$ line and mass accretion rates in all HAeBes.

The final set of 78 mass accretion rates (21+57) based on the INT and
X-Shooter data is the largest, and arguably the best such
  collection to date as they were determined in a homogeneous and
  consistent manner.  To arrive at the largest sample
possible however, we expand the sample using data provided in the
comprehensive study by V18.  Adding 85 objects from V18 results in a
large sample of 163 Herbig Ae/Be stars with mass accretion
rates\footnote{When drafting this manuscript, a paper by
  \cite{Arun2019} was published reporting on the accretion rates using
       {\em Gaia} data of a somewhat
       smaller sample than here. Notable differences are that we use
       homogeneously determined stellar parameters for a large
       fraction of the sample and the fact that their sample is not
       selected for parallax quality and therefore includes faulty
       {\em Gaia} parallax measurements affecting the derived
       distances. It would also appear that their determination of the
       H$\alpha$ line luminosity does not account for underlying
       absorption, which will affect especially the weakly emitting
       sources. As the authors do not provide all relevant datatables, it proved
       hard to investigate and assess further differences.}

\section{Analysis}

We have obtained new spectroscopic data of 30 northern Herbig Ae/Be
stars which are used to determine their stellar parameters, extinction
and H$\alpha$ emission line fluxes. Combined with {\em Gaia} DR2
parallaxes, this led to the determination of the mass accretion rates
for 21 of these objects. This dataset complements the large southern
sample of F15, for which we re-determined stellar parameters
  using the {\em Gaia} parallaxes and extinctions in the same manner
  as for the northern sample. This full sample contains 78 objects
with homogeneously determined values. To this we add 85 objects from
the V18 study for which literature values have been adapted and new
H$\alpha$ line measurements have been added using data from
archives. This led to a total of 163 objects for which we have stellar
parameters, distances, H$\alpha$ emission line fluxes and mass
accretion rates derived from these using the MA paradigm available. In
the following we will investigate the dependence of the MA-derived
accretion rate on stellar mass, and find that there is a break in
properties around 4M$_{\odot}$.

\subsection{Accretion luminosity as a function of stellar luminosity}

Before addressing the mass accretion rate, let us first discuss the
accretion luminosity, as that is less dependent on the stellar
parameters. In Figure~\ref{fig:LLacc_wfv} we show the accretion luminosity
versus the stellar luminosity for the entire sample. The accretion
luminosity increases monotonically with stellar luminosity, which
confirms earlier reports (F15, \citealt{Mendigutia2011b}). However,
the sample under consideration is much larger than previously.

It would appear that the slope decreases, while the scatter in the
relationship increases, with mass. To investigate whether there is a
significant difference in accretion luminosities between high- and low
mass objects, and if so, to determine the mass where there is a
turn-over, we split the data into a low mass and a high mass sample,
with a varying turnover mass.  We then fitted a straight line to the
data from the lowest mass to this intermediate mass, and a straight line
from the intermediate mass to the largest mass. The intermediate mass
was varied from the second smallest to the second largest mass.

This resulted in values of the slopes and their statistical
uncertainty for a range of masses. When taking the difference in
slopes and expressing this in terms of the respective uncertainties in
the fit, we arrive at a statistical assessment whether the low- and
high mass samples have a different slope. This approach takes
  into account the issue that the absolute value of the
  difference-in-slopes may not be a reliable indicator of the
  turn-over point. This is because each slope can vary depending on
  both the number and spread of data points used. For example, at
both the high and low mass ends, the slopes will have a large
uncertainty simply because of small number statistics.

We quantify the difference in slopes by combining the uncertainties on
both low and high mass gradients, $\sigma$, and compare this to the
difference in slopes, $\Delta$(slope). This is shown in the top panel
of Figure~\ref{fig:LLacc_wfv}. The $\Delta$(slope) reaches its
maximum significance of 4$\sigma$ for a luminosity of 194 L$_{\odot}$,
which was determined by a triple-Gaussian fit to the curve. For the lower
mass HAeBes, the linear best fit provides the empirical calibration of 
$\log\left( \frac{L_{\rm acc}}{\rm L_{\sun}} \right)=(-0.87 \pm 0.11)+(1.03 \pm 0.08) \times \log\left( \frac{L_{*}}{\rm L_{\sun}} \right)$.  
This is in agreement with the best fit for
low-mass stars in the work of \cite{Mendigutia2011b} ($L_{\rm acc}
\propto L^{1.2}_*$). For the higher mass HAeBes, the best fit provides the expression of  $\log\left( \frac{L_{\rm acc}}{\rm L_{\sun}} \right)=(0.19 \pm 0.27)+(0.60 \pm 0.08) \times \log\left( \frac{L_{*}}{\rm L_{\sun}} \right)$.

\subsection{Mass accretion rate as a function of stellar mass}

One of the main questions regarding the formation of massive stars is
at which mass the mass accretion mechanism changes from magnetically
controlled accretion to another mechanism, which could be direct
accretion from the disk onto the star. Above we saw that there is
  a difference in the accretion luminosity to stellar luminosity
  relationships for different masses. So, we now move to look at the
  mass accretion rates and investigate various subsamples
  individually.

\begin{landscape}
  \begin{table}
    \caption{The equivalent width measurements and accretion rates. Column
  2 provides the $\rm H\alpha$ emission line profile classification
  scheme according to \protect\cite{Reipurth1996}: single-peaked (I);
  double-peaked and the secondary peak rises above half strength of
  the primary peak (II); double-peaked and the secondary peak rises
  below half strength of the primary peak (III), when the secondary
  peak is located blueward or redward of the primary peak, class II
  and III are label with B or R respectively; regular P-Cygni profile
  (IV B); and inverse P-Cygni profile (IV R). 
  The classifications are based on the emission lines in Figure~\ref{fig:B1_EW} corrected for absorption. 
  Columns 3--10 present
  observed equivalent width, intrinsic equivalent width, corrected
  equivalent width, continuum flux density at central wavelength of the $\rm
  H\alpha$ profile, line flux, line luminosity, accretion luminosity
  and mass accretion rate respectively. This table is available in electronic form at \url{https://cdsarc.unistra.fr/viz-bin/cat/J/MNRAS/493/234}.}
\label{tab:EW_Mdot}
\begin{tabular}{lccccccccc}
\hline
Name&	$\rm H\alpha$&	$EW_{\rm obs}$&	$EW_{\rm int}$&	$EW_{\rm cor}$&	$F_{\lambda}$& 	$F_{\rm line}$&	$\log(L_{\rm line})$&	$\log(L_{\rm acc})$&	$\log(\dot M_{\rm acc})$\\[3pt]
&	profile&	(\AA)&	(\AA)&	(\AA)&	(W\,m$^{-2}$\,\AA$^{-1}$)&	(W\,m$^{-2}$)&		[$\rm L_{\sun}$]&	[$\rm L_{\sun}$]&	[$\rm M_{\sun}$\,$\rm yr^{-1}$]\\
\hline
V594~Cas& IV B& $-81.79\pm1.23$& $8.21\pm0.19$& $-90.00\pm1.24$& $1.11\times10^{-15}$& $(9.96\pm0.14)\times10^{-14}$& $0.00^{+0.03}_{-0.03}$& $2.09^{+0.09}_{-0.09}$& $-5.36^{+0.09}_{-0.10}$\\[3pt]
PDS~144S& I& $-14.55\pm0.53$& $12.82\pm0.14$& $-27.37\pm0.55$& $4.70\times10^{-17}$& $(1.29\pm0.03)\times10^{-15}$& -& -& -\\[3pt]
HD~141569& II R& $6.38\pm0.01$& $14.80\pm0.26$& $-8.42\pm0.26$& $4.17\times10^{-15}$& $(3.51\pm0.11)\times10^{-14}$& $-1.87^{+0.02}_{-0.02}$& $0.22^{+0.17}_{-0.18}$& $-7.35^{+0.18}_{-0.15}$\\[3pt]
HD~142666& IV R& $2.46\pm0.08$& $11.01\pm0.17$& $-8.54\pm0.19$& $1.84\times10^{-15}$& $(1.57\pm0.04)\times10^{-14}$& $-1.97^{+0.02}_{-0.02}$& $0.12^{+0.18}_{-0.18}$& $-7.24^{+0.17}_{-0.17}$\\[3pt]
HD~145718& I& $10.83\pm0.42$& $14.18\pm0.25$& $-3.35\pm0.49$& $1.50\times10^{-15}$& $(5.03\pm0.73)\times10^{-15}$& $-2.44^{+0.08}_{-0.09}$& $-0.35^{+0.26}_{-0.27}$& $-7.79^{+0.26}_{-0.29}$\\[3pt]
HD~150193& III B& $-0.29\pm0.05$& $14.86\pm0.18$& $-15.15\pm0.19$& $3.85\times10^{-15}$& $(5.83\pm0.07)\times10^{-14}$& $-1.38^{+0.02}_{-0.02}$& $0.71^{+0.15}_{-0.15}$& $-6.75^{+0.15}_{-0.18}$\\[3pt]
PDS~469& I& $4.65\pm0.17$& $13.50\pm0.17$& $-8.85\pm0.24$& $1.02\times10^{-16}$& $(9.04\pm0.24)\times10^{-16}$& -& -& -\\[3pt]
HD~163296& I& $-11.04\pm0.87$& $15.48\pm0.13$& $-26.51\pm0.88$& $5.11\times10^{-15}$& $(1.35\pm0.05)\times10^{-13}$& $-1.36^{+0.03}_{-0.03}$& $0.73^{+0.16}_{-0.16}$& $-6.79^{+0.15}_{-0.16}$\\[3pt]
MWC~297& I& $-455.84\pm1.41$& $3.50\pm0.02$& $-459.34\pm1.41$& $4.81\times10^{-14}$& $(2.21\pm0.01)\times10^{-11}$& $1.99^{+0.05}_{-0.04}$& $4.08^{+0.21}_{-0.20}$& $-3.61^{+0.19}_{-0.20}$\\[3pt]
VV~Ser& II R& $-39.07\pm0.10$& $8.04\pm0.07$& $-47.11\pm0.12$& $1.31\times10^{-15}$& $(6.15\pm0.02)\times10^{-14}$& -& -& -\\[3pt]
MWC~300& III B& $-140.04\pm0.40$& $1.95\pm0.06$& $-141.99\pm0.40$& $1.35\times10^{-15}$& $(1.91\pm0.01)\times10^{-13}$& $1.07^{+0.14}_{-0.11}$& $3.16^{+0.26}_{-0.21}$& $-4.56^{+0.25}_{-0.24}$\\[3pt]
AS~310& I& $0.83\pm0.04$& $4.13\pm0.03$& $-3.29\pm0.05$& $6.60\times10^{-16}$& $(2.17\pm0.03)\times10^{-15}$& $-0.52^{+0.14}_{-0.11}$& $1.57^{+0.22}_{-0.20}$& $-6.23^{+0.21}_{-0.22}$\\[3pt]
PDS~543& I& $0.16\pm0.01$& $2.66\pm0.02$& $-2.50\pm0.02$& $1.42\times10^{-14}$& $(3.55\pm0.02)\times10^{-14}$& $0.34^{+0.14}_{-0.11}$& $2.43^{+0.22}_{-0.18}$& $-5.33^{+0.15}_{-0.13}$\\[3pt]
HD~179218& I& $-1.70\pm0.05$& $12.43\pm0.13$& $-14.14\pm0.14$& $3.12\times10^{-15}$& $(4.41\pm0.05)\times10^{-14}$& $-1.01^{+0.02}_{-0.02}$& $1.08^{+0.13}_{-0.13}$& $-6.31^{+0.12}_{-0.12}$\\[3pt]
HD~190073& IV B& $-24.79\pm0.50$& $10.60\pm0.19$& $-35.39\pm0.53$& $1.99\times10^{-15}$& $(7.05\pm0.11)\times10^{-14}$& $0.22^{+0.10}_{-0.08}$& $2.31^{+0.18}_{-0.15}$& $-4.97^{+0.18}_{-0.14}$\\[3pt]
V1685~Cyg& II B& $-107.70\pm0.67$& $3.63\pm0.03$& $-111.33\pm0.67$& $2.63\times10^{-15}$& $(2.93\pm0.02)\times10^{-13}$& $0.88^{+0.05}_{-0.04}$& $2.97^{+0.15}_{-0.14}$& $-4.77^{+0.09}_{-0.15}$\\[3pt]
LkHA~134& I& $-68.47\pm0.36$& $9.64\pm0.08$& $-78.11\pm0.37$& $6.45\times10^{-16}$& $(5.04\pm0.02)\times10^{-14}$& $0.05^{+0.04}_{-0.03}$& $2.14^{+0.10}_{-0.10}$& $-5.28^{+0.11}_{-0.10}$\\[3pt]
HD~200775& II R& $-66.71\pm0.23$& $5.21\pm0.06$& $-71.92\pm0.24$& $1.36\times10^{-14}$& $(9.76\pm0.04)\times10^{-13}$& -& -& -\\[3pt]
LkHA~324& II R& $-2.54\pm0.03$& $7.74\pm0.13$& $-10.28\pm0.13$& $7.60\times10^{-16}$& $(7.81\pm0.10)\times10^{-15}$& $-1.05^{+0.03}_{-0.03}$& $1.04^{+0.14}_{-0.14}$& $-6.48^{+0.13}_{-0.15}$\\[3pt]
HD~203024& I& $8.88\pm0.04$& $14.42\pm0.20$& $-5.54\pm0.20$& $8.72\times10^{-16}$& $(4.83\pm0.18)\times10^{-15}$& -& -& -\\[3pt]
V645~Cyg& I& $-99.65\pm1.32$& $2.15\pm0.05$& $-101.80\pm1.33$& $4.39\times10^{-16}$& $(4.47\pm0.06)\times10^{-14}$& -& -& -\\[3pt]
V361~Cep& I& $-30.74\pm0.24$& $5.11\pm0.06$& $-35.85\pm0.25$& $1.05\times10^{-15}$& $(3.75\pm0.03)\times10^{-14}$& $-0.03^{+0.04}_{-0.03}$& $2.06^{+0.10}_{-0.10}$& $-5.54^{+0.10}_{-0.10}$\\[3pt]
V373~Cep& III B& $-55.39\pm0.39$& $7.40\pm0.15$& $-62.80\pm0.42$& $5.10\times10^{-16}$& $(3.21\pm0.02)\times10^{-14}$& -& -& -\\[3pt]
V1578~Cyg& III B& $-19.29\pm0.06$& $10.30\pm0.09$& $-29.59\pm0.11$& $7.77\times10^{-16}$& $(2.30\pm0.01)\times10^{-14}$& $-0.37^{+0.04}_{-0.03}$& $1.72^{+0.11}_{-0.11}$& $-5.67^{+0.10}_{-0.10}$\\[3pt]
LkHA~257& I& $-5.40\pm0.00$& $12.92\pm0.25$& $-18.32\pm0.25$& $8.56\times10^{-17}$& $(1.57\pm0.02)\times10^{-15}$& $-1.51^{+0.03}_{-0.02}$& $0.58^{+0.16}_{-0.16}$& $-6.95^{+0.17}_{-0.18}$\\[3pt]
SV~Cep& III R& $2.18\pm0.16$& $14.89\pm0.45$& $-12.71\pm0.48$& $2.08\times10^{-16}$& $(2.65\pm0.10)\times10^{-15}$& $-2.01^{+0.03}_{-0.03}$& $0.08^{+0.19}_{-0.19}$& $-7.46^{+0.19}_{-0.17}$\\[3pt]
V375~Lac& IV B& $-18.88\pm0.21$& $12.37\pm0.02$& $-31.24\pm0.21$& $9.02\times10^{-17}$& $(2.82\pm0.02)\times10^{-15}$& -& -& -\\[3pt]
HD~216629& II B& $-17.02\pm0.09$& $3.68\pm0.04$& $-20.70\pm0.10$& $5.84\times10^{-15}$& $(1.21\pm0.01)\times10^{-13}$& $0.39^{+0.03}_{-0.03}$& $2.48^{+0.12}_{-0.11}$& $-5.17^{+0.10}_{-0.09}$\\[3pt]
V374~Cep& II B& $-32.95\pm0.31$& $4.62\pm0.05$& $-37.56\pm0.31$& $3.09\times10^{-15}$& $(1.16\pm0.01)\times10^{-13}$& $0.44^{+0.04}_{-0.04}$& $2.53^{+0.13}_{-0.12}$& $-4.95^{+0.11}_{-0.10}$\\[3pt]
V628~Cas& IV B& $-124.89\pm2.44$& $2.10\pm0.01$& $-126.99\pm2.44$& $6.52\times10^{-15}$& $(8.28\pm0.16)\times10^{-13}$& -& -& -\\[3pt]
\hline
\end{tabular}
\end{table}
\end{landscape}

\begin{figure*}
\centering
\includegraphics[width=17.5cm]{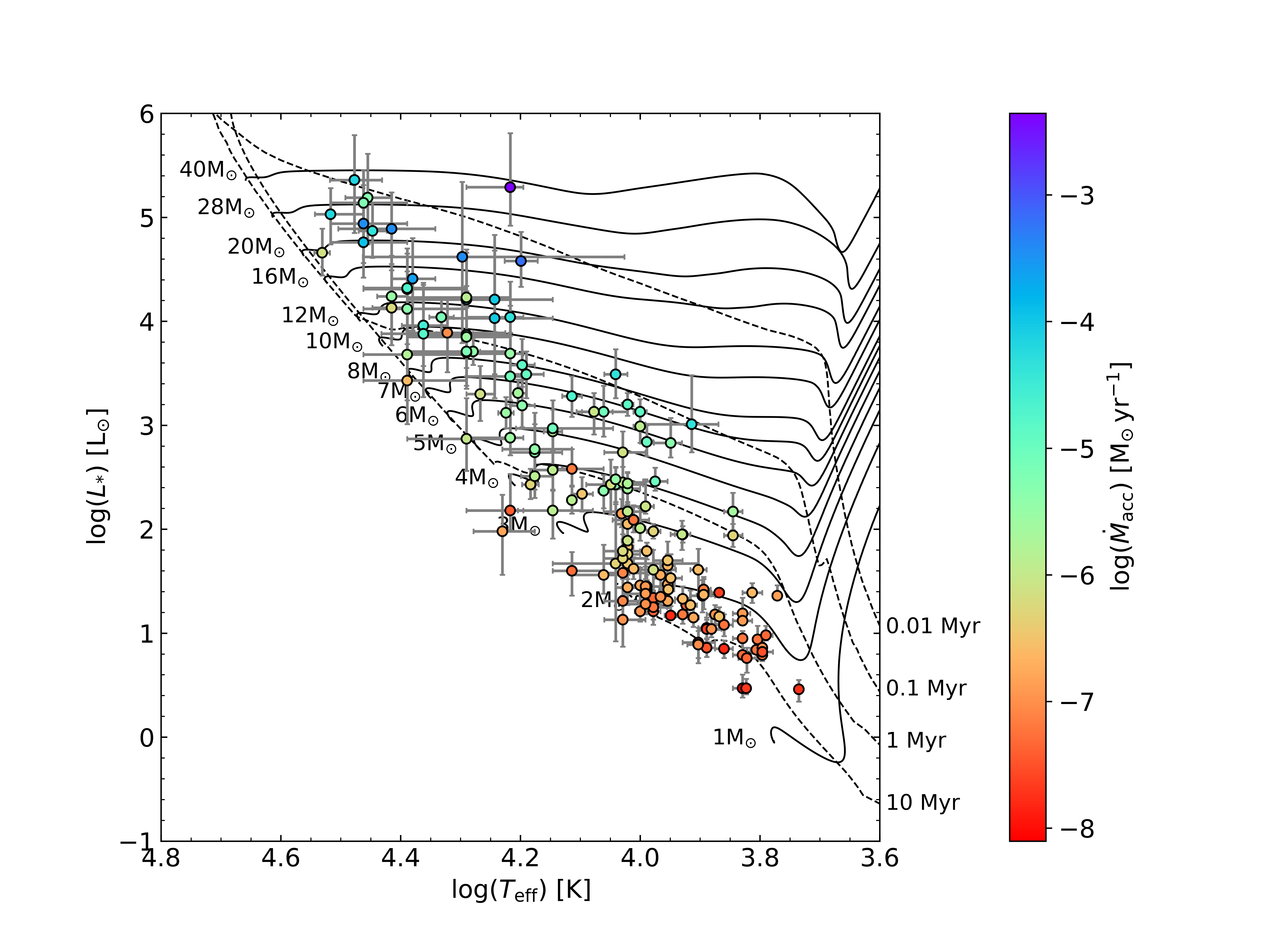}
\caption{The placement of all 163 HAeBes in the HR diagram. The PMS tracks with initial mass from 1 to 40\,$\rm M_{\sun}$ \protect\citep{Bressan2012,Tang2014} and isochrones of 0.01, 0.1, 1 and 10\,Myr \protect\citep{Marigo2017} are plotted as solid and dashed lines respectively. The colour map denotes the accretion rate.}
\label{fig:HR_Mdot}
\end{figure*}

\begin{figure}
\centering
\includegraphics[width=8cm]{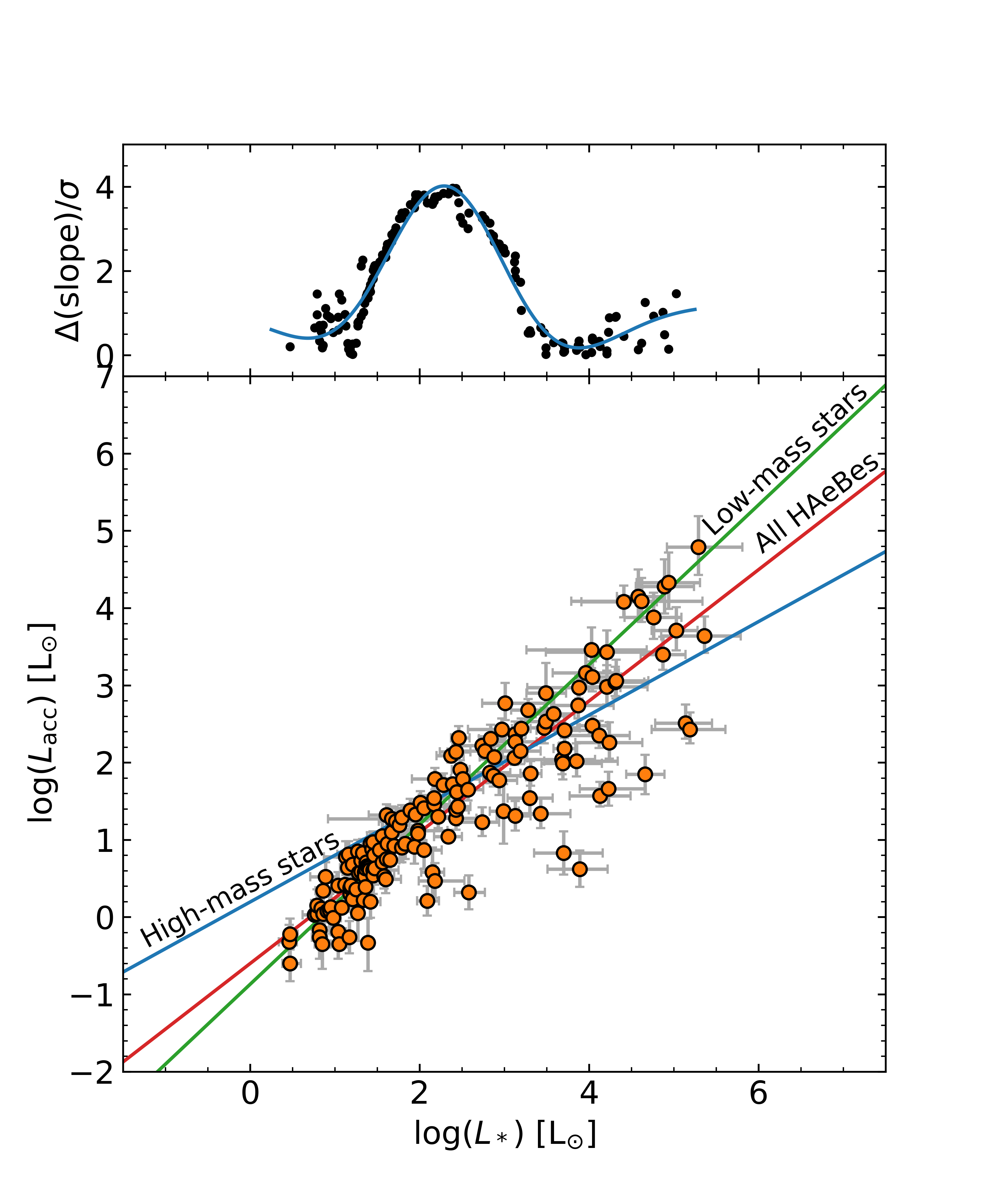}
\caption{The logarithmic accretion luminosities versus stellar luminosities  for the full, 
  163 stars, sample. Also shown are linear fits to the full sample and
  to the low- and high-mass stars respectively. As can be seen in the
  top panel, the difference in slopes between the low- and high-mass
  objects is most significant at at $L_*>194\,\rm L_{\sun}$ (see text
  for details). The gradients of best fits for the whole sample,
  low-mass and high-mass HAeBes are $0.85\pm0.03$, $1.03\pm0.08$ and
  $0.60\pm0.08$ respectively.
} 
\label{fig:LLacc_wfv}
\end{figure}

\begin{figure*}
\centering
\centering
\includegraphics[width=5.8cm]{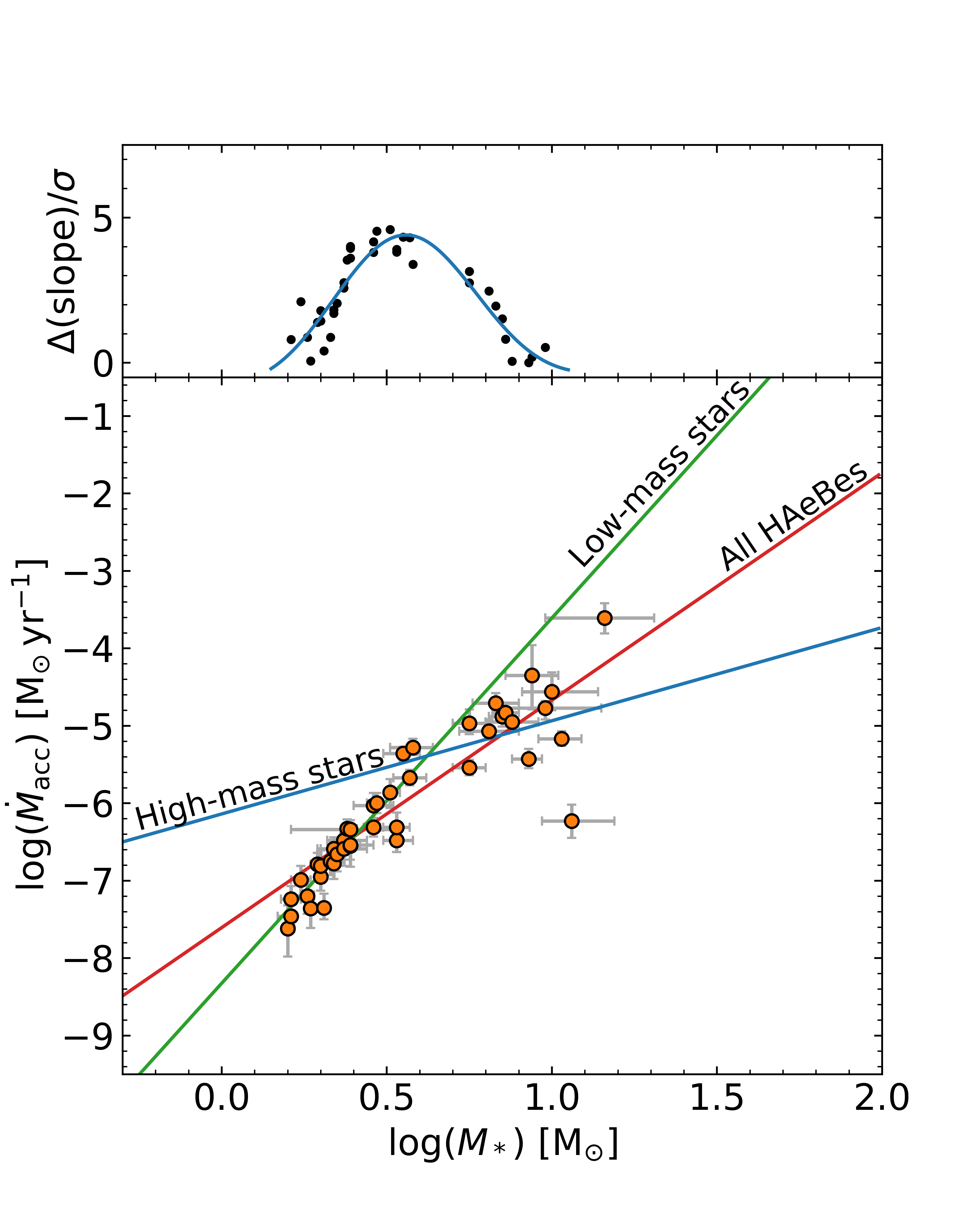}
\centering
\includegraphics[width=5.8cm]{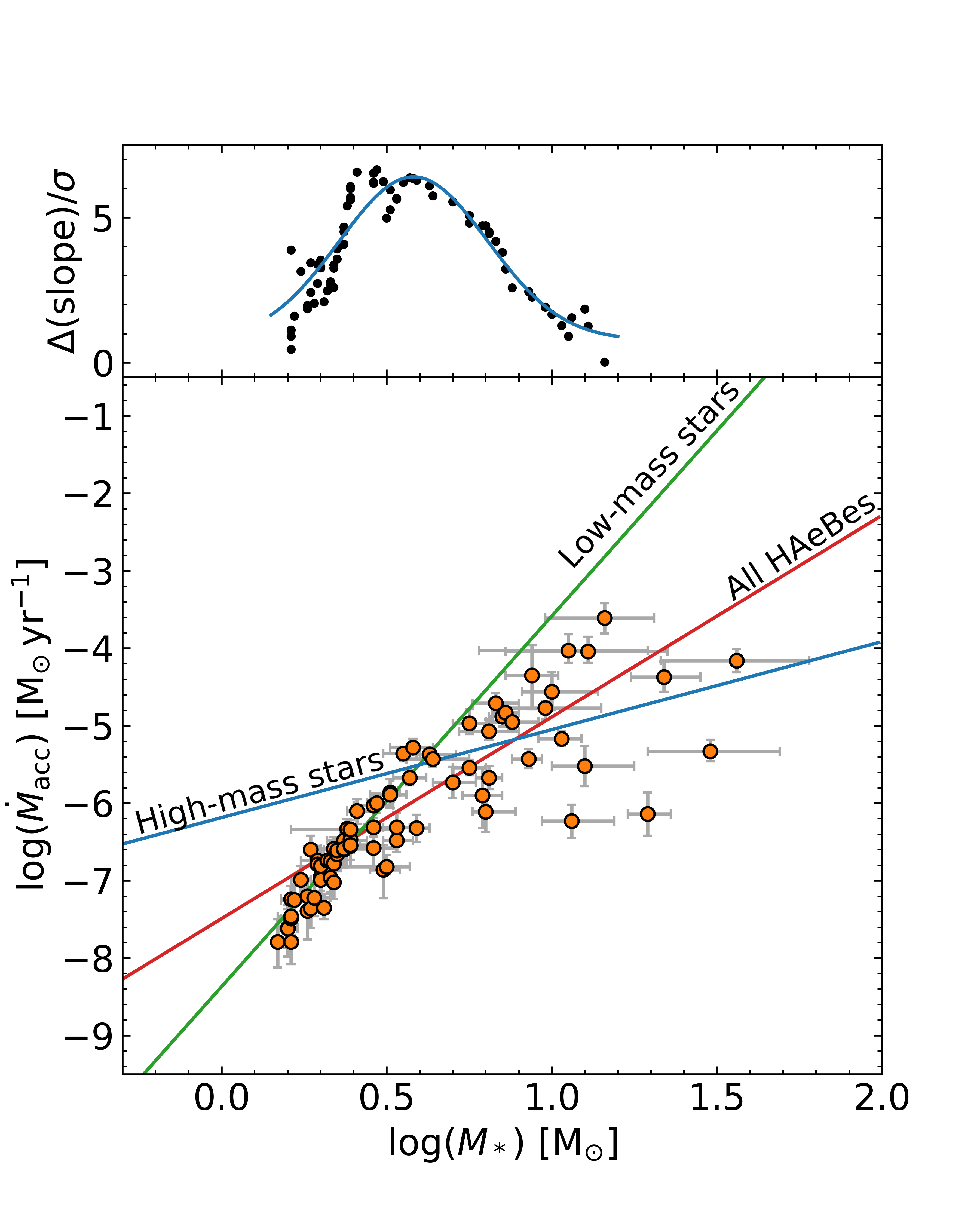}
\centering
\includegraphics[width=5.8cm]{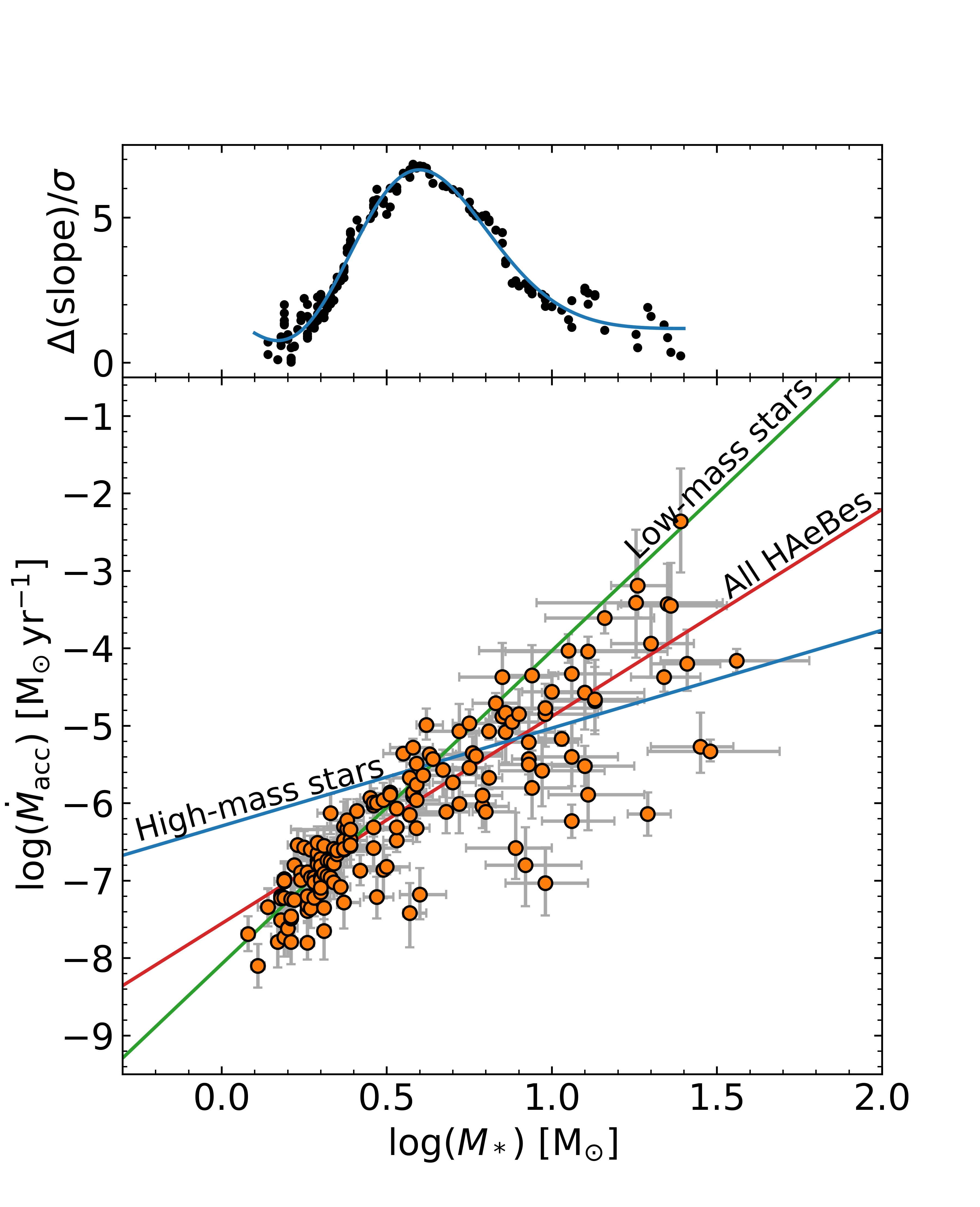}
\caption{
  The difference of the gradient of the best fits between the low-mass and 
  the high-mass HAeBes in terms of the uncertainty of the gradient difference (top panel) 
  and the mass accretion rates versus stellar masses (bottom panel).
  Left: 43 stars from this work sample and the sample of \protect\cite{Fairlamb2015} presents in table 1 of \protect\cite{The1994}. 
  Objects are separated into the low-mass HAeBes, $M_*<3.61\,\rm M_{\sun}$, and the high-mass HAeBes,
  $M_*>3.61\,\rm M_{\sun}$. The slopes of best fits for the whole
  sample, low-mass and high-mass HAeBes are $2.94\pm0.24$,
  $4.72\pm0.49$ and $1.20\pm0.65$ respectively.
  Middle: 78 objects from this work sample and the sample of \protect\cite{Fairlamb2015}.
  The break is defined at $M_*=3.81\,\rm M_{\sun}$. The gradients of best fits for the whole
  sample, low-mass and high-mass HAeBes are $2.60\pm0.19$,
  $4.78\pm0.34$ and $1.14\pm0.46$ respectively. 
  Right: 163 objects are separated into the low-mass HAeBes and the high-mass HAeBes at $M_*=3.98\,\rm M_{\sun}$. The gradients of best fits for the whole sample, low-mass and high-mass HAeBes are $2.67\pm0.13$, $4.05\pm0.24$ and $1.26\pm0.34$ respectively.}
\label{fig:MMdot_wft_wf_wfv}
\end{figure*}

Let us first consider the sample with homogeneously derived mass
accretion rates, before investigating the full sample.  The
relationship between mass accretion rate and stellar mass of the
combined sample from this work and \cite{Fairlamb2015,Fairlamb2017}
that are present in \citet[table 1]{The1994}, is shown in the left
hand panel of Figure~\ref{fig:MMdot_wft_wf_wfv}. The sample of
\cite{The1994}, contains the best established HAeBes, and is
thus least contaminated by possible misclassified objects. It can be
seen that
the mass accretion rate increases with stellar mass,
and that there is a different behaviour for objects with masses below
and above the mass range with $\log(M_*)=0.4$--0.6. This break is
consistent with the major finding in F15 that the relationship between
mass accretion rate and stellar mass shows a break around the boundary
between Herbig Ae and Herbig Be stars. In particular, they found that
the relationship between mass accretion rate and mass has a different
slope for the lower and higher mass objects respectively. With our
improved sample in hand, we can now revisit this finding and we
determine the mass at which the break occurs with higher precision in
the same way as we found the turnover luminosity earlier.

We found that the maximum slope difference is at the $4.4\,\sigma$ level for the
Th\'e et al. sample at $\log(M_*)=0.56^{+0.14}_{-0.14}$ or
$M_*=3.61^{+1.38}_{-0.98}\,\rm M_{\sun}$.
The uncertainty in the mass is decided where the difference in slope is
1$\sigma$ smaller than at maximum.
If we focus on the total combined sample of 78 objects from this work
and F15, the break is established at 
$\log(M_*)=0.58^{+0.14}_{-0.14}$ or $M_*=3.81^{+1.46}_{-1.05}\,\rm
M_{\sun}$ with the maximum gradient difference about $6.4\,\sigma$. as
shown in the middle panel of Figure~\ref{fig:MMdot_wft_wf_wfv}.  It
can be seen that this plot shows the same relationship as  the one in the
left-hand panel.
When considering the full, combined sample of 163 objects from this
work, F15 and V18 whose mass accretion rates are shown in the
right-hand panel of Figure~\ref{fig:MMdot_wft_wf_wfv}, the break is
now found at the $\log(M_*)=0.60^{+0.13}_{-0.12}$ or
$M_*=3.98^{+1.37}_{-0.94}\,\rm M_{\sun}$ with the maximum difference
of gradient about $6.6\,\sigma$.

We note that both the significance of the difference in slopes as well
as the mass at which the break occurs increases with the number of
objects. This could be due to the number of objects; especially as the \cite{The1994}
sample is sparsely populated at the high mass end, which could
increase the uncertainty on the resulting slope and thus lower the
significance of the difference in slopes between high and low mass
objects. Alternatively, it could be affected by a larger contamination
of non-Herbig Be stars in the full sample. It is notoriously hard to
 differentiate between a regular Be star and a Herbig Be star.
Given the low number statistics of the \cite{The1994}
sample and the similarity of the break in mass of the other two samples, 
we will proceed with a break in slope at 4 M$_{\odot}$.

\subsubsection{Literature comparisons with T Tauri stars}

\begin{figure}
\centering
\includegraphics[width=\columnwidth]{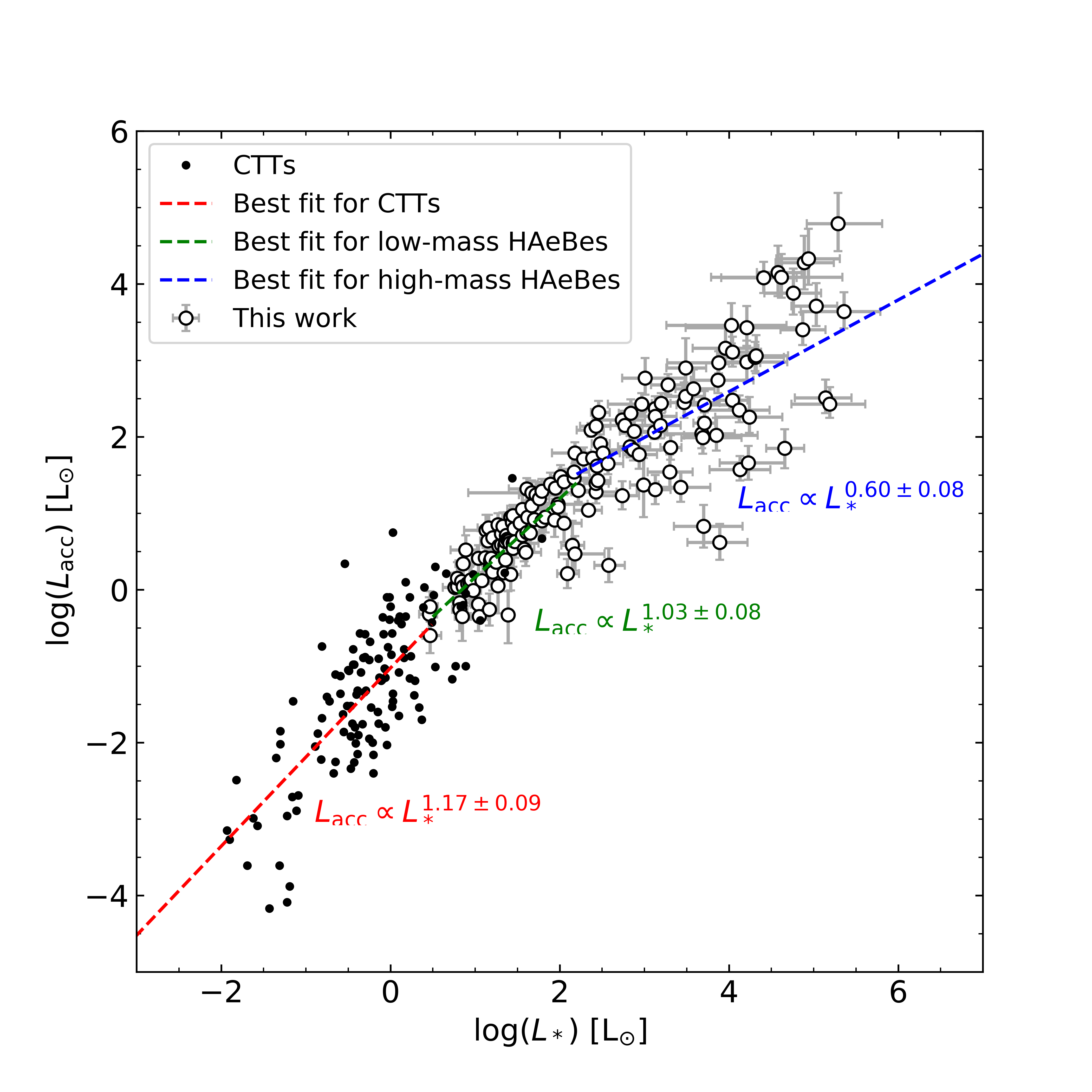}
\caption{Accretion luminosity versus stellar luminosity for 163 HAeBes
  in this work and classical T Tauri stars from \protect\cite{Hartmann1998},
  \protect\cite{White2003}, \protect\cite{Calvet2004} and \protect\cite{Natta2006}. Red, green
  and blue dashed lines are the best fit for T Tauri stars ($L_{\rm
    acc}\propto L^{1.17\pm0.09}_*$), low-mass HAeBes ($L_{\rm
    acc}\propto L^{1.03\pm0.08}_*$) and high-mass HAeBes ($L_{\rm
    acc}\propto L^{0.60\pm0.08}_*$) respectively.}
\label{fig:LLacc_CTTs}
\end{figure}

It is interesting to see how the accretion luminosities compare to
those of T Tauri stars at the low mass range. With the large caveat
that no {\it Gaia} DR2 study of T Tauri stars and their stellar
parameters and accretion rates exists at the moment, we show in
Figure~\ref{fig:LLacc_CTTs} the  $L_{\rm acc}$ against $L_*$ for the whole
sample of 163 HAeBes in this work compared to classical T Tauri stars
of which all luminosity values are taken from \cite{Hartmann1998},
\cite{White2003}, \cite{Calvet2004} and \cite{Natta2006}.  On one
hand, the gradient of  best fit for the classical T Tauri stars is
$1.17\pm0.09$ which is close, and well within the errorbars, to
$1.03\pm0.08$ for low-mass HAeBes.  On the other hand, high-mass
HAeBes shows a flatter relationship of $L_{\rm acc} \propto
L^{0.60\pm0.08}_*$.  It would seem that the Herbig Ae and Be stars
with masses up until 4\,M$_{\odot}$ behave similarly to the T Tauri
stars.  Again, stellar parameters of these classical T Tauri stars
were not derived in the same manner as HAeBes in this work.   This may
have an effect on our conclusion.

\subsection{Mass accretion rate as a function of stellar age}

\begin{figure}
\centering
\centering
\includegraphics[width=\columnwidth]{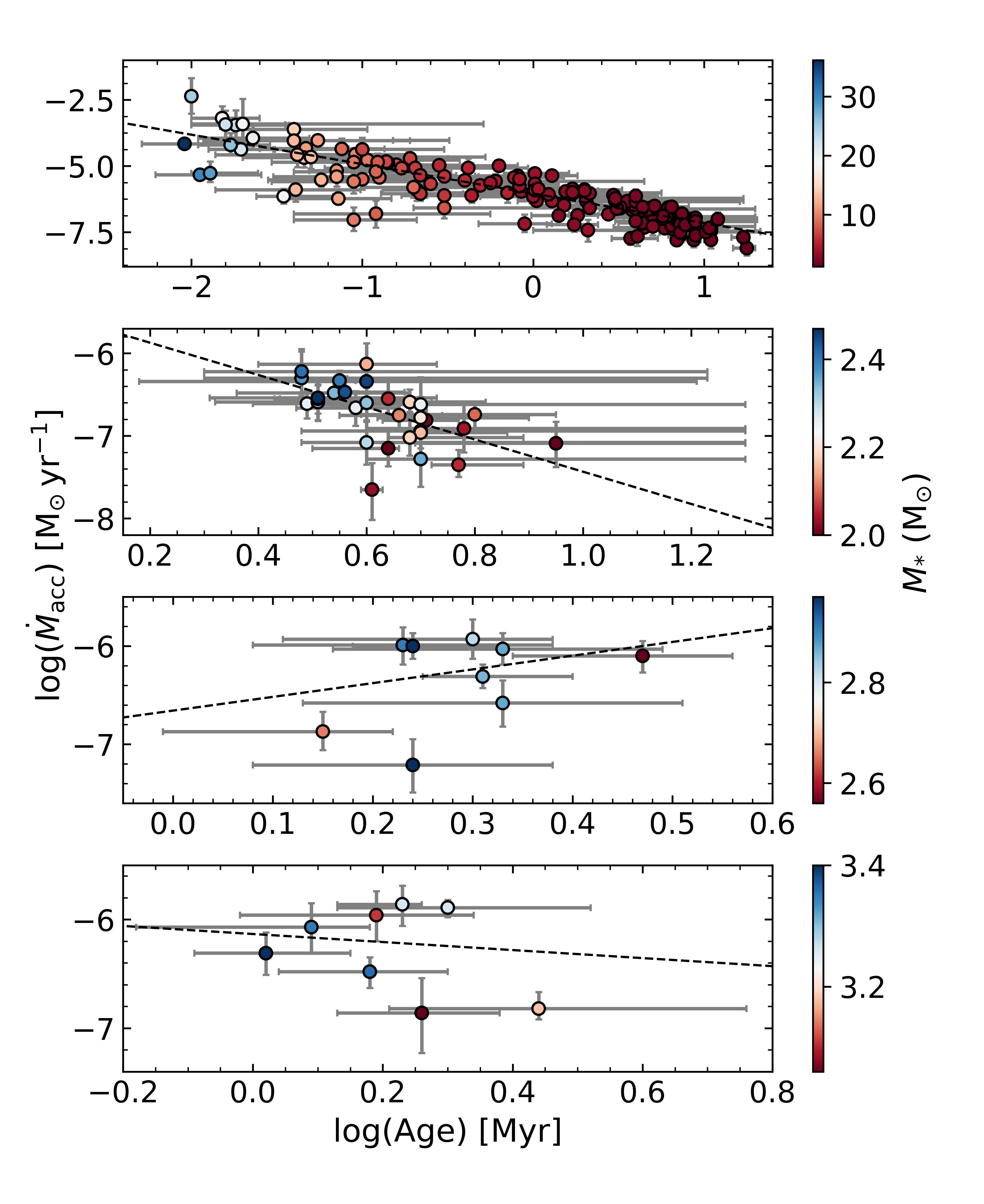}
\caption{
Ages versus mass accretion rates. From top to bottom all sample, the mass range 2.0-$2.5\,\rm M_{\sun}$, 2.5-$3.0\,\rm M_{\sun}$ and 3.0-$3.5\,\rm M_{\sun}$. The best fit is shown in the dashed line where  $\dot M_{\rm acc}\propto Age^{-1.11\pm0.05}$, $\dot M_{\rm acc}\propto Age^{-1.95\pm0.49}$, $\dot M_{\rm acc}\propto Age^{1.40\pm1.47}$ and $\dot M_{\rm acc}\propto Age^{-0.37\pm1.25}$ respectively. The colour map indicates stellar mass of each mass bin.
} 
\label{fig:Mdot_age}
\end{figure}

In Figure~\ref{fig:HR_Mdot} we illustrate the mass accretion rates
across the HR-diagram. As expected, it can be seen that the accretion
rates are largest for the brightest and most massive objects. However,
the sample may allow an investigation into the
evolution of the mass accretion rate as a function of stellar age as
provided by the evolutionary models. In general it can be stated that,
in terms of evolutionary age, the younger objects have higher
accretion rates than older objects.  As illustration, we show the mass
accretion rate as a function of age in the top panel of
Figure~\ref{fig:Mdot_age}. It is clear that the accretion rate
decreases with age (as also shown in F15); the Pearson correlation
coefficient is -0.87.

However, when considering any properties as a function of age, there
is always the issue that higher mass stars evolve faster than lower
mass stars. Hence, the observed fact that higher mass objects have
higher accretion rates could be the main underlying reason for a trend
of decreasing accretion rate with time. Indeed, in the top panel of
Figure~\ref{fig:Mdot_age} the range in accretion rates covers the
entire accretion range of the accretion vs. mass relation, so it is
hard to disentangle from this graph whether there is also an accretion
rate - age relation.

This can in principle be circumvented when selecting a sample of
objects in a small mass range, so that the spread in mass accretion
rates is minimized. This needs to be offset against the number of
objects under consideration. In the lower panels of
Figure~\ref{fig:Mdot_age} we show how the accretion rate change with
the age of stars in 3 different mass bins (2.0-$2.5\,\rm M_{\sun}$,
2.5-$3.0\,\rm M_{\sun}$ and 3.0-$3.5\,\rm M_{\sun}$.  A best fit taken into account the uncertainties in both the mass accretion rate and age to
the 30 objects where $2.0\,\rm M_{\sun}<M_*<2.5\,\rm M_{\sun}$ shows a
$\dot M_{\rm acc}\propto Age^{-1.95\pm0.49}$ with a Pearson's
correlation coefficient -0.55.  The 9 stars with masses $2.5\,\rm
M_{\sun}<M_*<3.0\,\rm M_{\sun}$ have a gradient of a best fit is
$-1.40\pm1.47$ (correlation coefficient 0.40).  The error of the best
fit gradient becomes slightly larger than its own value, which is due
to the smaller number of objects involved.  The 8 objects with
$3.0\,\rm M_{\sun}<M_*<3.5\,\rm M_{\sun}$ in the bottom panel of
Figure~\ref{fig:Mdot_age} have a gradient of a best fit is
$-0.37\pm1.25$ with a linear correlation -0.36. It can be seen that the error of 
the best fit gets larger than 3 times of its own absolute value. This is more 
likely to be due to the lack of HAeBes when the stellar mass increases.

The 2.0-2.5\,M$_{\odot}$ sample is best suited to study any trend in
accretion rate with age in the sense that the accretion rates change
by 0.4\,dex across this mass range (cf. the slope of 4.05 found for
the low mass objects in Sec. 5.2). The spread in accretion rates in
the graph is twice that, and if the additional decrease in mass
accretion rate is due to evolution, we find $\dot M_{\rm acc}\propto
Age^{-1.95\pm0.49}$. This value is the only such determination
  for Herbig Ae/Be stars in a narrow mass range, other determinations
  were based on full samples of HAeBe stars which contain many
  different masses and suffer from a mass-age degeneracy (e.g. F15,
  \cite{Mendigutia2012}).  Despite the relatively large errorbars, we
  note that our determination is remarkably close to the observed
  values for T Tauri stars by \cite{Hartmann1998}, who find an $\eta$
  between 1.5-2.8, while theoretically these authors predict $\eta$ to
  be larger than 1.5 for a single $\alpha$ viscous disk. We also refer
  the reader to the discussion in \cite{Mendigutia2012}, who,
  remarkably, find a similar value of the exponent to ours:
  1.8$^{+1.4}_{-0.7}$ as determined for their full sample.

\section{Discussion}

In the above we have worked out the mass accretion rates for a
  large sample of 163 Herbig Ae/Be stars. To this end new and
  spectroscopic archival data were employed. A large fraction of the
  sample now has homogeneously determined astrophysical parameters,
  while {\it Gaia} parallaxes were used to arrive at updated luminosities
  for the sample objects.   We compared the mass accretion rates
derived using the MA paradigm for Herbig Ae/Be stars with various
properties.  We find that:

\begin{itemize}

\item The mass accretion rate increases with stellar mass, but the
  sample can be split into two subsets depending on their masses.

\item The low mass Herbig Ae stars' accretion rates have a steeper
  dependence on mass than the higher mass Herbig Be stars.

\item The above findings corroborate previous reports in the
  literature. The larger sample and improved data allows us to determine
  the mass where the largest difference in slopes between high and low
  mass objects occurs. We find it is 4 M$_{\odot}$. 

\item Bearing in mind the caveat that T Tauri stars do not yet have
  {\it Gaia} based accretion rates, it appears that the Herbig Ae
  stars' accretion properties display a similar dependence on
  luminosity as the T Tauri stars.
  
\item In general, younger objects have larger mass accretion rates,
  but it proves not trivial to disentangle a mass dependence from an
  age dependence. We present the first attempt to do so. A small
  subset in a narrow mass range leads us to suggest that the accretion
  rate decreases with time. 
    
\end{itemize}

In the following we aim to put these results into context, but we
start by discussing the various assumptions that had to be made to
arrive at these results.

\subsection{On the viability of the MA model to determine accretion rates of Herbig Ae/Be stars}

In the above we have determined accretion luminosities and accretion
rates to young stars. However, the question remains whether the
magnetic accretion shock modelling that underpins these computations
should be applicable. After all, A and B-type stars typically do not
have magnetic field detections, and indeed, are not expected to
harbour magnetic fields as their radiative envelopes would not create
a dynamo that in turn can create a magnetic field. This was verified
in a study of Intermediate Mass T Tauri stars (IMTTS) by
\citet{Villebrun2019} who found that the fraction of pre-Main Sequence
objects with a magnetic field detection drops markedly as they become
hotter (for similar masses). These authors suspect that any B-field
detections in Herbig Ae/Be stars could be fossil fields from the
evolutionary stage immediately preceding them. The fields would not
only be weaker, but also more complex, hampering their direct
detection.

 \citet{Muzerolle2004} showed that the MA models provided
  qualitative agreement with the observed emission line profiles of
  the Herbig Ae star UX~Ori. In the process, they also found that the
  B-field geometry should be more complex than the usual dipole field
  in T Tauri stars and the expected field strength would be much below
  the usual T Tauri detections, in line with the \citet{Villebrun2019}
  findings.

Since, based on statistical studies and targeted individual
investigations evidence has emerged that Herbig Ae stars are similar
to the T Tauri stars. \citet{Garcia2006}
showed that the accretion rate properties of the Herbig Ae stars
constitute a natural extension to the T Tauri stars, while F15
demonstrated that the MA model can reproduce the observed UV excesses
towards most of their Herbig Ae/Be stars with realistic parameters for
the shocked regions. A small number of objects, all Herbig Be stars,
could not be explained with the MA model. They exhibit such a large
UV-excess that they would require shock covering factors larger than 
100\%, which is clearly unphysical.

\citet{Vink2002} and \citet{Vink2005} were the first to point out the
remarkable similarity in the observed linear spectropolarimetric
properties of the H$\alpha$ line in Herbig Ae stars and T Tauri stars.
The polarimetric line effects in these types of objects can be
explained with geometries consistent with light scattering off
(magnetically) truncated disks. In contrast, the very different
effects observed towards the Herbig Be stars were more consistent with
disks reaching onto the central star - hinting at a different
accretion mechanism in those. The large sample by \citet{Ababakr2017}
allowed them to identify the transition region to be around spectral
type B7/8.

\citet{Cauley2014} studied the He~{\sc i} 1.083 $\mu$m line profiles of
a large sample of Herbig stars and found that both T Tauri and Herbig
Ae stars could be explained with MA, while the Herbig Be stars could
not.
\citet{Reiter2018} did not detect a difference in line morphology
between the few (5) magnetic Herbig Ae/Be stars and the rest of the
sample. They argued that Herbig Ae stars may therefore not accrete
similarly to T Tauri stars. However, based on Poisson statistics
alone, even if all magnetic objects showed either a P Cygni or Inverse
P Cygni profile, a sample of 5 would not be sufficient to conclusively
demonstrate that the magnetic objects are, or are not, different from
the non-magnetic objects. Clearly, more work needs to be done in this
area.  \citet{Costigan2014} investigated the H$\alpha$ line
variability properties of Herbig Ae/Be stars and T Tauri stars and
found that both the timescales and amplitude of the variability were
similar for the Herbig Ae and T Tauri, which also led these authors to
suggest that the mode of accretion of these objects are similar. A
notion also implied by our result in Figure~\ref{fig:LLacc_CTTs} that
the accretion luminosities of both Herbig Ae and T Tauri stars appear
to have the same dependence on the stellar luminosity. Finally,
\citet{Mendigutia2011a} found that the H$\alpha$ line width
variability of Herbig Be stars is considerably smaller than for Herbig
Ae stars, which is in turn smaller than for TTs (see Figure~37 in
\citealt{Fang2013}).  This may suggest smaller line emitting
regions/magnetospheres as the stellar mass increases, and thus a
eventual transition from MA to some other mechanism responsible for
accretion.

To summarize this section, various studies of different observational
properties have confirmed the many similarities between T Tauri and
Herbig Ae stars, which in turn hint at a similar accretion mechanism,
the magnetically controlled accretion. In turn, this would validate
our use of the MA shock modelling to derive mass accretion rates for
at least the lower mass end of the Herbig Ae/Be star range. Before we
discuss the accretion mechanisms for low- and high mass stars, we
address the use of line luminosities to arrive at accretion
luminosities and rates below.

\subsection{On the use of line emission as accretion rate diagnostic}

The only ``direct'' manner to determine the mass accretion rate is to
measure the accretion luminosity, which is essentially the amount of
gravitational potential energy that is converted into radiation at
ultra-violet wavelengths. This can then be turned into a mass
accretion rate once the stellar mass and radius are known. The
determination of the contribution of the accretion shock to the total
UV emission for HAeBe objects is not straightforward due to the fact
that these stars intrinsically emit many UV photons by virtue of their
higher temperatures.

\citet{Calvet2004} showed that the Br$\gamma$ emission strengths
observed towards a sample of IMTTS correlated with the accretion
luminosity as determined by the UV excess for a large mass range
extending to 3.7 M$_{\odot}$ (the mass of their most massive target,
GW Ori, with spectral type G0 - cool enough to unambiguously determine
the UV excess emission). This was already known for lower mass
objects, but, significantly, these authors extended it to higher
masses.  In addition, these authors also showed that the mass
accretion rate correlated with the stellar mass over this mass
range. Therefore they also demonstrated that line emission, in this
case the hydrogen recombination Br$\gamma$, can be used to measure
accretion rates to masses of at least up to $\sim$4 M$_{\odot}$.
\citet{Mendigutia2011b} measured the UV-excess of Herbig Ae/Be
  stars using UV-blue spectroscopy and photometry respectively and
  noted the correlation with emission line strengths (see also
  \citealt{Donehew2011}).  F15 and F17 took this further and
  demonstrated the strong correlation between emission line strength
  and accretion luminosity for a much larger sample and a mass range
  going up to 10--15 M$_{\odot}$. It extended to an accretion
  luminosity of 10$^4$ L$_{\odot}$ and was shown to hold for a large
  number of different emission lines.

One should keep in mind the caveat, also pointed out by F17, that high
spatial resolution optical and near-infrared studies of the hydrogen line
emission do not necessarily identify the line emitting regions of
Herbig Ae/Be stars with the magnetospheric accretion channels. Already
in 2008, \citet{Kraus2008} reported that their milli-arcsec resolution
AMBER interferometric data indicated different Br$\gamma$ line forming
regions which were consistent with the MA scenario for some objects
but with disk-winds for others. Further studies by e.g.
\citet{Mendigutia2015b} find the line emission region consistent with
a rotating disk. Similarly, \citet{Tam2016} and \citet{Kreplin2018}
find the disk-wind a more likely explanation for the emitting region
than the compact accretion channels. Even higher resolution H$\alpha$
CHARA data discussed by \citet{Mendigutia2017} present a similar
diversity.

There thus remains
the question whether we can use emission lines to probe accretion if
they are apparently not related to the accretion process itself, or
indeed, why the line luminosities correlate with the accretion
luminosity at all. It is well known that accretion onto young stars
drives jets and outflows, where the mass ejection rates are of order
10\% of the accretion rates (see e.g. \citealt{Purser2016}). If this
is also the case for Herbig Ae/Be stars, then one could expect the
emission lines to be correlated with the accretion rates.  On the
other hand, \citet{Mendigutia2015a} pointed out that while the physical
origin of the lines may not be related to the accretion process {\it
  per se}, it is the underlying correlation between accretion
luminosity and stellar luminosity that gives rise to a correlation
between the line strengths and accretion.  Hence, although the lines
may not necessarily be directly accretion-related, the empirical
correlations between emission line luminosities and accretion
luminosities are strong enough to validate their use as accretion
tracers.

When we revisit the various studies mentioned above, it is
  notable that a distinction between Herbig Ae and Herbig Be stars is
  found, spectroscopically \citep{Cauley2014}, spectropolarimetrically
  \citep{Vink2005}, due to spectral variability \citep{Costigan2014},
  and even when considering the accretion rates from UV-excesses
  (F15).

Based on the break in accretion rates, we can move this boundary to
a critical mass of 4M$_{\odot}$, remarkably close, but not perfectly
so, to the boundary of around B7/8 put forward by \citet{Ababakr2017}.
It is thus implied that there is a transition from magnetically
controlled accretion in low-mass HAeBes to another accretion mechanism
in high-mass HAeBes at around $4\,\rm M_{\sun}$.  The remaining
question is what this mechanism should be.

\subsection{If  MA does not operate in massive objects, what then?}

The spectropolarimetric finding of a disk reaching onto the star is
reminiscent of the Boundary Layer (BL) accretion mechanism that was
found to be a natural consequence of a viscous circumstellar disk
around a stellar object (\citealt{Lynden1974}).
The BL is a thin annulus close to the star in which the material
reduces its (Keplerian) velocity to the slow rotation of the star when
it reaches the stellar surface. It is here that kinetic energy and
angular momentum will be dissipated. The BL mechanism was originally
used to explain the observed UV excesses of low mass pre-Main Sequence
stars (\citealt{Bertout1988}) until observations led to strong support
for the magnetic accretion scenario instead
(\citealt{Bouvier2007}). One of the difficulties the BL mechanism
  had was the smaller redshifted absorption linewidths predicted from
  the Keplerian widths expected from the BL than from freefall in the
  case of MA (\citealt{Bertout1988}).  It had been suggested in the
past to operate in Herbig Ae/Be objects (eg \citealt{Blondel2006} who
studied Herbig Ae stars; \citealt{Mendigutia2011b};
\citealt{Cauley2014}) but has never been adapted and tested for
  masses of Herbig Be stars and greater.

It is very much beyond the scope of this paper to address the BL
theoretically, but let us suffice with a basic consideration to see
whether we would be able to expect a different slope in the derived
accretion rate or luminosity for an object undergoing MA or BL
accretion.  In both situations infalling material converts energy into
radiation.
In MA the energy released by a mass $dM$ falling onto a star
with mass and radius $M_*$ and $R_*$ respectively will be the
gravitational potential energy of the infalling material,
$\frac{GM_*dM}{R_*}$, times a factor close to 1 accounting for the
fact that material is not falling from infinity.

How would this amount of released energy compare to that in the BL
scenario for infalling material of the same mass $dM$? For the
BL case, the accretion energy will be at most the kinetic energy of the
rotating material prior to being decelerated in the very thin layer
close to the stellar surface. As the material rotates Keplerian, we
know that the centripetal force equals the force of gravity,
$\frac{dMv^2}{R_*} = \frac{GM_*dM}{R_*^2}$. We thus find that the
kinetic energy $\frac{1}{2}dMv^2 = \frac{GM_*dM}{2R_*}$, which is half
the gravitational potential energy for the same mass.

In other words, the energy released in the BL scenario will be {\it
  less} than that released by MA for the same mass. Therefore if the
mass accretion rate is the same, we obtain a lower accretion
luminosity for objects accreting material through a Boundary Layer
than through Magnetospheric Accretion. This also means that the
accretion rate has to be larger in BL to arrive at the same accretion
luminosity. In turn, this has as implication that if we derive the
mass accretion rates using the MA paradigm, while the accretion
  is due to BL, then the resulting accretion luminosities and rates
  would have been underestimated.

Earlier, we found that the (MA derived) accretion luminosities
  have the same dependence of the accretion luminosity on the stellar
  luminosity for T Tauri stars and Herbig Ae stars, while we find a
  smaller gradient for the more massive stars. If we would assume
that this dependence would hold for more massive stars too, then it
could be concluded that the accretion rates have been
  underestimated for massive Herbig Be stars. If the BL scenario was
  the acting mechanism for the Herbig Be stars then the accretion
luminosities would be larger - possibly resulting in the same
relationship between accretion and stellar luminosity as the lower
mass stars, and no break would be visible.

Could that be the case here? The accretion onto the stars likely
depends on the rates the accretion disks are fed and mass needs to be
transported through the disk, so it may be reasonable to assume that
this process is less dependent on the stellar parameters and a simple
correlation between accretion onto the star and stellar luminosity (or
mass) would be expected. It is intriguing that the BL scenario for
more massive stars might explain why we obtain lower accretion rates
when we assume MA for the more massive objects resulting in a break at
around  $4\,\rm M_{\sun}$. An in-depth investigation into the BL is
certainly warranted.

\section{Final Remarks}

In this paper we presented an analysis of a new set of optical
spectroscopy of 30 northern Herbig Ae/Be stars. This was combined with
our data and analysis of southern objects in F15 resulting in a set of
temperatures, gravities, extinctions and luminosities that were
derived in a consistent manner. As such this constitutes the largest
homogeneously analysed sample of 78 Herbig Ae/Be stars to date.  To
these we added 85 objects from the {\em Gaia} DR2 study of V18. The
total sample of 163 objects allowed us to derive the accretion
luminosities and mass accretion rates using the empirical power-law
relationship between accretion luminosity and line luminosity as
derived under the MA paradigm.
  
We identified a subset in the total sample as being the strongest
Herbig Ae/Be star candidates known. The set contains 60 per cent of
the objects in Table~1 from the \cite{The1994} catalogue. All
  trends found in the large sample are also present in this
  subsample. This implies that the large sample likely has a low
  contamination and is therefore a good representation of the Herbig
  Ae/Be class.

\begin{itemize}
\item We find that the mass accretion rate increases with stellar mass, and that the lower mass Herbig Ae stars' accretion rates have a steeper dependence on mass than the higher mass Herbig Be stars. This confirms previous findings, but the large sample allows us to determine the mass where this break occurs. This is found to be $4\,\rm M_{\sun}$.

\item A comparison of accretion luminosities of the Herbig Ae/Be stars with those of T Tauri stars from the literature indicates that the Herbig Ae stars' accretion rates display a similar dependence on luminosity as the T Tauri stars. This provides further evidence that Herbig Ae stars may accrete in a similar fashion as the T Tauri stars. We do caution however that T Tauri stars do not yet have {\em Gaia}-based accretion rates.

\item We also find that in general, younger objects have larger mass accretion rates.  However, it is not trivial to disentangle a mass and age dependence from each other: More massive stars have larger accretion rates, but have much smaller ages as well. A small subset selected in a narrow mass range leads us to suggest that the accretion rate does indeed decrease with time. The best value could be determined
    for the mass range 2.0-2.5\,M$_{\odot}$. We find $\dot M_{\rm
      acc}\propto Age^{-1.95\pm0.49}$.

\end{itemize}

Finally, we discussed the similarities and differences between the accretion properties of lower mass and higher mass Herbig pre-Main Sequence stars. In particular we discuss the various lines of evidence that suggest they accrete in different fashions. In addition, from linear spectropolarimetric studies, the Herbig Be stars are found to have disks reaching onto the stellar surface while the Herbig Ae stars (with the break around 4\,M$_{\sun}$) have disks with inner holes, similar to the T Tauri stars.

We therefore put forward the Boundary Layer mechanism as a viable manner for the accretion onto the stellar surface of massive pre-Main Sequence stars. More work needs to be done, but an initial, crude, estimate of the accretion luminosity dependency on mass -- assuming a global correlation between mass accretion rate and stellar mass for all masses -- can explain the observed break in accretion properties.

\section*{Acknowledgements}
The authors would like to thank to the anonymous reviewer who provided constructive comments which helped improve the manuscript.
CW thanks Thammasat University for financial support in the form of a PhD scholarship at the University of Leeds.
IM acknowledges the funds from a ``Talento'' Fellowship (2016-T1/TIC-1890, Government of Comunidad Aut\'{o}noma de Madrid, Spain).
M. Vioque was funded through the STARRY project which received funding from the European Union's Horizon 2020 research and innovation programme under MSCA ITN-EID grant agreement No 676036.
This work has made use of data from the European Space Agency (ESA)
mission {\it Gaia} (\url{https://www.cosmos.esa.int/gaia}), processed
by the {\it Gaia} Data Processing and Analysis Consortium (DPAC,
\url{https://www.cosmos.esa.int/web/gaia/dpac/consortium}). Funding
for the DPAC has been provided by national institutions, in particular
the institutions participating in the {\it Gaia} Multilateral
Agreement.  This research has made use of {\small IRAF} which is
distributed by the National Optical Astronomy Observatory, which is
operated by the Association of Universities for Research in Astronomy
(AURA) under a cooperative agreement with the National Science
Foundation. This publication has made use of SIMBAD data base,
operated at CDS, Strasbourg, France.  It has also made use of NASA's
Astrophysics Data System and the services of the ESO Science Archive
Facility.  Based on observations collected at the European Southern
Observatory under ESO programmes 60.A-9022(C), 073.D-0609(A),
075.D-0177(A), 076.B-0055(A), 082.A-9011(A), 082.C-0831(A),
082.D-0061(A), 083.A-9013(A), 084.A-9016(A) and 085.A-9027(B).
This research was made possible through the use of the AAVSO Photometric 
All-Sky Survey (APASS), funded by the Robert Martin Ayers Sciences Fund 
and NSF AST-1412587.

%%%%%%%%%%%%%%%%%%%%%%%%%%%%%%%%%%%%%%%%%%%%%%%%%%



\bibliographystyle{mnras}
\bibliography{references}


%\begin{thebibliography}{99}
%\end{thebibliography}

\section{Supporting Information}
Supplementary appendices are available for download alongside this article on the MNRAS website (https://academic.oup.com/mnras).

Table 2, 3, C1, C2 and D are available in electronic form at \url{https://cdsarc.unistra.fr/viz-bin/cat/J/MNRAS/493/234}.

% Don't change these lines
\bsp	% typesetting comment
\label{lastpage}

%%%%%%%%%%%%%%%%%%%%%%%%%%%%%%%%%%%%%%%%%%%%%%%%%%

%%%%%%%%%%%%%%%%% APPENDICES %%%%%%%%%%%%%%%%%%%%%


%\appendix

%\section{Standard Stars}
\begin{landscape}
\begin{table}
\centering
\renewcommand{\thetable}{A1}
\caption{Log of observations of standard stars. Column 1 gives the object name. Columns 2 and 3 are right ascension (RA) in the units of time (\fh \, \fm \, \fs) and declination (DEC) in the units of angle (\fdg \, \farcm \, \farcs) respectively, column 4 lists the observation dates. Columns 5--7 present the exposure times for each grating. The signal-to-noise ratios for each grating are given in columns 8--10. Spectral type is listed along with references in columns 11--12.}
\label{tab:A1_standard_stars}
\begin{tabular}{lcclcccccccc}
\hline
Name& 				RA& 				DEC& 				Obs Date& 	\multicolumn{3}{c}{Exposure Time (s)}&	\multicolumn{3}{c}{SNR}&	\multicolumn{2}{c}{Spectral Type}\\[3pt]	
 & 				 	(J2000)& 		(J2000)& 			(June 2013)& R1200B& 	R1200Y& 	R1200R& 				R1200B& 	R1200Y& 	R1200R	&	&					Ref.\\
\hline
Feige~98& 		14:38:15.8&	+27:29:33.0& 	23& 				-& 			300& 		-&							-&				94&			-&				sdB9III:He1&	1\\[3pt]	
HD~130109& 	14:46:14.9&	+01:53:34.4& 	21; 23&			60; 20&		20& 			-&							179&			420&			-&				A0 IIInn&		2\\[3pt]	
BD+26\,2606& 	14:49:02.4&	+25:42:09.2&		23&				-&				150&			-&							-&				124&			-&				A5&				3\\[3pt]	
HD~147394&		16:19:44.4&	+46:18:48.1&		21; 23&			20&			20; 10&		-&							319&			376&			-&				B5 IV&			4\\[3pt]	
HD~152614&		16:54:00.5&	+10:09:55.3&		20; 22; 24&	30&			15&			30; 15&					257&			339&			179&			B8 V&			5\\[3pt]	
HD~153653&		17:00:29.4&	+06:35:01.3&		20; 22; 24&	60&			40&			15; 30&					36&			247&			69&			A7 V&			6\\[3pt]	
HD~157741&		17:24:33.8&	+15:36:21.8&		22; 23; 24&	-&				40; 30&		40&						-&				534&			157&			B9 V&			6\\[3pt]	
HD~158485&		17:26:04.8&	+58:39:06.8&		21; 23&			60&			60&			-&							52&			439&			-&				A4 V&			6\\[3pt]	
HD~160762&		17:39:27.9&	+46:00:22.8&		21; 23&			20&			10; 5&		-&							247&			302&			-&				B3 IV&			7\\[3pt]	
HD~161056&		17:43:47.0&	-07:04:46.6&		23&				-&				30&			-&							-&				464&			-&				B3 Vn&			8\\[3pt]	
HD~175426&		18:53:43.6&	+36:58:18.2&		20; 22; 24&	60&			40&			40&						245&			390&			204&			B2 V&			9\\[3pt]	
HD~177724&		19:05:24.6&	+13:51:48.5&		21; 23; 25&	20; 15&		10&			10&						184&			224&			395&			A0 IV-Vnn&	10\\[3pt]	
HD~186307&		19:41:57.6&	+40:15:14.9&		25&				-&				-&				60&						-&				-&				125&			A6 V&			6\\[3pt]	
HD~196504&		20:37:04.7&	+26:27:43.0&		25&				-&				-&				40&						-&				-&				235&			B8 V&			11\\[3pt]	
HD~199081&		20:53:14.8&	+44:23:14.1&		21; 23; 25&	20&			20&			20; 60&					234&			492&			118&			B5 V&			7\\[3pt]	
\hline
\end{tabular}
\\[3pt]
\begin{flushleft}
References: (1) \citet{Drilling2013}; (2) \citet{Abt1995}; (3) \citet{Fulbright2000}; (4) \citet{Morgan1973}; (5) \citet{Cowley1972}; (6) \citet{Cowley1969}; (7) \citet{Lesh1968}; (8) \citet{deVaucouleurs1957}; (9) \citet{Hill1977}; (10) \citet{Gray2003}; (11) \citet{Hube1970}.
\end{flushleft}
\end{table}
\end{landscape}

%\section{The $\rm H\alpha$ profiles}
\begin{figure*}
\centering
\renewcommand{\thefigure}{B1}
\includegraphics[width=17 cm]{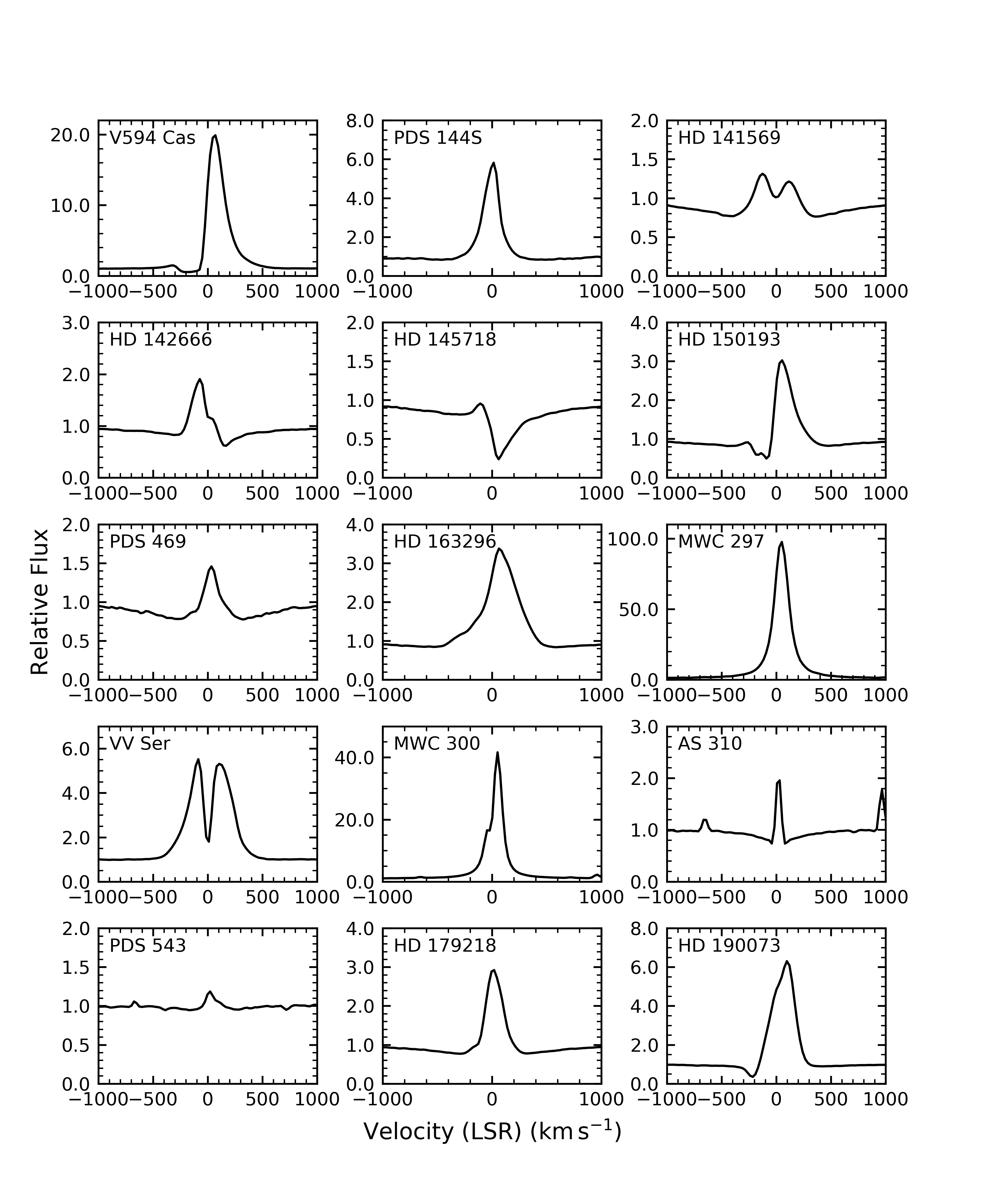}
\caption{The $\rm H\alpha$ line profiles for the 30 objects listed in Table 1.}
\label{fig:B1_EW}
\end{figure*}

\begin{figure*}
\centering
\renewcommand{\thefigure}{B1}
\includegraphics[width=17 cm]{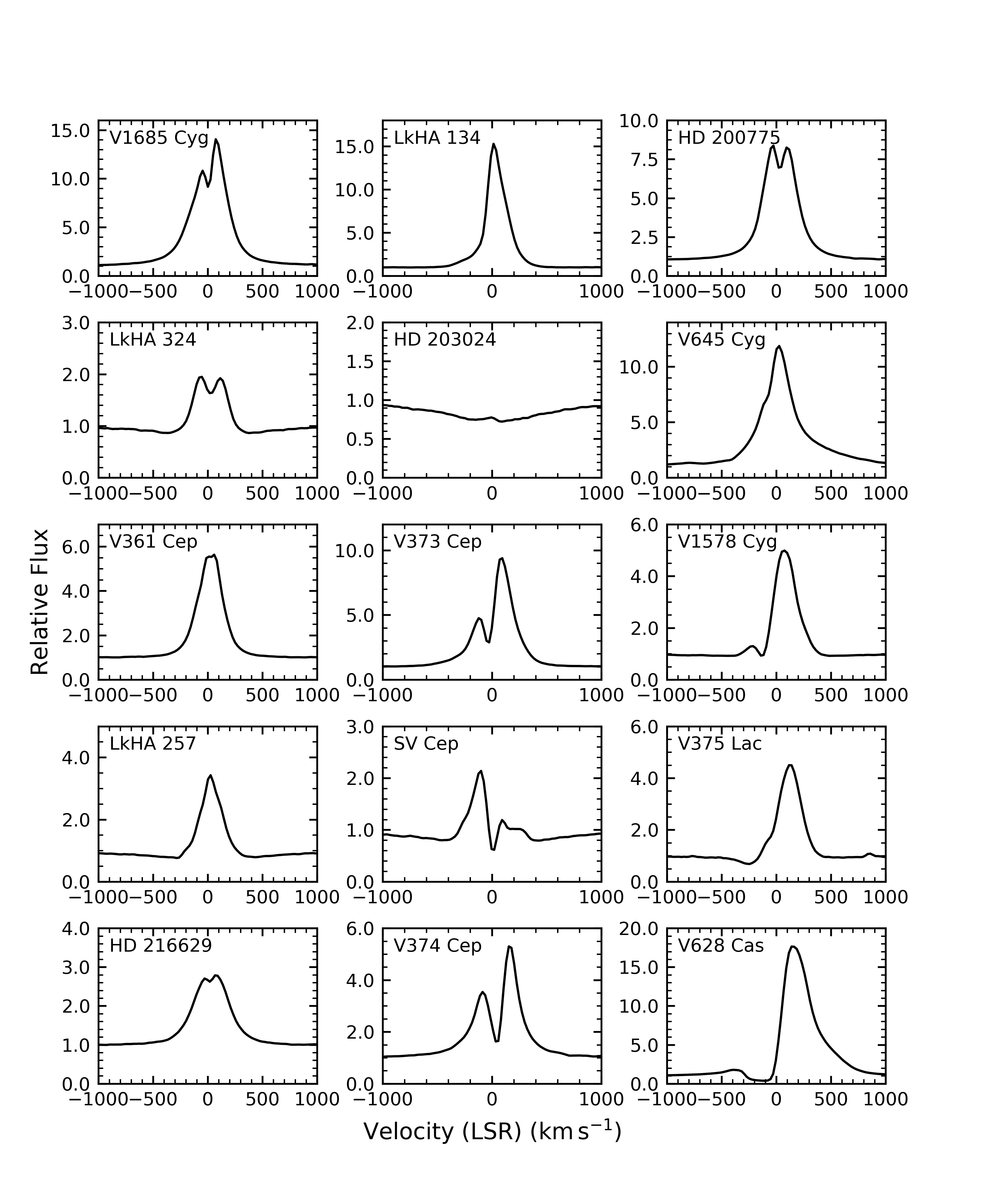}
\contcaption{.}
\label{fig:B1_EW2}
\end{figure*}

%\section{91 Herbig stars in the Fairlamb et al. (2015) Sample}
\begin{landscape}
\begin{table}
\renewcommand{\thetable}{C1}
\caption{Stellar parameters of 91 Herbig Ae/Be stars in the \protect\cite{Fairlamb2015} sample. Column 1 gives the object name. Columns 2 and 3 are right ascension (RA) in the units of time (\fh \, \fm \, \fs) and declination (DEC) in the units of angle (\fdg \, \farcm \, \farcs) respectively.
Columns 4--11 are effective temperature, surface gravity, visual extinction, distance, radius, luminosity, mass and age respectively.
Temperature and surface gravity are taken from \protect\cite{Fairlamb2015}. Distance is obtained from \protect\cite{Vioque2018}. The rest of parameters are redetermined in this work. This table is available in electronic form at \url{https://cdsarc.unistra.fr/viz-bin/cat/J/MNRAS/493/234}.}
\label{tab:C1_Fairlamb_para}
\begin{tabular}{lllcccccccc}
\hline
Name&	RA&	DEC&	$T_{\rm eff}$&	$\log(g)$&	$A_V$&		D&	$R_*$&	$\log(L_*)$&	$M_*$&		Age\\[3pt]
&	(J2000)&	(J2000)&	(K)&	[cm\,s$^{-2}$]&		(mag)&		(pc)&			($\rm R_{\sun}$)&		[$\rm L_{\sun}$]&		($\rm M_{\sun}$)&		(Myr)\\	
\hline
UX~Ori& 05:04:29.9& -03:47:14.2& $8500^{+250}_{-250}$& $3.90^{+0.25}_{-0.25}$& $0.99^{+0.04}_{-0.03}$& $324.9^{+9.3}_{-8.4}$& $1.80^{+0.09}_{-0.08}$& $1.18^{+0.09}_{-0.09}$& $1.82^{+0.07}_{-0.08}$& $7.08^{+0.51}_{-0.47}$\\[3pt]
PDS~174& 05:06:55.5& -03:21:13.3& $17000^{+2000}_{-2000}$& $4.10^{+0.40}_{-0.40}$& $3.36^{+0.29}_{-0.39}$& $398^{+10}_{-9}$& $1.12^{+0.23}_{-0.23}$& $1.98^{+0.35}_{-0.42}$& $3.18^{+0.57}_{-0.87}$& $2.75^{+3.01}_{-1.13}$\\[3pt]
V1012~Ori& 05:11:36.5& -02:22:48.4& $8500^{+250}_{-250}$& $4.38^{+0.15}_{-0.15}$& $1.15^{+0.05}_{-0.06}$& -$^{d}$& -& -& -& -\\[3pt]
HD~34282& 05:16:00.4& -09:48:35.3& $9500^{+250}_{-250}$& $4.40^{+0.15}_{-0.15}$& $0.55^{+0.03}_{-0.02}$& $311.5^{+7.9}_{-7.2}$& $1.48^{+0.06}_{-0.05}$& $1.21^{+0.08}_{-0.08}$& $1.87^{+0.11}_{-0.04}$& $8.91^{+2.89}_{-2.15}$\\[3pt]
HD~287823& 05:24:08.0& +02:27:46.8& $8375^{+125}_{-125}$& $4.23^{+0.11}_{-0.15}$& $0.39^{+0.03}_{-0.03}$& $359^{+12}_{-11}$& $2.06^{+0.11}_{-0.10}$& $1.27^{+0.07}_{-0.07}$& $1.80^{+0.07}_{-0.04}$& $6.61^{+0.64}_{-0.58}$\\[3pt]
HD~287841& 05:24:42.8& +01:43:48.2& $7750^{+250}_{-250}$& $4.27^{+0.12}_{-0.12}$& $0.33^{+0.03}_{-0.03}$& $366^{+10}_{-9}$& $1.85^{+0.08}_{-0.07}$& $1.04^{+0.09}_{-0.09}$& $1.60^{+0.07}_{-0.02}$& $8.91^{+0.64}_{-1.15}$\\[3pt]
HD~290409& 05:27:05.4& +00:25:07.6& $9750^{+500}_{-500}$& $4.25^{+0.25}_{-0.25}$& $0.41^{+0.03}_{-0.03}$& $455^{+31}_{-25}$& $1.87^{+0.16}_{-0.13}$& $1.45^{+0.16}_{-0.16}$& $2.18^{+0.01}_{-0.24}$& $4.79^{+2.97}_{-0.42}$\\[3pt]
HD~35929& 05:27:42.7& -08:19:38.4& $7000^{+250}_{-250}$& $3.47^{+0.11}_{-0.11}$& $0.35^{+0.05}_{-0.04}$& $387^{+13}_{-12}$& $6.39^{+0.38}_{-0.34}$& $1.94^{+0.11}_{-0.11}$& $3.40^{+0.31}_{-0.33}$& $1.05^{+0.36}_{-0.24}$\\[3pt]
HD~290500& 05:29:48.0& -00:23:43.5& $9500^{+500}_{-500}$& $3.80^{+0.40}_{-0.40}$& $1.03^{+0.06}_{-0.07}$ $^{b}$& $438^{+24}_{-20}$& $1.56^{+0.14}_{-0.12}$& $1.25^{+0.16}_{-0.17}$& $1.89^{+0.21}_{-0.11}$& $8.13^{+4.77}_{-3.00}$\\[3pt]
HD~244314& 05:30:19.0& +11:20:20.0& $8500^{+250}_{-250}$& $4.15^{+0.11}_{-0.15}$& $0.51^{+0.06}_{-0.06}$& $432^{+19}_{-17}$& $2.14^{+0.17}_{-0.15}$& $1.33^{+0.12}_{-0.12}$& $1.86^{+0.16}_{-0.07}$& $6.17^{+0.91}_{-1.27}$\\[3pt]
HK~Ori& 05:31:28.0& +12:09:10.1& $8500^{+500}_{-500}$& $4.22^{+0.13}_{-0.13}$& $1.31^{+0.20}_{-0.25}$& -$^{e}$& -& -& -& -\\[3pt]
HD~244604& 05:31:57.2& +11:17:41.3& $9000^{+250}_{-250}$& $3.99^{+0.15}_{-0.13}$& $0.59^{+0.04}_{-0.05}$& $421^{+19}_{-17}$& $2.77^{+0.19}_{-0.18}$& $1.65^{+0.11}_{-0.11}$& $2.26^{+0.22}_{-0.16}$& $3.80^{+0.77}_{-0.85}$\\[3pt]
UY~Ori& 05:32:00.3& -04:55:53.9& $9750^{+250}_{-250}$& $4.30^{+0.20}_{-0.20}$& $1.37^{+0.13}_{-0.14}$& -$^{d}$& -& -& -& -\\[3pt]
HD~245185& 05:35:09.6& +10:01:51.4& $10000^{+500}_{-500}$& $4.25^{+0.25}_{-0.25}$& $0.40^{+0.03}_{-0.02}$& $429^{+37}_{-29}$& $1.78^{+0.19}_{-0.14}$& $1.46^{+0.17}_{-0.16}$& $2.21^{+0.04}_{-0.24}$& $5.01^{+2.93}_{-0.84}$\\[3pt]
T~Ori& 05:35:50.4& -05:28:34.9& $9000^{+500}_{-500}$& $3.60^{+0.30}_{-0.30}$& $1.78^{+0.12}_{-0.13}$& $408^{+13}_{-11}$& $2.91^{+0.30}_{-0.27}$& $1.70^{+0.18}_{-0.18}$& $2.34^{+0.38}_{-0.27}$& $3.47^{+1.32}_{-1.18}$\\[3pt]
V380~Ori& 05:36:25.4& -06:42:57.6& $9750^{+750}_{-750}$& $4.00^{+0.35}_{-0.35}$& $2.05^{+0.26}_{-0.35}$& -$^{d}$& -& -& -& -\\[3pt]
HD~37258& 05:36:59.2& -06:09:16.3& $9750^{+500}_{-500}$& $4.25^{+0.25}_{-0.25}$& $0.54^{+0.03}_{-0.04}$& -$^{d}$& -& -& -& -\\[3pt]
HD~290770& 05:37:02.4& -01:37:21.3& $10500^{+250}_{-250}$& $4.20^{+0.30}_{-0.30}$& $0.34^{+0.05}_{-0.05}$& $399^{+21}_{-18}$& $2.08^{+0.17}_{-0.15}$& $1.67^{+0.11}_{-0.11}$& $2.44^{+0.05}_{-0.14}$& $3.63^{+1.05}_{-0.16}$\\[3pt]
BF~Ori& 05:37:13.2& -06:35:00.5& $9000^{+250}_{-250}$& $3.97^{+0.15}_{-0.13}$& $0.74^{+0.06}_{-0.07}$& $389^{+14}_{-12}$& $2.25^{+0.16}_{-0.15}$& $1.47^{+0.11}_{-0.11}$& $2.01^{+0.15}_{-0.08}$& $5.13^{+0.76}_{-0.86}$\\[3pt]
HD~37357& 05:37:47.0& -06:42:30.2& $9500^{+250}_{-250}$& $4.10^{+0.10}_{-0.10}$& $0.40^{+0.03}_{-0.03}$& -$^{d}$& -& -& -& -\\[3pt]
HD~290764& 05:38:05.2& -01:15:21.6& $7875^{+375}_{-375}$& $3.90^{+0.17}_{-0.15}$& $0.58^{+0.07}_{-0.08}$& $398^{+18}_{-15}$& $2.58^{+0.22}_{-0.20}$& $1.36^{+0.15}_{-0.16}$& $1.94^{+0.24}_{-0.20}$& $5.25^{+1.83}_{-1.36}$\\[3pt]
HD~37411& 05:38:14.5& -05:25:13.3& $9750^{+250}_{-250}$& $4.35^{+0.15}_{-0.15}$& $0.59^{+0.07}_{-0.07}$& -$^{e}$& -& -& -& -\\[3pt]
V599~Ori& 05:38:58.6& -07:16:45.6& $8000^{+250}_{-250}$& $3.72^{+0.13}_{-0.12}$& $4.92^{+0.25}_{-0.35}$& $410^{+12}_{-11}$& $3.34^{+0.60}_{-0.62}$& $1.61^{+0.20}_{-0.23}$& $2.34^{+0.41}_{-0.39}$& $3.24^{+2.01}_{-1.15}$\\[3pt]
V350~Ori& 05:40:11.7& -09:42:11.0& $9000^{+250}_{-250}$& $4.18^{+0.11}_{-0.16}$& $1.06^{+0.08}_{-0.08}$& -$^{d}$& -& -& -& -\\[3pt]
HD~250550& 06:01:59.9& +16:30:56.7& $11000^{+500}_{-500}$& $3.80^{+0.40}_{-0.40}$& $1.55^{+0.10}_{-0.13}$ $^{b}$& $697^{+94}_{-64}$& $15.35^{+3.09}_{-2.38}$& $3.49^{+0.24}_{-0.23}$& $8.81^{+1.83}_{-1.51}$& $0.08^{+0.06}_{-0.03}$\\[3pt]
V791~Mon& 06:02:14.8& -10:00:59.5& $15000^{+1500}_{-1500}$& $4.30^{+0.16}_{-0.16}$& $1.48^{+0.11}_{-0.13}$& $887^{+53}_{-44}$& $3.47^{+0.44}_{-0.39}$& $2.74^{+0.27}_{-0.29}$& $4.31^{+0.92}_{-0.40}$& $0.81^{+0.42}_{-0.36}$\\[3pt]
PDS~124& 06:06:58.4& -05:55:06.7& $10250^{+250}_{-250}$& $4.30^{+0.20}_{-0.20}$& $1.79^{+0.02}_{-0.01}$& $853^{+64}_{-51}$& $2.06^{+0.18}_{-0.14}$& $1.62^{+0.11}_{-0.10}$& $2.33^{+0.02}_{-0.04}$& $3.98^{+0.59}_{-0.26}$\\[3pt]
LkHa~339& 06:10:57.8& -06:14:39.6& $10500^{+250}_{-250}$& $4.20^{+0.20}_{-0.20}$& $3.11^{+0.36}_{-0.54}$& $857^{+33}_{-29}$& $2.50^{+0.67}_{-0.67}$& $1.83^{+0.25}_{-0.31}$& $2.47^{+0.46}_{-0.27}$& $3.24^{+2.13}_{-1.20}$\\[3pt]
VY~Mon& 06:31:06.9& +10:26:04.9& $12000^{+4000}_{-4000}$& $3.75^{+0.50}_{-0.50}$& $5.62^{+0.34}_{-0.51}$& -$^{d}$& -& -& -& -\\[3pt]
R~Mon& 06:39:09.9& +08:44:09.5& $12000^{+2000}_{-2000}$& $4.00^{+0.11}_{-0.24}$& $2.47^{+0.23}_{-0.29}$& -$^{e}$& -& -& -& -\\[3pt]
V590~Mon& 06:40:44.6& +09:48:02.1& $12500^{+1000}_{-1000}$& $4.20^{+0.30}_{-0.30}$& $1.03^{+0.19}_{-0.24}$& -$^{d}$& -& -& -& -\\[3pt]
PDS~24& 06:48:41.6& -16:48:05.6& $10500^{+500}_{-500}$& $4.20^{+0.30}_{-0.30}$& $1.45^{+0.10}_{-0.12}$& $1130^{+42}_{-37}$& $1.59^{+0.15}_{-0.15}$& $1.44^{+0.16}_{-0.17}$& $2.10^{+0.32}_{-0.12}$& $6.31^{+2.60}_{-2.42}$\\[3pt]
\hline
\end{tabular}
\end{table}
\end{landscape}

\begin{landscape}
\begin{table}
\renewcommand{\thetable}{C1}
\contcaption{.}
\label{tab:C1_Fairlamb_para2}
\begin{tabular}{lllcccccccc}
\hline
Name&	RA&	DEC&	$T_{\rm eff}$&	$\log(g)$&	$A_V$&		D&	$R_*$&	$\log(L_*)$&	$M_*$&		Age\\[3pt]
&	(J2000)&	(J2000)&	(K)&	[cm\,s$^{-2}$]&		(mag)&		(pc)&			($\rm R_{\sun}$)&		[$\rm L_{\sun}$]&		($\rm M_{\sun}$)&		(Myr)\\	
\hline
PDS~130& 06:49:58.5& -07:38:52.2& $10500^{+250}_{-250}$& $3.90^{+0.20}_{-0.20}$& $2.38^{+0.13}_{-0.15}$& $1316^{+62}_{-53}$& $2.68^{+0.34}_{-0.31}$& $1.89^{+0.14}_{-0.15}$& $2.56^{+0.28}_{-0.17}$& $2.95^{+0.68}_{-0.76}$\\[3pt]
PDS~229& 06:55:40.0& -03:09:50.5& $12500^{+250}_{-250}$& $4.20^{+0.20}_{-0.20}$& $2.24^{+0.15}_{-0.17}$& -$^{d}$& -& -& -& -\\[3pt]
GU~CMa& 07:01:49.5& -11:18:03.3& $22500^{+1500}_{-1500}$& $3.90^{+0.40}_{-0.40}$& $0.80^{+0.10}_{-0.12}$& -$^{e}$& -& -& -& -\\[3pt]
HT~CMa& 07:02:42.5& -11:26:11.8& $10500^{+500}_{-500}$& $4.00^{+0.20}_{-0.20}$& $1.61^{+0.09}_{-0.09}$ $^{b}$& $1121^{+83}_{-66}$& $3.19^{+0.41}_{-0.33}$& $2.05^{+0.19}_{-0.18}$& $2.87^{+0.44}_{-0.36}$& $2.14^{+0.95}_{-0.69}$\\[3pt]
Z~CMa& 07:03:43.1& -11:33:06.2& $8500^{+500}_{-500}$& $2.53^{+0.17}_{-0.17}$& $3.09^{+0.23}_{-0.28}$ $^{b}$& -$^{d}$& -& -& -& -\\[3pt]
HU~CMa& 07:04:06.7& -11:26:08.5& $13000^{+250}_{-250}$& $4.20^{+0.20}_{-0.20}$& $1.29^{+0.07}_{-0.07}$& $1170^{+100}_{-80}$& $2.72^{+0.35}_{-0.28}$& $2.28^{+0.14}_{-0.13}$& $3.26^{+0.26}_{-0.02}$& $1.70^{+0.12}_{-0.35}$\\[3pt]
HD~53367& 07:04:25.5& -10:27:15.7& $29500^{+1000}_{-1000}$& $4.25^{+0.25}_{-0.25}$& $2.12^{+0.13}_{-0.14}$& -$^{d}$& -& -& -& -\\[3pt]
PDS~241& 07:08:38.7& -04:19:04.8& $26000^{+1500}_{-1500}$& $4.00^{+0.30}_{-0.30}$& $2.92^{+0.27}_{-0.37}$& $2890^{+600}_{-400}$& $6.51^{+2.63}_{-1.90}$& $4.24^{+0.39}_{-0.40}$& $12.60^{+5.09}_{-2.58}$& $0.06^{+0.06}_{-0.03}$\\[3pt]
NX~Pup& 07:19:28.2& -44:35:11.2& $7000^{+250}_{-250}$& $3.78^{+0.13}_{-0.13}$& $0.26^{+0.14}_{-0.17}$& -$^{d}$& -& -& -& -\\[3pt]
PDS~27& 07:19:35.9& -17:39:17.9& $17500^{+3500}_{-3500}$& $3.16^{+0.27}_{-0.27}$& $4.96^{+0.33}_{-0.49}$& $2550^{+460}_{-310}$& $13.84^{+5.77}_{-4.42}$& $4.21^{+0.62}_{-0.72}$& $12.89^{+9.62}_{-5.67}$& $0.04^{+0.15}_{-0.03}$\\[3pt]
PDS~133& 07:25:04.9& -25:45:49.6& $14000^{+2000}_{-2000}$& $4.08^{+0.12}_{-0.11}$& $1.92^{+0.09}_{-0.10}$& $1475^{+92}_{-76}$& $2.09^{+-0.11}_{-0.01}$& $2.18^{+0.19}_{-0.27}$& $3.26^{+0.40}_{-0.39}$& $2.00^{+1.31}_{-0.65}$\\[3pt]
HD~59319$^{c}$& 07:28:36.7& -21:57:49.2& $12500^{+500}_{-500}$& $3.50^{+0.20}_{-0.20}$& $0.06^{+0.02}_{-0.01}$& $668^{+39}_{-33}$& $4.00^{+0.28}_{-0.22}$& $2.54^{+0.13}_{-0.12}$& $3.89^{+0.41}_{-0.36}$& $1.00^{+0.32}_{-0.26}$\\[3pt]
PDS~134& 07:32:26.6& -21:55:35.7& $14000^{+500}_{-500}$& $3.40^{+0.30}_{-0.30}$& $1.70^{+0.06}_{-0.07}$& $2550^{+370}_{-260}$& $5.00^{+0.92}_{-0.69}$& $2.94^{+0.21}_{-0.19}$& $5.05^{+0.94}_{-0.67}$& $0.49^{+0.25}_{-0.20}$\\[3pt]
HD~68695& 08:11:44.5& -44:05:08.7& $9250^{+250}_{-250}$& $4.40^{+0.15}_{-0.15}$& $0.37^{+0.05}_{-0.05}$& $396^{+10}_{-9}$& $1.86^{+0.10}_{-0.09}$& $1.35^{+0.09}_{-0.09}$& $1.98^{+0.11}_{-0.03}$& $5.76^{+0.70}_{-0.39}$\\[3pt]
HD~72106& 08:29:34.8& -38:36:21.1& $8750^{+250}_{-250}$& $3.89^{+0.13}_{-0.12}$& $0.00^{+0.03}_{-0.00}$& -$^{d}$& -& -& -& -\\[3pt]
TYC~8581-2002-1$^{c}$& 08:44:23.6& -59:56:57.8& $9750^{+250}_{-250}$& $4.00^{+0.10}_{-0.10}$& $1.54^{+0.05}_{-0.07}$& $558^{+12}_{-11}$& $1.98^{+0.10}_{-0.11}$& $1.50^{+0.09}_{-0.09}$& $2.14^{+0.11}_{-0.03}$& $4.79^{+0.22}_{-0.32}$\\[3pt]
PDS~33& 08:48:45.6& -40:48:21.0& $9750^{+250}_{-250}$& $4.40^{+0.15}_{-0.15}$& $1.03^{+0.03}_{-0.02}$& $951^{+43}_{-37}$& $1.79^{+0.11}_{-0.09}$& $1.41^{+0.10}_{-0.09}$& $2.16^{+0.04}_{-0.19}$& $5.01^{+2.24}_{-0.33}$\\[3pt]
HD~76534& 08:55:08.7& -43:27:59.8& $19000^{+500}_{-500}$& $4.10^{+0.20}_{-0.20}$& $0.95^{+0.07}_{-0.08}$& $911^{+55}_{-46}$& $6.61^{+0.69}_{-0.60}$& $3.71^{+0.13}_{-0.13}$& $8.49^{+0.90}_{-0.88}$& $0.13^{+0.05}_{-0.03}$\\[3pt]
PDS~281$^{c}$& 08:55:45.9& -44:25:14.1& $16000^{+1500}_{-1500}$& $3.50^{+0.30}_{-0.30}$& $2.32^{+0.09}_{-0.09}$& $932^{+47}_{-40}$& $10.14^{+1.03}_{-0.90}$& $3.78^{+0.24}_{-0.25}$& $9.31^{+2.03}_{-1.70}$& $0.09^{+0.07}_{-0.04}$\\[3pt]
PDS~286& 09:06:00.0& -47:18:58.1& $30000^{+3000}_{-3000}$& $4.25^{+0.16}_{-0.16}$& $6.44^{+0.35}_{-0.56}$& $1820^{+210}_{-150}$& $17.72^{+6.33}_{-5.58}$& $5.36^{+0.43}_{-0.51}$& $36.13^{+24.39}_{-14.82}$& $0.01^{+0.01}_{-0.00}$\\[3pt]
PDS~297& 09:42:40.3& -56:15:34.1& $10750^{+250}_{-250}$& $4.00^{+0.20}_{-0.20}$& $1.23^{+0.06}_{-0.06}$& $1590^{+140}_{-110}$& $3.44^{+0.43}_{-0.34}$& $2.15^{+0.14}_{-0.13}$& $3.06^{+0.35}_{-0.29}$& $1.82^{+0.58}_{-0.47}$\\[3pt]
HD~85567& 09:50:28.5& -60:58:02.9& $13000^{+500}_{-500}$& $3.50^{+0.30}_{-0.30}$& $1.00^{+0.16}_{-0.20}$& $1023^{+53}_{-45}$& $8.60^{+1.29}_{-1.22}$& $3.28^{+0.19}_{-0.20}$& $6.79^{+1.16}_{-1.04}$& $0.19^{+0.13}_{-0.07}$\\[3pt]
HD~87403& 10:02:51.4& -59:16:54.6& $10000^{+250}_{-250}$& $3.30^{+0.10}_{-0.10}$& $0.34^{+0.04}_{-0.05}$& $1910^{+280}_{-190}$& $10.49^{+1.81}_{-1.30}$& $2.99^{+0.18}_{-0.16}$& $6.22^{+0.98}_{-0.77}$& $0.21^{+0.11}_{-0.08}$\\[3pt]
PDS~37& 10:10:00.3& -57:02:07.3& $17500^{+3500}_{-3500}$& $2.94^{+0.35}_{-0.35}$& $5.64^{+0.38}_{-0.60}$& $1930^{+360}_{-230}$& $11.33^{+5.23}_{-4.04}$& $4.03^{+0.65}_{-0.77}$& $11.19^{+8.37}_{-5.14}$& $0.06^{+0.27}_{--0.10}$\\[3pt]
HD~305298& 10:33:04.9& -60:19:51.3& $34000^{+1000}_{-1000}$& $4.31^{+0.16}_{-0.16}$& $1.51^{+0.12}_{-0.14}$& $4040^{+630}_{-440}$& $6.19^{+1.46}_{-1.07}$& $4.66^{+0.23}_{-0.22}$& $19.41^{+3.25}_{-2.58}$& $0.03^{+0.03}_{-0.01}$\\[3pt]
HD~94509& 10:53:27.2& -58:25:24.5& $11500^{+1000}_{-1000}$& $2.90^{+0.40}_{-0.40}$& $0.35^{+0.01}_{-0.02}$& $1830^{+210}_{-150}$& $9.32^{+1.14}_{-0.87}$& $3.13^{+0.25}_{-0.24}$& $6.45^{+1.40}_{-1.17}$& $0.20^{+0.18}_{-0.09}$\\[3pt]
HD~95881& 11:01:57.6& -71:30:48.3& $10000^{+250}_{-250}$& $3.20^{+0.10}_{-0.10}$& $0.72^{+0.03}_{-0.04}$ $^{b}$& $1168^{+82}_{-66}$& $12.29^{+1.10}_{-0.95}$& $3.13^{+0.12}_{-0.11}$& $7.02^{+0.62}_{-0.67}$& $0.14^{+0.05}_{-0.03}$\\[3pt]
HD~96042$^{c}$& 11:03:40.5& -59:25:59.0& $25500^{+1500}_{-1500}$& $3.80^{+0.20}_{-0.20}$& $1.13^{+0.08}_{-0.08}$& $3100^{+510}_{-350}$& $15.68^{+3.38}_{-2.37}$& $4.97^{+0.27}_{-0.25}$& $24.10^{+7.71}_{-5.24}$& $0.01^{+0.01}_{-0.01}$\\[3pt]
HD~97048& 11:08:03.3& -77:39:17.4& $10500^{+500}_{-500}$& $4.30^{+0.20}_{-0.20}$& $1.37^{+0.05}_{-0.06}$& $184.8^{+2.2}_{-2.1}$& $2.28^{+0.09}_{-0.10}$& $1.76^{+0.12}_{-0.12}$& $2.40^{+0.15}_{-0.03}$& $3.55^{+0.25}_{-0.60}$\\[3pt]
HD~98922& 11:22:31.6& -53:22:11.4& $10500^{+250}_{-250}$& $3.60^{+0.10}_{-0.10}$& $0.45^{+0.07}_{-0.07}$& $689^{+28}_{-25}$& $12.02^{+0.99}_{-0.86}$& $3.20^{+0.11}_{-0.11}$& $7.17^{+0.67}_{-0.69}$& $0.14^{+0.05}_{-0.03}$\\[3pt]
HD~100453& 11:33:05.5& -54:19:28.5& $7250^{+250}_{-250}$& $4.08^{+0.15}_{-0.13}$& $0.12^{+0.04}_{-0.04}$& $104.2^{+0.7}_{-0.69}$& $1.68^{+0.04}_{-0.05}$& $0.85^{+0.08}_{-0.09}$& $1.48^{+0.03}_{-0.02}$& $11.00^{+0.80}_{-1.00}$\\[3pt]
HD~100546& 11:33:25.4& -70:11:41.2& $9750^{+500}_{-500}$& $4.34^{+0.06}_{-0.06}$& $0.17^{+0.02}_{-0.03}$& $110^{+1}_{-1}$& $1.87^{+0.04}_{-0.05}$& $1.45^{+0.11}_{-0.11}$& $2.18^{+0.02}_{-0.17}$& $4.79^{+1.82}_{-0.22}$\\[3pt]
HD~101412& 11:39:44.4& -60:10:27.7& $9750^{+250}_{-250}$& $4.30^{+0.20}_{-0.20}$& $0.69^{+0.03}_{-0.03}$& $411.3^{+8.1}_{-7.6}$& $2.75^{+0.10}_{-0.10}$& $1.79^{+0.08}_{-0.08}$& $2.44^{+0.16}_{-0.13}$& $3.24^{+0.48}_{-0.55}$\\[3pt]
PDS~344& 11:40:32.8& -64:32:05.7& $15250^{+500}_{-500}$& $4.30^{+0.20}_{-0.20}$& $1.27^{+0.07}_{-0.07}$& $2440^{+170}_{-140}$& $2.36^{+0.26}_{-0.22}$& $2.43^{+0.15}_{-0.14}$& $3.85^{+0.33}_{-0.20}$& $1.29^{+0.53}_{-0.34}$\\[3pt]
\hline
\end{tabular}
\end{table}
\end{landscape}

\begin{landscape}
\begin{table}
\renewcommand{\thetable}{C1}
\contcaption{.}
\label{tab:C1_Fairlamb_para3}
\begin{tabular}{lllcccccccc}
\hline
Name&	RA&	DEC&	$T_{\rm eff}$&	$\log(g)$&	$A_V$&		D&	$R_*$&	$\log(L_*)$&	$M_*$&		Age\\[3pt]
&	(J2000)&	(J2000)&	(K)&	[cm\,s$^{-2}$]&		(mag)&		(pc)&			($\rm R_{\sun}$)&		[$\rm L_{\sun}$]&		($\rm M_{\sun}$)&		(Myr)\\	
\hline
HD~104237& 12:00:05.0& -78:11:34.5& $8000^{+250}_{-250}$& $3.89^{+0.12}_{-0.12}$& $0.29^{+0.05}_{-0.06}$& -$^{d}$& -& -& -& -\\[3pt]
V1028~Cen& 13:01:17.8& -48:53:18.7& $14000^{+500}_{-500}$& $3.80^{+0.30}_{-0.30}$& $0.81^{+0.11}_{-0.13}$& -$^{d}$& -& -& -& -\\[3pt]
PDS~361& 13:03:21.4& -62:13:26.2& $18500^{+1000}_{-1000}$& $3.80^{+0.30}_{-0.30}$& $2.20^{+0.13}_{-0.15}$& $2950^{+430}_{-300}$& $4.37^{+1.01}_{-0.75}$& $3.30^{+0.27}_{-0.26}$& $6.26^{+1.37}_{-0.56}$& $0.30^{+0.14}_{-0.13}$\\[3pt]
HD~114981& 13:14:40.6& -38:39:05.6& $16000^{+500}_{-500}$& $3.60^{+0.20}_{-0.20}$& $0.15^{+0.02}_{-0.03}$& $705^{+57}_{-44}$& $5.91^{+0.55}_{-0.46}$& $3.31^{+0.13}_{-0.12}$& $6.46^{+0.68}_{-0.62}$& $0.25^{+0.09}_{-0.07}$\\[3pt]
PDS~364& 13:20:03.5& -62:23:54.0& $12500^{+1000}_{-1000}$& $4.20^{+0.20}_{-0.20}$& $2.02^{+0.18}_{-0.22}$& -$^{d}$& -& -& -& -\\[3pt]
PDS~69& 13:57:44.1& -39:58:44.2& $15000^{+2000}_{-2000}$& $4.00^{+0.35}_{-0.35}$& $1.65^{+0.19}_{-0.23}$& $643^{+33}_{-28}$& $3.60^{+0.60}_{-0.56}$& $2.77^{+0.35}_{-0.39}$& $4.39^{+1.29}_{-0.74}$& $0.78^{+0.73}_{-0.42}$\\[3pt]
DG~Cir& 15:03:23.7& -63:22:58.8& $11000^{+3000}_{-3000}$& $4.41^{+0.18}_{-0.18}$& $4.00^{+0.25}_{-0.32}$& $833^{+52}_{-43}$& $1.89^{+0.41}_{-0.38}$& $1.67^{+0.59}_{-0.75}$& $2.45^{+0.84}_{-0.83}$& $3.98^{+12.22}_{-2.47}$\\[3pt]
HD~132947$^{c}$& 15:04:56.0& -63:07:52.6& $10250^{+250}_{-250}$& $3.90^{+0.10}_{-0.10}$& $0.29^{+0.03}_{-0.02}$& $382^{+15}_{-13}$& $2.33^{+0.13}_{-0.11}$& $1.73^{+0.09}_{-0.08}$& $2.35^{+0.11}_{-0.04}$& $3.72^{+0.26}_{-0.48}$\\[3pt]
HD~135344B& 15:15:48.4& -37:09:16.0& $6375^{+125}_{-125}$& $3.94^{+0.12}_{-0.12}$& $0.53^{+0.18}_{-0.22}$& $135.8^{+2.4}_{-2.3}$& $2.42^{+0.28}_{-0.28}$& $0.94^{+0.13}_{-0.14}$& $1.67^{+0.18}_{-0.16}$& $6.46^{+1.86}_{-1.56}$\\[3pt]
HD~139614& 15:40:46.3& -42:29:53.5& $7750^{+250}_{-250}$& $4.31^{+0.12}_{-0.12}$& $0.21^{+0.04}_{-0.04}$& $134.7^{+1.6}_{-1.6}$& $1.49^{+0.05}_{-0.05}$& $0.86^{+0.08}_{-0.09}$& $1.61^{+0.02}_{-0.11}$& $10.20^{+11.20}_{-0.65}$\\[3pt]
PDS~144S$^{a}$& 15:49:15.3& -26:00:54.7& $7750^{+250}_{-250}$& $4.13^{+0.14}_{-0.16}$& $1.02^{+0.07}_{-0.07}$& -$^{d}$& -& -& -& -\\[3pt]
HD~141569$^{a}$& 15:49:57.7& -03:55:16.3& $9750^{+250}_{-250}$& $4.35^{+0.15}_{-0.15}$& $0.45^{+0.03}_{-0.03}$& $110.63^{+0.91}_{-0.88}$& $1.78^{+0.05}_{-0.04}$& $1.41^{+0.07}_{-0.07}$& $2.16^{+0.04}_{-0.15}$& $5.01^{+1.60}_{-0.33}$\\[3pt]
HD~141926& 15:54:21.7& -55:19:44.3& $28000^{+1500}_{-1500}$& $3.75^{+0.25}_{-0.25}$& $2.61^{+0.17}_{-0.20}$& $1340^{+150}_{-110}$& $11.61^{+2.59}_{-2.05}$& $4.87^{+0.27}_{-0.26}$& $21.92^{+6.34}_{-4.51}$& $0.02^{+0.01}_{-0.01}$\\[3pt]
HD~142666$^{a}$& 15:56:40.0& -22:01:40.0& $7500^{+250}_{-250}$& $4.13^{+0.11}_{-0.16}$& $0.95^{+0.08}_{-0.08}$& $148.3^{+2}_{-1.9}$& $2.19^{+0.12}_{-0.12}$& $1.14^{+0.10}_{-0.11}$& $1.69^{+0.13}_{-0.11}$& $7.41^{+1.50}_{-1.38}$\\[3pt]
HD~142527& 15:56:41.8& -42:19:23.2& $6500^{+250}_{-250}$& $3.93^{+0.08}_{-0.08}$& $1.20^{+0.04}_{-0.05}$ $^{b}$& $157.3^{+2}_{-1.9}$& $3.92^{+0.13}_{-0.15}$& $1.39^{+0.09}_{-0.10}$& $2.26^{+0.20}_{-0.13}$& $3.09^{+0.46}_{-0.63}$\\[3pt]
HD~144432& 16:06:57.9& -27:43:09.7& $7500^{+250}_{-250}$& $4.05^{+0.17}_{-0.14}$& $0.48^{+0.04}_{-0.05}$& $155.4^{+2.4}_{-2.2}$& $2.32^{+0.09}_{-0.09}$& $1.18^{+0.09}_{-0.09}$& $1.74^{+0.13}_{-0.10}$& $6.76^{+1.18}_{-1.14}$\\[3pt]
HD~144668& 16:08:34.2& -39:06:18.3& $8500^{+250}_{-250}$& $3.75^{+0.13}_{-0.12}$& $0.87^{+0.03}_{-0.03}$& $161.1^{+3.1}_{-2.9}$& $4.34^{+0.16}_{-0.15}$& $1.95^{+0.08}_{-0.08}$& $2.97^{+0.17}_{-0.21}$& $1.74^{+0.40}_{-0.23}$\\[3pt]
HD~145718$^{a}$& 16:13:11.5& -22:29:06.6& $8000^{+250}_{-250}$& $4.37^{+0.15}_{-0.15}$& $1.23^{+0.06}_{-0.06}$& $152.5^{+3.2}_{-3}$& $1.85^{+0.10}_{-0.09}$& $1.10^{+0.10}_{-0.10}$& $1.67^{+0.06}_{-0.02}$& $8.13^{+0.58}_{-0.88}$\\[3pt]
PDS~415& 16:18:37.2& -24:05:18.1& $6250^{+250}_{-250}$& $4.47^{+0.15}_{-0.15}$& $1.54^{+0.12}_{-0.13}$& -$^{d}$& -& -& -& -\\[3pt]
HD~150193$^{a}$& 16:40:17.9& -23:53:45.1& $9000^{+250}_{-250}$& $4.27^{+0.17}_{-0.17}$& $1.80^{+0.15}_{-0.17}$& $150.8^{+2.7}_{-2.5}$& $2.32^{+0.24}_{-0.23}$& $1.50^{+0.13}_{-0.14}$& $2.05^{+0.19}_{-0.12}$& $4.90^{+0.99}_{-1.01}$\\[3pt]
AK~Sco& 16:54:44.8& -36:53:18.5& $6250^{+250}_{-250}$& $4.26^{+0.10}_{-0.10}$& $0.51^{+0.03}_{-0.02}$& $140.6^{+2.1}_{-2}$& $2.21^{+0.07}_{-0.06}$& $0.82^{+0.09}_{-0.09}$& $1.57^{+0.11}_{-0.12}$& $7.25^{+1.87}_{-1.08}$\\[3pt]
PDS~431& 16:54:59.1& -43:21:49.7& $10500^{+500}_{-500}$& $3.70^{+0.20}_{-0.20}$& $2.10^{+0.12}_{-0.14}$& $1810^{+160}_{-120}$& $3.20^{+0.52}_{-0.43}$& $2.05^{+0.21}_{-0.21}$& $2.87^{+0.52}_{-0.40}$& $2.14^{+1.10}_{-0.79}$\\[3pt]
KK~Oph& 17:10:08.1& -27:15:18.8& $8500^{+500}_{-500}$& $4.38^{+0.15}_{-0.15}$& $1.57^{+0.10}_{-0.12}$ $^{b}$& -$^{d}$& -& -& -& -\\[3pt]
HD~163296$^{a}$& 17:56:21.2& -21:57:21.8& $9250^{+250}_{-250}$& $4.30^{+0.20}_{-0.20}$& $0.37^{+0.04}_{-0.05}$& $101.5^{+2}_{-1.9}$& $1.86^{+0.08}_{-0.08}$& $1.36^{+0.08}_{-0.09}$& $2.04^{+0.07}_{-0.10}$& $5.50^{+0.67}_{-0.13}$\\[3pt]
MWC~297$^{a}$& 18:27:39.5& -03:49:52.1& $24500^{+1500}_{-1500}$& $4.00^{+0.30}_{-0.30}$& $7.87^{+0.41}_{-0.64}$& $375^{+22}_{-18}$& $9.12^{+3.09}_{-3.00}$& $4.43^{+0.36}_{-0.46}$& $14.79^{+5.59}_{-4.61}$& $0.04^{+0.06}_{-0.02}$\\[3pt]
\hline
\end{tabular}
\\
\begin{flushleft}
{\bf Notes.}
$^{(a)}$ Stars which are also in this work's sample.
$^{(b)}$ Stars of which photometry were used for the photometry fitting different from values listed in \citet[table A1]{Fairlamb2015} (see text for details). 
$^{(c)}$ Stars which have small emissions.
$^{(d)}$ Stars which have low quality parallaxes in the {\em Gaia} DR2 Catalogue (see the text for discussion).
$^{(e)}$ Stars which do not have parallaxes in the {\em Gaia} DR2 Catalogue.
\end{flushleft}
\end{table}
\end{landscape}

\begin{landscape}
\begin{table}
\renewcommand{\thetable}{C2}
\caption{The equivalent width measurements and accretion rates of 91 Herbig Ae/Be stars in the \protect\cite{Fairlamb2015} sample. 
Columns 2--9 present observed equivalent width, intrinsic equivalent width, corrected equivalent width, continuum flux density at central wavelength of the $\rm H\alpha$ profile, line flux, line luminosity, accretion luminosity and mass accretion rate respectively.
Equivalent widths are taken from \protect\cite{Fairlamb2017}. The rest of parameters are redetermined in this work. This table is available in electronic form at \url{https://cdsarc.unistra.fr/viz-bin/cat/J/MNRAS/493/234}.}
\label{tab:C2_Fairlamb_Mdot}
\begin{tabular}{lcccccccc}
\hline
Name&	$EW_{\rm obs}$&	$EW_{\rm int}$&	$EW_{\rm cor}$&	$F_{\lambda}$&		$F_{\rm line}$&	$\log(L_{\rm line})$&	$\log(L_{\rm acc})$&		$\log(\dot M_{\rm acc})$\\[3pt]
&	(\AA)&		(\AA)&		(\AA)&	(W\,m$^{-2}$\,\AA$^{-1}$)&		(W\,m$^{-2}$)&		[$\rm L_{\sun}$]&	[$\rm L_{\sun}$]&	[$\rm M_{\sun}$\,$\rm yr^{-1}$]\\	
\hline
UX~Ori& $2.39\pm0.34$& $14.65\pm1.47$& $-12.26\pm1.51$& $4.03\times10^{-16}$& $(4.94\pm0.61)\times10^{-15}$& $-1.79^{+0.07}_{-0.08}$& $0.30^{+0.22}_{-0.23}$& $-7.20^{+0.23}_{-0.23}$\\[3pt]
PDS~174& $-54.19\pm1.34$& $6.46\pm0.65$& $-60.65\pm1.49$& $3.59\times10^{-16}$& $(2.18\pm0.05)\times10^{-14}$& $-0.97^{+0.03}_{-0.03}$& $1.12^{+0.14}_{-0.14}$& $-6.82^{+0.15}_{-0.10}$\\[3pt]
V1012~Ori& $4.74\pm0.55$& $16.36\pm1.64$& $-11.62\pm1.73$& $1.06\times10^{-16}$& $(1.23\pm0.18)\times10^{-15}$& -& -& -\\[3pt]
HD~34282& $4.18\pm0.40$& $16.23\pm1.62$& $-12.05\pm1.67$& $3.83\times10^{-16}$& $(4.61\pm0.64)\times10^{-15}$& $-1.86^{+0.08}_{-0.09}$& $0.23^{+0.23}_{-0.24}$& $-7.36^{+0.22}_{-0.25}$\\[3pt]
HD~287823& $10.18\pm0.50$& $15.58\pm1.56$& $-5.40\pm1.64$& $4.18\times10^{-16}$& $(2.26\pm0.69)\times10^{-15}$& $-2.04^{+0.14}_{-0.18}$& $0.05^{+0.30}_{-0.36}$& $-7.39^{+0.30}_{-0.37}$\\[3pt]
HD~287841& $8.36\pm0.43$& $13.22\pm1.32$& $-4.86\pm1.39$& $2.61\times10^{-16}$& $(1.27\pm0.36)\times10^{-15}$& $-2.28^{+0.13}_{-0.17}$& $-0.19^{+0.30}_{-0.35}$& $-7.62^{+0.30}_{-0.36}$\\[3pt]
HD~290409& $0.35\pm0.55$& $15.00\pm1.50$& $-14.65\pm1.60$& $2.97\times10^{-16}$& $(4.36\pm0.48)\times10^{-15}$& $-1.55^{+0.10}_{-0.10}$& $0.54^{+0.23}_{-0.24}$& $-7.02^{+0.27}_{-0.22}$\\[3pt]
HD~35929& $1.54\pm0.69$& $8.69\pm0.87$& $-7.15\pm1.11$& $2.00\times10^{-15}$& $(1.43\pm0.22)\times10^{-14}$& $-1.18^{+0.09}_{-0.10}$& $0.91^{+0.21}_{-0.22}$& $-6.31^{+0.19}_{-0.20}$\\[3pt]
HD~290500& $-1.46\pm0.25$& $13.14\pm1.31$& $-14.60\pm1.33$& $2.14\times10^{-16}$& $(3.12\pm0.28)\times10^{-15}$& $-1.73^{+0.08}_{-0.08}$& $0.36^{+0.23}_{-0.23}$& $-7.22^{+0.22}_{-0.24}$\\[3pt]
HD~244314& $-14.04\pm0.61$& $15.54\pm1.55$& $-29.58\pm1.67$& $3.23\times10^{-16}$& $(9.56\pm0.54)\times10^{-15}$& $-1.26^{+0.06}_{-0.06}$& $0.83^{+0.18}_{-0.19}$& $-6.60^{+0.18}_{-0.20}$\\[3pt]
HK~Ori& $-61.86\pm0.72$& $15.79\pm1.58$& $-77.65\pm1.74$& $2.17\times10^{-16}$& $(1.68\pm0.04)\times10^{-14}$& -& -& -\\[3pt]
HD~244604& $2.32\pm0.51$& $14.79\pm1.48$& $-12.47\pm1.57$& $6.52\times10^{-16}$& $(8.14\pm1.02)\times10^{-15}$& $-1.35^{+0.09}_{-0.09}$& $0.74^{+0.21}_{-0.23}$& $-6.66^{+0.20}_{-0.22}$\\[3pt]
UY~Ori& $4.98\pm0.26$& $15.28\pm1.53$& $-10.30\pm1.55$& $5.81\times10^{-17}$& $(5.99\pm0.90)\times10^{-16}$& -& -& -\\[3pt]
HD~245185& $-13.56\pm0.50$& $14.48\pm1.45$& $-28.04\pm1.53$& $3.22\times10^{-16}$& $(9.03\pm0.49)\times10^{-15}$& $-1.29^{+0.10}_{-0.09}$& $0.80^{+0.21}_{-0.21}$& $-6.78^{+0.25}_{-0.20}$\\[3pt]
T~Ori& $-4.15\pm0.43$& $13.07\pm1.31$& $-17.22\pm1.38$& $7.65\times10^{-16}$& $(1.32\pm0.11)\times10^{-14}$& $-1.17^{+0.06}_{-0.06}$& $0.92^{+0.18}_{-0.18}$& $-6.48^{+0.15}_{-0.17}$\\[3pt]
V380~Ori& $-81.88\pm0.48$& $13.61\pm1.36$& $-95.49\pm1.44$& $9.35\times10^{-16}$& $(8.93\pm0.14)\times10^{-14}$& -& -& -\\[3pt]
HD~37258& $-0.33\pm0.39$& $15.00\pm1.50$& $-15.33\pm1.55$& $4.66\times10^{-16}$& $(7.14\pm0.72)\times10^{-15}$& -& -& -\\[3pt]
HD~290770& $-24.08\pm0.35$& $12.98\pm1.30$& $-37.06\pm1.35$& $5.55\times10^{-16}$& $(2.06\pm0.07)\times10^{-14}$& $-0.99^{+0.06}_{-0.06}$& $1.10^{+0.17}_{-0.17}$& $-6.47^{+0.19}_{-0.17}$\\[3pt]
BF~Ori& $-0.02\pm0.46$& $14.70\pm1.47$& $-14.72\pm1.54$& $5.03\times10^{-16}$& $(7.41\pm0.78)\times10^{-15}$& $-1.46^{+0.07}_{-0.08}$& $0.63^{+0.20}_{-0.21}$& $-6.81^{+0.20}_{-0.22}$\\[3pt]
HD~37357& $4.76\pm0.41$& $14.64\pm1.46$& $-9.88\pm1.52$& $8.69\times10^{-16}$& $(8.58\pm1.32)\times10^{-15}$& -& -& -\\[3pt]
HD~290764& $-2.38\pm0.43$& $13.31\pm1.33$& $-15.69\pm1.40$& $4.48\times10^{-16}$& $(7.02\pm0.63)\times10^{-15}$& $-1.46^{+0.08}_{-0.07}$& $0.63^{+0.20}_{-0.21}$& $-6.74^{+0.19}_{-0.20}$\\[3pt]
HD~37411& $-1.07\pm0.47$& $15.55\pm1.56$& $-16.62\pm1.63$& $4.29\times10^{-16}$& $(7.13\pm0.70)\times10^{-15}$& -& -& -\\[3pt]
V599~Ori& $1.69\pm0.57$& $13.35\pm1.34$& $-11.66\pm1.46$& $7.44\times10^{-16}$& $(8.68\pm1.08)\times10^{-15}$& $-1.34^{+0.08}_{-0.08}$& $0.75^{+0.20}_{-0.21}$& $-6.59^{+0.20}_{-0.22}$\\[3pt]
V350~Ori& $3.09\pm0.33$& $15.66\pm1.57$& $-12.57\pm1.60$& $2.70\times10^{-16}$& $(3.39\pm0.43)\times10^{-15}$& -& -& -\\[3pt]
HD~250550& $-48.83\pm0.39$& $10.03\pm1.00$& $-58.86\pm1.07$& $1.08\times10^{-14}$& $(6.35\pm0.12)\times10^{-13}$& $0.98^{+0.12}_{-0.09}$& $3.07^{+0.23}_{-0.20}$& $-4.18^{+0.23}_{-0.19}$\\[3pt]
V791~Mon& $-88.08\pm0.25$& $7.94\pm0.79$& $-96.02\pm0.83$& $5.71\times10^{-16}$& $(5.49\pm0.05)\times10^{-14}$& $0.13^{+0.05}_{-0.05}$& $2.22^{+0.12}_{-0.11}$& $-5.37^{+0.09}_{-0.12}$\\[3pt]
PDS~124& $-13.83\pm1.37$& $14.13\pm1.41$& $-27.96\pm1.97$& $1.13\times10^{-16}$& $(3.16\pm0.22)\times10^{-15}$& $-1.14^{+0.09}_{-0.09}$& $0.95^{+0.20}_{-0.21}$& $-6.60^{+0.24}_{-0.23}$\\[3pt]
LkHa~339& $-5.39\pm1.37$& $12.98\pm1.30$& $-18.37\pm1.89$& $1.74\times10^{-16}$& $(3.20\pm0.33)\times10^{-15}$& $-1.14^{+0.08}_{-0.08}$& $0.95^{+0.19}_{-0.20}$& $-6.54^{+0.22}_{-0.28}$\\[3pt]
VY~Mon& $-17.81\pm1.76$& $8.45\pm0.84$& $-26.26\pm1.95$& $2.75\times10^{-15}$& $(7.21\pm0.54)\times10^{-14}$& -& -& -\\[3pt]
R~Mon& $-114.51\pm1.76$& $9.35\pm0.93$& $-123.86\pm1.99$& $3.69\times10^{-16}$& $(4.57\pm0.07)\times10^{-14}$& -& -& -\\[3pt]
V590~Mon& $-60.10\pm0.52$& $9.63\pm0.96$& $-69.73\pm1.09$& $5.06\times10^{-17}$& $(3.53\pm0.06)\times10^{-15}$& -& -& -\\[3pt]
PDS~24& $-25.47\pm1.79$& $12.98\pm1.30$& $-38.45\pm2.21$& $4.03\times10^{-17}$& $(1.55\pm0.09)\times10^{-15}$& $-1.21^{+0.06}_{-0.05}$& $0.88^{+0.17}_{-0.18}$& $-6.74^{+0.15}_{-0.19}$\\[3pt]
\hline
\end{tabular}
\end{table}
\end{landscape}

\begin{landscape}
\begin{table}
\renewcommand{\thetable}{C2}
\contcaption{.}
\label{tab:C2_Fairlamb_Mdot2}
\begin{tabular}{lcccccccc}
\hline
Name&	$EW_{\rm obs}$&	$EW_{\rm int}$&	$EW_{\rm cor}$&	$F_{\lambda}$&		$F_{\rm line}$&	$\log(L_{\rm line})$&	$\log(L_{\rm acc})$&		$\log(\dot M_{\rm acc})$\\[3pt]
&	(\AA)&		(\AA)&		(\AA)&	(W\,m$^{-2}$\,\AA$^{-1}$)&		(W\,m$^{-2}$)&		[$\rm L_{\sun}$]&	[$\rm L_{\sun}$]&	[$\rm M_{\sun}$\,$\rm yr^{-1}$]\\	
\hline
PDS~130& $-31.17\pm0.70$& $11.41\pm1.14$& $-42.58\pm1.34$& $8.43\times10^{-17}$& $(3.59\pm0.11)\times10^{-15}$& $-0.71^{+0.05}_{-0.05}$& $1.38^{+0.15}_{-0.15}$& $-6.10^{+0.15}_{-0.17}$\\[3pt]
PDS~229& $7.41\pm0.82$& $9.63\pm0.96$& $-2.22\pm1.26$& $9.37\times10^{-17}$& $(2.08\pm1.18)\times10^{-16}$& -& -& -\\[3pt]
GU~CMa& $-14.86\pm0.48$& $4.75\pm0.47$& $-19.61\pm0.67$& $9.95\times10^{-15}$& $(1.95\pm0.07)\times10^{-13}$& -& -& -\\[3pt]
HT~CMa& $-20.94\pm0.35$& $11.87\pm1.19$& $-32.81\pm1.24$& $1.65\times10^{-16}$& $(5.40\pm0.20)\times10^{-15}$& $-0.68^{+0.08}_{-0.07}$& $1.41^{+0.17}_{-0.17}$& $-6.03^{+0.16}_{-0.16}$\\[3pt]
Z~CMa& $-63.55\pm0.99$& $9.97\pm1.00$& $-73.52\pm1.41$& $1.65\times10^{-14}$& $(1.21\pm0.02)\times10^{-12}$& -& -& -\\[3pt]
HU~CMa& $-52.00\pm0.40$& $9.09\pm0.91$& $-61.09\pm0.99$& $1.62\times10^{-16}$& $(9.87\pm0.16)\times10^{-15}$& $-0.38^{+0.08}_{-0.07}$& $1.71^{+0.15}_{-0.15}$& $-5.86^{+0.17}_{-0.20}$\\[3pt]
HD~53367& $-7.62\pm0.52$& $4.02\pm0.40$& $-11.64\pm0.66$& $2.27\times10^{-14}$& $(2.64\pm0.15)\times10^{-13}$& -& -& -\\[3pt]
PDS~241& $-8.36\pm0.29$& $4.21\pm0.42$& $-12.57\pm0.51$& $4.57\times10^{-16}$& $(5.75\pm0.23)\times10^{-15}$& $0.17^{+0.18}_{-0.15}$& $2.26^{+0.26}_{-0.21}$& $-5.52^{+0.26}_{-0.26}$\\[3pt]
NX~Pup& $-37.01\pm0.45$& $8.74\pm0.87$& $-45.75\pm0.98$& $4.73\times10^{-16}$& $(2.16\pm0.05)\times10^{-14}$& -& -& -\\[3pt]
PDS~27& $-73.20\pm0.73$& $4.40\pm0.44$& $-77.60\pm0.85$& $1.39\times10^{-15}$& $(1.08\pm0.01)\times10^{-13}$& $1.34^{+0.15}_{-0.12}$& $3.43^{+0.28}_{-0.24}$& $-4.04^{+0.19}_{-0.15}$\\[3pt]
PDS~133& $-103.11\pm3.91$& $7.84\pm0.78$& $-110.95\pm3.99$& $6.73\times10^{-17}$& $(7.47\pm0.27)\times10^{-15}$& $-0.30^{+0.07}_{-0.06}$& $1.79^{+0.14}_{-0.14}$& $-5.89^{+0.07}_{-0.09}$\\[3pt]
HD~59319& $5.84\pm0.07$& $7.13\pm0.71$& $-1.29\pm0.71$& $1.00\times10^{-15}$& -& -& -& -\\[3pt]
PDS~134& $-12.22\pm0.41$& $5.98\pm0.60$& $-18.20\pm0.73$& $1.30\times10^{-16}$& $(2.36\pm0.09)\times10^{-15}$& $-0.32^{+0.13}_{-0.11}$& $1.77^{+0.20}_{-0.19}$& $-5.73^{+0.20}_{-0.20}$\\[3pt]
HD~68695& $0.48\pm0.48$& $16.51\pm1.65$& $-16.03\pm1.72$& $3.51\times10^{-16}$& $(5.63\pm0.60)\times10^{-15}$& $-1.56^{+0.07}_{-0.07}$& $0.53^{+0.20}_{-0.21}$& $-6.99^{+0.20}_{-0.23}$\\[3pt]
HD~72106& $8.78\pm0.63$& $14.60\pm1.46$& $-5.82\pm1.59$& $8.58\times10^{-16}$& $(4.99\pm1.36)\times10^{-15}$& -& -& -\\[3pt]
TYC~8581-2002-1& $11.05\pm0.42$& $13.61\pm1.36$& $-2.56\pm1.42$& $2.22\times10^{-16}$& -& -& -& -\\[3pt]
PDS~33& $-3.10\pm0.48$& $15.83\pm1.58$& $-18.93\pm1.65$& $6.26\times10^{-17}$& $(1.18\pm0.10)\times10^{-15}$& $-1.48^{+0.07}_{-0.07}$& $0.61^{+0.20}_{-0.21}$& $-6.96^{+0.22}_{-0.19}$\\[3pt]
HD~76534& $-11.00\pm0.34$& $5.84\pm0.58$& $-16.84\pm0.67$& $2.80\times10^{-15}$& $(4.72\pm0.19)\times10^{-14}$& $0.09^{+0.07}_{-0.06}$& $2.18^{+0.14}_{-0.12}$& $-5.43^{+0.13}_{-0.12}$\\[3pt]
PDS~281& $4.30\pm0.49$& $5.42\pm0.54$& $-1.12\pm0.73$& $4.89\times10^{-15}$& -& -& -& -\\[3pt]
PDS~286& $-26.84\pm0.42$& $3.93\pm0.39$& $-30.77\pm0.57$& $1.12\times10^{-14}$& $(3.45\pm0.06)\times10^{-13}$& $1.55^{+0.10}_{-0.08}$& $3.64^{+0.25}_{-0.22}$& $-4.16^{+0.15}_{-0.15}$\\[3pt]
PDS~297& $7.42\pm0.41$& $11.36\pm1.14$& $-3.94\pm1.21$& $9.95\times10^{-17}$& $(3.92\pm1.21)\times10^{-16}$& $-1.51^{+0.19}_{-0.22}$& $0.58^{+0.32}_{-0.37}$& $-6.86^{+0.32}_{-0.37}$\\[3pt]
HD~85567& $-50.25\pm0.47$& $6.76\pm0.68$& $-57.01\pm0.83$& $2.11\times10^{-15}$& $(1.20\pm0.02)\times10^{-13}$& $0.59^{+0.05}_{-0.05}$& $2.68^{+0.14}_{-0.13}$& $-4.71^{+0.13}_{-0.13}$\\[3pt]
HD~87403& $6.79\pm0.63$& $9.76\pm0.98$& $-2.97\pm1.17$& $5.60\times10^{-16}$& $(1.66\pm0.65)\times10^{-15}$& $-0.72^{+0.26}_{-0.31}$& $1.37^{+0.35}_{-0.42}$& $-5.90^{+0.35}_{-0.42}$\\[3pt]
PDS~37& $-119.89\pm0.47$& $3.87\pm0.39$& $-123.76\pm0.61$& $1.63\times10^{-15}$& $(2.01\pm0.01)\times10^{-13}$& $1.37^{+0.15}_{-0.11}$& $3.46^{+0.29}_{-0.24}$& $-4.03^{+0.21}_{-0.16}$\\[3pt]
HD~305298& $0.01\pm0.41$& $3.25\pm0.32$& $-3.24\pm0.52$& $3.53\times10^{-16}$& $(1.14\pm0.18)\times10^{-15}$& $-0.24^{+0.19}_{-0.18}$& $1.85^{+0.25}_{-0.26}$& $-6.14^{+0.28}_{-0.28}$\\[3pt]
HD~94509& $-16.84\pm0.53$& $6.33\pm0.63$& $-23.17\pm0.82$& $6.27\times10^{-16}$& $(1.45\pm0.05)\times10^{-14}$& $0.18^{+0.11}_{-0.09}$& $2.27^{+0.18}_{-0.15}$& $-5.07^{+0.15}_{-0.11}$\\[3pt]
HD~95881& $-12.53\pm0.39$& $9.35\pm0.94$& $-21.88\pm1.02$& $2.06\times10^{-15}$& $(4.50\pm0.21)\times10^{-14}$& $0.28^{+0.08}_{-0.07}$& $2.37^{+0.16}_{-0.14}$& $-4.88^{+0.16}_{-0.13}$\\[3pt]
HD~96042& $3.20\pm0.80$& $3.95\pm0.39$& $-0.75\pm0.89$& $2.23\times10^{-15}$& -& -& -& -\\[3pt]
HD~97048& $-24.43\pm0.34$& $13.54\pm1.35$& $-37.97\pm1.39$& $3.11\times10^{-15}$& $(1.18\pm0.04)\times10^{-13}$& $-0.90^{+0.03}_{-0.03}$& $1.19^{+0.13}_{-0.13}$& $-6.33^{+0.12}_{-0.15}$\\[3pt]
HD~98922& $-14.64\pm0.44$& $10.04\pm1.00$& $-24.68\pm1.09$& $6.20\times10^{-15}$& $(1.53\pm0.07)\times10^{-13}$& $0.35^{+0.05}_{-0.05}$& $2.44^{+0.13}_{-0.13}$& $-4.83^{+0.13}_{-0.12}$\\[3pt]
HD~100453& $5.27\pm0.77$& $10.23\pm1.02$& $-4.96\pm1.28$& $2.14\times10^{-15}$& $(1.06\pm0.27)\times10^{-14}$& $-2.44^{+0.11}_{-0.14}$& $-0.35^{+0.28}_{-0.32}$& $-7.79^{+0.29}_{-0.33}$\\[3pt]
HD~100546& $-23.96\pm0.42$& $15.50\pm1.55$& $-39.46\pm1.61$& $5.12\times10^{-15}$& $(2.02\pm0.08)\times10^{-13}$& $-1.12^{+0.03}_{-0.03}$& $0.97^{+0.14}_{-0.14}$& $-6.59^{+0.15}_{-0.12}$\\[3pt]
HD~101412& $-0.15\pm0.69$& $15.28\pm1.53$& $-15.43\pm1.68$& $7.90\times10^{-16}$& $(1.22\pm0.13)\times10^{-14}$& $-1.19^{+0.06}_{-0.07}$& $0.90^{+0.18}_{-0.19}$& $-6.55^{+0.17}_{-0.18}$\\[3pt]
PDS~344& $-22.23\pm0.45$& $7.80\pm0.78$& $-30.03\pm0.90$& $3.59\times10^{-17}$& $(1.08\pm0.03)\times10^{-15}$& $-0.70^{+0.07}_{-0.06}$& $1.39^{+0.16}_{-0.16}$& $-6.32^{+0.17}_{-0.18}$\\[3pt]
\hline
\end{tabular}
\end{table}
\end{landscape}

\begin{landscape}
\begin{table}
\renewcommand{\thetable}{C2}
\contcaption{.}
\label{tab:C2_Fairlamb_Mdot3}
\begin{tabular}{lcccccccc}
\hline
Name&	$EW_{\rm obs}$&	$EW_{\rm int}$&	$EW_{\rm cor}$&	$F_{\lambda}$&		$F_{\rm line}$&	$\log(L_{\rm line})$&	$\log(L_{\rm acc})$&		$\log(\dot M_{\rm acc})$\\[3pt]
&	(\AA)&		(\AA)&		(\AA)&	(W\,m$^{-2}$\,\AA$^{-1}$)&		(W\,m$^{-2}$)&		[$\rm L_{\sun}$]&	[$\rm L_{\sun}$]&	[$\rm M_{\sun}$\,$\rm yr^{-1}$]\\	
\hline
HD~104237& $-13.74\pm0.98$& $13.75\pm1.38$& $-27.49\pm1.69$& $7.42\times10^{-15}$& $(2.04\pm0.13)\times10^{-13}$& -& -& -\\[3pt]
V1028~Cen& $-101.61\pm0.52$& $7.01\pm0.70$& $-108.62\pm0.87$& $2.50\times10^{-16}$& $(2.71\pm0.02)\times10^{-14}$& -& -& -\\[3pt]
PDS~361& $-3.97\pm0.45$& $5.35\pm0.53$& $-9.32\pm0.70$& $1.12\times10^{-16}$& $(1.05\pm0.08)\times10^{-15}$& $-0.55^{+0.15}_{-0.13}$& $1.54^{+0.23}_{-0.22}$& $-6.11^{+0.23}_{-0.26}$\\[3pt]
HD~114981& $-7.60\pm0.46$& $5.64\pm0.56$& $-13.24\pm0.72$& $2.90\times10^{-15}$& $(3.84\pm0.21)\times10^{-14}$& $-0.23^{+0.09}_{-0.08}$& $1.86^{+0.16}_{-0.16}$& $-5.67^{+0.15}_{-0.15}$\\[3pt]
PDS~364& $-78.38\pm0.74$& $9.63\pm0.96$& $-88.01\pm1.21$& $5.77\times10^{-17}$& $(5.07\pm0.07)\times10^{-15}$& -& -& -\\[3pt]
PDS~69& $-69.42\pm1.30$& $7.00\pm0.70$& $-76.42\pm1.48$& $1.17\times10^{-15}$& $(8.96\pm0.17)\times10^{-14}$& $0.06^{+0.05}_{-0.05}$& $2.15^{+0.12}_{-0.11}$& $-5.43^{+0.07}_{-0.10}$\\[3pt]
DG~Cir& $-47.98\pm0.83$& $13.10\pm1.31$& $-61.08\pm1.55$& $1.14\times10^{-16}$& $(6.98\pm0.18)\times10^{-15}$& $-0.82^{+0.06}_{-0.06}$& $1.27^{+0.16}_{-0.16}$& $-6.34^{+0.12}_{-0.08}$\\[3pt]
HD~132947& $10.38\pm0.33$& $11.96\pm1.20$& $-1.58\pm1.24$& $7.22\times10^{-16}$& -& -& -& -\\[3pt]
HD~135344B& $-4.60\pm0.75$& $5.66\pm0.57$& $-10.26\pm0.94$& $1.69\times10^{-15}$& $(1.73\pm0.16)\times10^{-14}$& $-2.00^{+0.05}_{-0.06}$& $0.09^{+0.21}_{-0.22}$& $-7.25^{+0.21}_{-0.23}$\\[3pt]
HD~139614& $0.48\pm0.61$& $13.26\pm1.33$& $-12.78\pm1.46$& $1.24\times10^{-15}$& $(1.59\pm0.18)\times10^{-14}$& $-2.05^{+0.06}_{-0.06}$& $0.04^{+0.22}_{-0.23}$& $-7.49^{+0.23}_{-0.21}$\\[3pt]
PDS~144S& $-16.12\pm0.68$& $13.07\pm1.31$& $-29.19\pm1.48$& $4.68\times10^{-17}$& $(1.36\pm0.07)\times10^{-15}$& -& -& -\\[3pt]
HD~141569& $5.13\pm0.57$& $15.55\pm1.56$& $-10.42\pm1.66$& $4.57\times10^{-15}$& $(4.76\pm0.76)\times10^{-14}$& $-1.74^{+0.07}_{-0.08}$& $0.35^{+0.21}_{-0.23}$& $-7.23^{+0.22}_{-0.21}$\\[3pt]
HD~141926& $-43.48\pm0.43$& $3.40\pm0.34$& $-46.88\pm0.55$& $7.76\times10^{-15}$& $(3.64\pm0.04)\times10^{-13}$& $1.31^{+0.10}_{-0.08}$& $3.40^{+0.23}_{-0.20}$& $-4.37^{+0.20}_{-0.19}$\\[3pt]
HD~142666& $5.14\pm0.68$& $11.70\pm1.17$& $-6.56\pm1.35$& $2.02\times10^{-15}$& $(1.32\pm0.27)\times10^{-14}$& $-2.04^{+0.09}_{-0.11}$& $0.05^{+0.25}_{-0.28}$& $-7.33^{+0.24}_{-0.27}$\\[3pt]
HD~142527& $-6.95\pm0.98$& $6.18\pm0.62$& $-13.13\pm1.16$& $3.56\times10^{-15}$& $(4.67\pm0.41)\times10^{-14}$& $-1.44^{+0.05}_{-0.05}$& $0.65^{+0.18}_{-0.19}$& $-6.61^{+0.15}_{-0.18}$\\[3pt]
HD~144432& $-0.98\pm0.68$& $11.65\pm1.17$& $-12.63\pm1.35$& $2.06\times10^{-15}$& $(2.60\pm0.28)\times10^{-14}$& $-1.71^{+0.06}_{-0.06}$& $0.38^{+0.20}_{-0.21}$& $-6.99^{+0.18}_{-0.20}$\\[3pt]
HD~144668& $-8.41\pm0.45$& $14.12\pm1.41$& $-22.53\pm1.48$& $9.59\times10^{-15}$& $(2.16\pm0.14)\times10^{-13}$& $-0.76^{+0.04}_{-0.05}$& $1.33^{+0.14}_{-0.15}$& $-6.00^{+0.13}_{-0.13}$\\[3pt]
HD~145718& $8.34\pm0.52$& $14.53\pm1.45$& $-6.19\pm1.54$& $1.66\times10^{-15}$& $(1.02\pm0.26)\times10^{-14}$& $-2.13^{+0.11}_{-0.14}$& $-0.04^{+0.28}_{-0.32}$& $-7.49^{+0.28}_{-0.33}$\\[3pt]
PDS~415& $3.10\pm1.30$& $5.04\pm0.50$& $-1.94\pm1.39$& $1.83\times10^{-16}$& $(3.55\pm2.55)\times10^{-16}$& -& -& -\\[3pt]
HD~150193& $-5.59\pm0.57$& $16.07\pm1.61$& $-21.66\pm1.71$& $3.57\times10^{-15}$& $(7.74\pm0.61)\times10^{-14}$& $-1.26^{+0.05}_{-0.05}$& $0.83^{+0.17}_{-0.18}$& $-6.61^{+0.17}_{-0.20}$\\[3pt]
AK~Sco& $-0.82\pm0.95$& $5.06\pm0.51$& $-5.88\pm1.08$& $1.23\times10^{-15}$& $(7.22\pm1.32)\times10^{-15}$& $-2.35^{+0.09}_{-0.10}$& $-0.26^{+0.26}_{-0.28}$& $-7.61^{+0.24}_{-0.26}$\\[3pt]
PDS~431& $1.27\pm0.50$& $10.49\pm1.05$& $-9.22\pm1.16$& $6.40\times10^{-17}$& $(5.90\pm0.74)\times10^{-16}$& $-1.22^{+0.13}_{-0.12}$& $0.87^{+0.24}_{-0.25}$& $-6.58^{+0.23}_{-0.24}$\\[3pt]
KK~Oph& $-25.15\pm0.31$& $16.36\pm1.64$& $-41.51\pm1.67$& $1.84\times10^{-15}$& $(7.64\pm0.31)\times10^{-14}$& -& -& -\\[3pt]
HD~163296& $-3.63\pm0.36$& $16.02\pm1.60$& $-19.65\pm1.64$& $5.38\times10^{-15}$& $(1.06\pm0.09)\times10^{-13}$& $-1.47^{+0.05}_{-0.05}$& $0.62^{+0.18}_{-0.19}$& $-6.91^{+0.19}_{-0.19}$\\[3pt]
MWC~297& $-590.00\pm0.90$& $4.52\pm0.45$& $-594.52\pm1.01$& $4.81\times10^{-14}$& $(2.86\pm0.01)\times10^{-11}$& $2.10^{+0.05}_{-0.04}$& $4.19^{+0.22}_{-0.21}$& $-3.52^{+0.21}_{-0.22}$\\[3pt]
\hline
\end{tabular}
\end{table}
\end{landscape}

\begin{landscape}
\begin{table}
\renewcommand{\thetable}{D1}
\caption{The equivalent width measurements and accretion rates of 144 Herbig Ae/Be stars in the \protect\cite{Vioque2018} sample.
Column 1 gives the object name. Columns 2 and 3 are right ascension (RA) in the units of time (\fh \, \fm \, \fs) and declination (DEC) in the units of angle (\fdg \, \farcm \, \farcs) respectively.
Observed equivalent width is listed along with references in columns 4--5. 
The rest of parameters are derived in this work.
Columns 6--12 present intrinsic equivalent width, corrected equivalent width, continuum flux density at central wavelength of the $\rm H\alpha$ profile, line flux, line luminosity, accretion luminosity and mass accretion rate respectively. This table is available in electronic form at \url{https://cdsarc.unistra.fr/viz-bin/cat/J/MNRAS/493/234}.}
\label{tab:D1_Vioque_Mdot}
\begin{adjustbox}{width=1.33\textwidth}
\begin{tabular}{lccccccccccc}
\hline
Name&		RA&			DEC&	\multicolumn{2}{c}{$EW_{\rm obs}$}&	$EW_{\rm int}$&	$EW_{\rm cor}$&	$F_{\lambda}$&	$F_{\rm line}$&	$\log(L_{\rm line})$&		$\log(L_{\rm acc})$&	$\log(\dot M_{\rm acc})$\\[3pt]	
&				(J2000)&	(J2000)&	(\AA)&		Ref.&		(\AA)&		(\AA)&	(W\,m$^{-2}$\,\AA$^{-1}$)&		(W\,m$^{-2}$)&		[$\rm L_{\sun}$]&		[$\rm L_{\sun}$]&			[$\rm M_{\sun}$\,$\rm yr^{-1}$]\\	
\hline		
HBC~1& 00:07:02.6& +65:38:38.3& $-31.00\pm3.10$& 1& $14.43\pm0.05$& $-45.43\pm3.10$& $1.15\times10^{-18}$& $(5.24\pm0.36)\times10^{-17}$& -$^{a}$& -& -\\[3pt]
HBC~324& 00:07:30.6& +65:39:52.6& $-16.90\pm0.85$& 2& $13.99\pm0.11$& $-30.89\pm0.85$& $2.42\times10^{-17}$& $(7.48\pm0.21)\times10^{-16}$& -$^{a}$& -& -\\[3pt]
MQ~Cas& 00:09:37.5& +58:13:10.7& -& -& $15.82\pm0.12$& -& $4.25\times10^{-17}$& -& -$^{a}$& -& -\\[3pt]
VX~Cas& 00:31:30.6& +61:58:50.9& $-22.10\pm0.66$& 3& $14.94\pm0.11$& $-37.04\pm0.67$& $1.16\times10^{-16}$& $(4.31\pm0.08)\times10^{-15}$& $-1.41^{+0.04}_{-0.03}$& $0.68^{+0.17}_{-0.17}$& $-6.96^{+0.22}_{-0.21}$\\[3pt]
HBC~7& 00:43:25.3& +61:38:23.3& $-41.60\pm2.08$& 2& $4.06\pm0.04$& $-45.66\pm2.08$& $4.17\times10^{-16}$& $(1.91\pm0.09)\times10^{-14}$& $0.65^{+0.09}_{-0.08}$& $2.74^{+0.19}_{-0.17}$& $-4.85^{+0.39}_{-0.42}$\\[3pt]
PDS~2& 01:17:43.4& -52:33:30.7& $-4.93\pm0.20$& 4& $8.37\pm0.34$& $-13.30\pm0.39$& $1.25\times10^{-16}$& $(1.66\pm0.05)\times10^{-15}$& $-2.05^{+0.03}_{-0.03}$& $0.04^{+0.19}_{-0.19}$& $-7.34^{+0.24}_{-0.21}$\\[3pt]
HD~9672& 01:34:37.7& -15:40:34.8& $12.28\pm0.04$& 5& $15.97\pm0.19$& $-3.69\pm0.19$& $1.19\times10^{-14}$& $(4.41\pm0.23)\times10^{-14}$& $-2.35^{+0.03}_{-0.03}$& $-0.26^{+0.21}_{-0.21}$& $-7.80^{+0.25}_{-0.22}$\\[3pt]
HBC~334& 02:16:30.1& +55:22:57& $-0.20\pm0.01$& 2& $7.65\pm0.08$& $-7.85\pm0.08$& $3.13\times10^{-17}$& $(2.45\pm0.02)\times10^{-16}$& $-1.62^{+0.09}_{-0.08}$& $0.47^{+0.23}_{-0.22}$& $-7.42^{+0.39}_{-0.44}$\\[3pt]
HD~17081& 02:44:07.3& -13:51:31.3& $5.95\pm0.05$& 6& $7.19\pm0.00$& $-1.24\pm0.05$& $3.89\times10^{-14}$& $(4.83\pm0.21)\times10^{-14}$& $-1.77^{+0.08}_{-0.07}$& $0.32^{+0.22}_{-0.22}$& $-7.18^{+0.34}_{-0.32}$\\[3pt]
BD+30~549& 03:29:19.7& +31:24:57.0& -& -& $11.64\pm0.16$& -& $6.12\times10^{-16}$& -& -& -& -\\[3pt]
PDS~4& 03:39:00.5& +29:41:45.7& $-9.00\pm0.45$& 7& $14.75\pm0.13$& $-23.75\pm0.47$& $2.59\times10^{-16}$& $(6.15\pm0.12)\times10^{-15}$& $-1.52^{+0.04}_{-0.04}$& $0.57^{+0.17}_{-0.17}$& $-7.02^{+0.25}_{-0.28}$\\[3pt]
IP~Per& 03:40:46.9& +32:31:53.7& $-21.40\pm1.07$& 2& $15.18\pm0.32$& $-36.58\pm1.12$& $2.48\times10^{-16}$& $(9.06\pm0.28)\times10^{-15}$& $-1.57^{+0.05}_{-0.05}$& $0.52^{+0.19}_{-0.19}$& $-7.01^{+0.26}_{-0.28}$\\[3pt]
XY~Per~A& 03:49:36.3& +38:58:55.4& $-9.80\pm0.29$& 3& $12.13\pm0.05$& $-21.93\pm0.30$& $1.04\times10^{-15}$& $(2.28\pm0.03)\times10^{-14}$& -$^{a}$& -& -\\[3pt]
V892~Tau& 04:18:40.6& +28:19:15.6& $-17.80\pm0.89$& 2& $9.43\pm0.14$& $-27.23\pm0.90$& $1.51\times10^{-16}$& $(4.10\pm0.14)\times10^{-15}$& -$^{a}$& -& -\\[3pt]
AB~Aur& 04:55:45.8& +30:33:04.2& $-45.00\pm4.50$& 8& $13.66\pm0.14$& $-58.66\pm4.50$& $3.51\times10^{-15}$& $(2.06\pm0.16)\times10^{-13}$& $-0.77^{+0.05}_{-0.05}$& $1.32^{+0.14}_{-0.15}$& $-6.13^{+0.25}_{-0.27}$\\[3pt]
HD~31648& 04:58:46.2& +29:50:36.9& $-19.40\pm0.58$& 3& $15.10\pm0.22$& $-34.50\pm0.62$& $2.07\times10^{-15}$& $(7.13\pm0.13)\times10^{-14}$& $-1.24^{+0.03}_{-0.03}$& $0.85^{+0.15}_{-0.15}$& $-6.57^{+0.21}_{-0.17}$\\[3pt]
HD~34700& 05:19:41.4& +05:38:42.7& $-2.40\pm0.07$& 3& $5.20\pm0.25$& $-7.60\pm0.26$& $6.69\times10^{-16}$& $(5.08\pm0.18)\times10^{-15}$& $-1.70^{+0.04}_{-0.04}$& $0.39^{+0.18}_{-0.19}$& $-6.87^{+0.20}_{-0.19}$\\[3pt]
HD~290380& 05:23:31.0& -01:04:23.6& $-7.00\pm0.70$& 9& $6.66\pm0.06$& $-13.66\pm0.70$& $1.98\times10^{-16}$& $(2.71\pm0.14)\times10^{-15}$& $-1.98^{+0.05}_{-0.05}$& $0.11^{+0.20}_{-0.21}$& $-7.23^{+0.24}_{-0.22}$\\[3pt]
HD~35187& 05:24:01.1& +24:57:37.5& $-3.50\pm0.20$& 10& $14.14\pm0.14$& $-17.64\pm0.25$& $2.29\times10^{-15}$& $(4.03\pm0.06)\times10^{-14}$& $-1.48^{+0.03}_{-0.03}$& $0.61^{+0.16}_{-0.16}$& $-6.94^{+0.29}_{-0.31}$\\[3pt]
CO~Ori& 05:27:38.3& +11:25:38.9& $-21.10\pm0.63$& 3& $6.28\pm0.08$& $-27.38\pm0.64$& $7.01\times10^{-16}$& $(1.92\pm0.05)\times10^{-14}$& -$^{a}$& -& -\\[3pt]
HD~36112& 05:30:27.5& +25:19:57.0& $-6.30\pm0.63$& 11& $12.85\pm0.19$& $-19.15\pm0.66$& $1.36\times10^{-15}$& $(2.61\pm0.09)\times10^{-14}$& $-1.68^{+0.03}_{-0.03}$& $0.41^{+0.17}_{-0.18}$& $-7.00^{+0.24}_{-0.22}$\\[3pt]
RY~Ori& 05:32:09.9& -02:49:46.7& $-15.80\pm0.47$& 3& $6.19\pm0.05$& $-21.99\pm0.48$& $1.94\times10^{-16}$& $(4.26\pm0.09)\times10^{-15}$& $-1.75^{+0.03}_{-0.03}$& $0.34^{+0.18}_{-0.18}$& $-6.98^{+0.20}_{-0.22}$\\[3pt]
HD~36408& 05:32:14.1& +17:03:29.2& $4.00\pm0.20$& 7& $6.80\pm0.03$& $-2.80\pm0.20$& $1.01\times10^{-14}$& $(2.82\pm0.20)\times10^{-14}$& $-0.78^{+0.10}_{-0.09}$& $1.31^{+0.19}_{-0.19}$& $-6.04^{+0.27}_{-0.28}$\\[3pt]
HD~288012& 05:33:04.7& +02:28:09.7& -& -& $13.09\pm0.09$& -& $6.26\times10^{-16}$& -& -& -& -\\[3pt]
HBC~442& 05:34:14.1& -05:36:54.1& $-1.30\pm0.07$& 2& $6.00\pm0.02$& $-7.30\pm0.07$& $2.35\times10^{-16}$& $(1.72\pm0.02)\times10^{-15}$& $-2.10^{+0.03}_{-0.02}$& $-0.01^{+0.19}_{-0.19}$& $-7.32^{+0.20}_{-0.20}$\\[3pt]
HD~36917& 05:34:46.9& -05:34:14.5& $-2.50\pm0.10$& 10& $8.82\pm0.03$& $-11.32\pm0.11$& $1.94\times10^{-15}$& $(2.20\pm0.02)\times10^{-14}$& $-0.81^{+0.06}_{-0.05}$& $1.28^{+0.15}_{-0.15}$& $-6.15^{+0.28}_{-0.28}$\\[3pt]
HD~36982& 05:35:09.8& -05:27:53.2& -& -& $5.58\pm0.02$& -& $1.64\times10^{-15}$& -& -& -& -\\[3pt]
NV~Ori& 05:35:31.3& -05:33:08.8& $-4.00\pm0.12$& 3& $8.45\pm0.35$& $-12.45\pm0.37$& $3.59\times10^{-16}$& $(4.47\pm0.13)\times10^{-15}$& $-1.68^{+0.05}_{-0.04}$& $0.41^{+0.19}_{-0.19}$& $-6.89^{+0.23}_{-0.23}$\\[3pt]
CQ~Tau& 05:35:58.4& +24:44:54.0& $-4.80\pm0.14$& 3& $8.28\pm0.34$& $-13.08\pm0.37$& $9.67\times10^{-16}$& $(1.26\pm0.04)\times10^{-14}$& -$^{a}$& -& -\\[3pt]
V1787~Ori& 05:38:09.3& -06:49:16.5& $-12.00\pm0.60$& 12& $15.29\pm0.27$& $-27.29\pm0.66$& $2.74\times10^{-16}$& $(7.49\pm0.18)\times10^{-15}$& $-1.45^{+0.04}_{-0.04}$& $0.64^{+0.17}_{-0.17}$& $-6.80^{+0.22}_{-0.22}$\\[3pt]
HD~37371& 05:38:09.9& -00:11:01.1& $-4.30\pm0.22$& 13& $9.51\pm0.16$& $-13.81\pm0.27$& $2.99\times10^{-15}$& $(4.12\pm0.08)\times10^{-14}$& $-0.66^{+0.05}_{-0.04}$& $1.43^{+0.14}_{-0.14}$& $-5.96^{+0.22}_{-0.24}$\\[3pt]
\hline
\end{tabular}
\end{adjustbox}
\end{table}
\end{landscape}

%\section{144 Herbig stars in the Vioque et al. (2018) Sample}
\begin{landscape}
\begin{table}
\renewcommand{\thetable}{D1}
\contcaption{.}
\label{tab:D1_Vioque_Mdot2}
\begin{adjustbox}{width=1.33\textwidth}
\begin{tabular}{lccccccccccc}
\hline
Name&		RA&			DEC&	\multicolumn{2}{c}{$EW_{\rm obs}$}&	$EW_{\rm int}$&	$EW_{\rm cor}$&	$F_{\lambda}$&	$F_{\rm line}$&	$\log(L_{\rm line})$&		$\log(L_{\rm acc})$&	$\log(\dot M_{\rm acc})$\\[3pt]	
&				(J2000)&	(J2000)&	(\AA)&		Ref.&		(\AA)&		(\AA)&	(W\,m$^{-2}$\,\AA$^{-1}$)&		(W\,m$^{-2}$)&		[$\rm L_{\sun}$]&		[$\rm L_{\sun}$]&			[$\rm M_{\sun}$\,$\rm yr^{-1}$]\\	
\hline	
HD~37490& 05:39:11.1& +04:07:17.2& $-3.50\pm0.18$& 14& $4.60\pm0.02$& $-8.10\pm0.18$& $3.27\times10^{-14}$& $(2.64\pm0.06)\times10^{-13}$& $-0.10^{+0.21}_{-0.14}$& $1.99^{+0.27}_{-0.21}$& $-5.50^{+0.38}_{-0.39}$\\[3pt]
RR~Tau& 05:39:30.5& +26:22:26.9& $-25.60\pm0.77$& 3& $11.70\pm0.02$& $-37.30\pm0.77$& $3.55\times10^{-16}$& $(1.33\pm0.03)\times10^{-14}$& $-0.61^{+0.06}_{-0.06}$& $1.48^{+0.15}_{-0.15}$& $-5.93^{+0.20}_{-0.20}$\\[3pt]
HD~245906& 05:39:30.4& +26:19:55.1& $-3.60\pm0.18$& 2& $13.23\pm0.20$& $-16.83\pm0.27$& $2.94\times10^{-16}$& $(4.94\pm0.08)\times10^{-15}$& -$^{a}$& -& -\\[3pt]
HD~37806& 05:41:02.2& -02:43:00.7& $-17.00\pm0.85$& 14& $10.40\pm0.08$& $-27.40\pm0.85$& $1.51\times10^{-15}$& $(4.13\pm0.13)\times10^{-14}$& $-0.63^{+0.05}_{-0.05}$& $1.46^{+0.14}_{-0.14}$& $-5.96^{+0.22}_{-0.24}$\\[3pt]
HD~38087& 05:43:00.5& -02:18:45.3& -& -& $8.06\pm0.05$& -& $1.42\times10^{-15}$& -& -& -& -\\[3pt]
HD~38120& 05:43:11.8& -04:59:49.8& $-37.00\pm1.85$& 7& $11.58\pm0.11$& $-48.58\pm1.85$& $5.74\times10^{-16}$& $(2.79\pm0.11)\times10^{-14}$& $-0.85^{+0.07}_{-0.06}$& $1.24^{+0.17}_{-0.17}$& $-6.30^{+0.32}_{-0.28}$\\[3pt]
V351~Ori& 05:44:18.7& +00:08:40.4& $-0.90\pm0.05$& 13& $13.64\pm0.03$& $-14.54\pm0.06$& $6.70\times10^{-16}$& $(9.74\pm0.04)\times10^{-15}$& $-1.45^{+0.02}_{-0.02}$& $0.64^{+0.16}_{-0.16}$& $-6.72^{+0.20}_{-0.17}$\\[3pt]
HD~39014& 05:44:46.3& -65:44:07.9& $8.52\pm0.34$& 15& $13.58\pm0.18$& $-5.05\pm0.39$& $4.19\times10^{-14}$& $(2.12\pm0.16)\times10^{-13}$& $-1.89^{+0.05}_{-0.05}$& $0.20^{+0.20}_{-0.21}$& $-7.15^{+0.24}_{-0.22}$\\[3pt]
PDS~123& 05:50:54.2& +20:14:50.0& -& -& $5.61\pm0.03$& -& $1.17\times10^{-16}$& -& -& -& -\\[3pt]
V1818~Ori& 05:53:42.5& -10:24:00.7& $-40.00\pm2.00$& 12& $6.56\pm0.06$& $-46.56\pm2.00$& $2.20\times10^{-15}$& $(1.02\pm0.04)\times10^{-13}$& -$^{a}$& -& -\\[3pt]
GSC~5360-1033& 05:57:49.4& -14:05:33.6& $-4.00\pm0.20$& 7& $6.24\pm0.04$& $-10.24\pm0.20$& $2.30\times10^{-17}$& $(2.36\pm0.05)\times10^{-16}$& -$^{a}$& -& -\\[3pt]
HD~249879& 05:58:55.7& +16:39:57.3& $-41.00\pm2.05$& 7& $11.58\pm0.07$& $-52.58\pm2.05$& $1.24\times10^{-16}$& $(6.54\pm0.26)\times10^{-15}$& $-1.04^{+0.10}_{-0.08}$& $1.05^{+0.21}_{-0.20}$& $-6.62^{+0.33}_{-0.36}$\\[3pt]
PDS~22& 06:03:37.0& -14:53:03.1& $-25.00\pm1.25$& 7& $12.66\pm0.13$& $-37.66\pm1.26$& $1.97\times10^{-16}$& $(7.43\pm0.25)\times10^{-15}$& -$^{a}$& -& -\\[3pt]
HD~41511& 06:04:59.1& -16:29:03.9& $0.24\pm0.31$& 16& $11.22\pm0.10$& $-10.98\pm0.33$& $3.57\times10^{-14}$& $(3.92\pm0.12)\times10^{-13}$& $-0.22^{+0.07}_{-0.07}$& $1.87^{+0.14}_{-0.14}$& $-5.35^{+0.21}_{-0.20}$\\[3pt]
GSC~1876-0892& 06:07:15.3& +29:57:55.0& $-5.00\pm0.25$& 12& $3.94\pm0.04$& $-8.94\pm0.25$& $3.36\times10^{-16}$& $(3.01\pm0.09)\times10^{-15}$& $-0.07^{+0.17}_{-0.13}$& $2.02^{+0.23}_{-0.20}$& $-5.58^{+0.43}_{-0.46}$\\[3pt]
LkHA~208& 06:07:49.5& +18:39:26.4& $-4.90\pm0.25$& 2& $13.80\pm0.13$& $-18.70\pm0.28$& $8.62\times10^{-17}$& $(1.61\pm0.02)\times10^{-15}$& -$^{a}$& -& -\\[3pt]
PDS~211& 06:10:17.3& +29:25:16.6& $-35.00\pm1.75$& 7& $11.53\pm0.01$& $-46.53\pm1.75$& $9.60\times10^{-17}$& $(4.47\pm0.17)\times10^{-15}$& $-0.80^{+0.07}_{-0.06}$& $1.29^{+0.16}_{-0.16}$& $-6.22^{+0.27}_{-0.30}$\\[3pt]
LkHA~338& 06:10:47.1& -06:12:50.6& $-51.00\pm2.55$& 2& $14.72\pm0.11$& $-65.72\pm2.55$& $3.07\times10^{-17}$& $(2.02\pm0.08)\times10^{-15}$& $-1.31^{+0.08}_{-0.07}$& $0.78^{+0.20}_{-0.20}$& $-6.96^{+0.34}_{-0.37}$\\[3pt]
PDS~126& 06:13:37.2& -06:25:01.6& $-5.00\pm0.25$& 7& $13.61\pm0.12$& $-18.61\pm0.28$& $1.01\times10^{-16}$& $(1.89\pm0.03)\times10^{-15}$& $-1.38^{+0.06}_{-0.05}$& $0.71^{+0.18}_{-0.18}$& $-6.65^{+0.23}_{-0.23}$\\[3pt]
MWC~137& 06:18:45.5& +15:16:52.2& $-370.00\pm37.00$& 8& $2.30\pm0.04$& $-372.30\pm37.00$& $1.78\times10^{-15}$& $(6.64\pm0.66)\times10^{-13}$& $2.24^{+0.20}_{-0.17}$& $4.33^{+0.39}_{-0.34}$& $-3.45^{+0.55}_{-0.48}$\\[3pt]
CPM~25& 06:23:56.3& +14:30:28.0& $-200.00\pm10.00$& 12& $5.50\pm0.06$& $-205.50\pm10.00$& $6.66\times10^{-17}$& $(1.37\pm0.07)\times10^{-14}$& -$^{a}$& -& -\\[3pt]
NSV~2968& 06:26:53.9& -10:15:34.9& $-40.00\pm2.00$& 7& $3.97\pm0.07$& $-43.97\pm2.00$& $6.85\times10^{-16}$& $(3.01\pm0.14)\times10^{-14}$& -$^{a}$& -& -\\[3pt]
HD~45677& 06:28:17.4& -13:03:11.1& $-57.00\pm5.70$& 8& $5.80\pm0.03$& $-62.80\pm5.70$& $1.28\times10^{-15}$& $(8.02\pm0.73)\times10^{-14}$& $-0.02^{+0.09}_{-0.09}$& $2.07^{+0.16}_{-0.15}$& $-5.57^{+0.26}_{-0.35}$\\[3pt]
HD~46060& 06:30:49.8& -09:39:14.7& $3.61\pm0.01$& 17& $4.00\pm0.02$& $-0.40\pm0.02$& $3.22\times10^{-15}$& $(1.27\pm0.06)\times10^{-15}$& $-1.47^{+0.11}_{-0.09}$& $0.62^{+0.24}_{-0.23}$& $-7.03^{+0.46}_{-0.42}$\\[3pt]
PDS~129& 06:31:03.6& +10:01:13.5& -& -& $7.13\pm0.12$& -& $6.91\times10^{-17}$& -& -& -& -\\[3pt]
LKHA~215& 06:32:41.7& +10:09:34.2& $-25.70\pm1.29$& 2& $6.82\pm0.07$& $-32.52\pm1.29$& $7.11\times10^{-16}$& $(2.31\pm0.09)\times10^{-14}$& $-0.44^{+0.07}_{-0.06}$& $1.65^{+0.14}_{-0.14}$& $-5.91^{+0.24}_{-0.26}$\\[3pt]
HD~259431& 06:33:05.1& +10:19:19.9& $-71.00\pm2.13$& 11& $6.24\pm0.08$& $-77.24\pm2.13$& $1.75\times10^{-15}$& $(1.35\pm0.04)\times10^{-13}$& $0.34^{+0.06}_{-0.06}$& $2.43^{+0.14}_{-0.13}$& $-5.07^{+0.35}_{-0.33}$\\[3pt]
HBC~217& 06:40:42.1& +09:33:37.4& $-10.30\pm0.52$& 2& $6.12\pm0.08$& $-16.42\pm0.52$& $4.60\times10^{-17}$& $(7.56\pm0.24)\times10^{-16}$& $-1.94^{+0.06}_{-0.05}$& $0.15^{+0.21}_{-0.21}$& $-7.20^{+0.24}_{-0.24}$\\[3pt]
HBC~222& 06:40:51.1& +09:44:46.1& $-1.20\pm0.06$& 2& $6.18\pm0.02$& $-7.38\pm0.06$& $4.79\times10^{-17}$& $(3.54\pm0.03)\times10^{-16}$& $-2.26^{+0.04}_{-0.04}$& $-0.17^{+0.21}_{-0.21}$& $-7.51^{+0.25}_{-0.25}$\\[3pt]
HD~50138& 06:51:33.3& -06:57:59.4& $-71.00\pm7.10$& 8& $11.09\pm0.06$& $-82.09\pm7.10$& $4.65\times10^{-15}$& $(3.82\pm0.33)\times10^{-13}$& $0.23^{+0.07}_{-0.07}$& $2.32^{+0.15}_{-0.14}$& $-4.99^{+0.21}_{-0.19}$\\[3pt]
HD~50083& 06:51:45.7& +05:05:03.8& $-43.00\pm1.29$& 11& $4.16\pm0.03$& $-47.16\pm1.29$& $5.99\times10^{-15}$& $(2.83\pm0.08)\times10^{-13}$& $1.02^{+0.09}_{-0.08}$& $3.11^{+0.20}_{-0.19}$& $-4.33^{+0.29}_{-0.35}$\\[3pt]
PDS~25& 06:54:27.8& -25:02:15.8& $-13.00\pm0.65$& 7& $16.11\pm0.28$& $-29.11\pm0.71$& $2.03\times10^{-17}$& $(5.91\pm0.14)\times10^{-16}$& -$^{a}$& -& -\\[3pt]
HD~56895B& 07:18:30.5& -11:11:53.8& -& -& $9.15\pm0.27$& -& $1.16\times10^{-15}$& -& -& -& -\\[3pt]
GSC~6546-3156& 07:24:17.5& -26:16:05.2& $-9.00\pm0.45$& 7& $14.31\pm0.12$& $-23.31\pm0.47$& $2.61\times10^{-17}$& $(6.08\pm0.12)\times10^{-16}$& $-1.43^{+0.06}_{-0.05}$& $0.66^{+0.19}_{-0.19}$& $-6.91^{+0.29}_{-0.29}$\\[3pt]
\hline
\end{tabular}
\end{adjustbox}
\end{table}
\end{landscape}

\begin{landscape}
\begin{table}
\renewcommand{\thetable}{D1}
\contcaption{.}
\label{tab:D1_Vioque_Mdot3}
\begin{adjustbox}{width=1.33\textwidth}
\begin{tabular}{lccccccccccc}
\hline
Name&		RA&			DEC&	\multicolumn{2}{c}{$EW_{\rm obs}$}&	$EW_{\rm int}$&	$EW_{\rm cor}$&	$F_{\lambda}$&	$F_{\rm line}$&	$\log(L_{\rm line})$&		$\log(L_{\rm acc})$&	$\log(\dot M_{\rm acc})$\\[3pt]	
&				(J2000)&	(J2000)&	(\AA)&		Ref.&		(\AA)&		(\AA)&	(W\,m$^{-2}$\,\AA$^{-1}$)&		(W\,m$^{-2}$)&		[$\rm L_{\sun}$]&		[$\rm L_{\sun}$]&			[$\rm M_{\sun}$\,$\rm yr^{-1}$]\\	
\hline	
GSC~6542-2339& 07:24:36.9& -24:34:47.4& $-25.00\pm1.25$& 7& $14.43\pm0.05$& $-39.43\pm1.25$& $2.00\times10^{-16}$& $(7.90\pm0.25)\times10^{-15}$& -$^{a}$& -& -\\[3pt]
HD~58647& 07:25:56.0& -14:10:43.5& $-11.40\pm0.30$& 3& $9.68\pm0.01$& $-21.08\pm0.30$& $5.05\times10^{-15}$& $(1.06\pm0.02)\times10^{-13}$& $-0.47^{+0.03}_{-0.02}$& $1.62^{+0.11}_{-0.11}$& $-5.76^{+0.14}_{-0.15}$\\[3pt]
GSC~5988-2257& 07:41:41.0& -20:00:13.4& $-15.00\pm0.75$& 12& $5.60\pm0.03$& $-20.60\pm0.75$& $2.24\times10^{-17}$& $(4.62\pm0.17)\times10^{-16}$& -$^{a}$& -& -\\[3pt]
GSC~8143-1225& 07:59:11.5& -50:22:46.8& $1.71\pm0.05$& 17& $8.07\pm0.29$& $-6.36\pm0.29$& $6.89\times10^{-17}$& $(4.38\pm0.20)\times10^{-16}$& $-2.69^{+0.03}_{-0.03}$& $-0.60^{+0.22}_{-0.23}$& $-8.10^{+0.28}_{-0.28}$\\[3pt]
PDS~277& 08:23:11.8& -39:07:01.6& $-3.00\pm0.15$& 7& $8.34\pm0.39$& $-11.34\pm0.42$& $2.58\times10^{-16}$& $(2.93\pm0.11)\times10^{-15}$& $-1.96^{+0.03}_{-0.03}$& $0.13^{+0.19}_{-0.19}$& $-7.22^{+0.22}_{-0.21}$\\[3pt]
V388~Vel& 08:42:16.5& -40:44:09.9& -& -& $11.09\pm0.05$& -& $1.06\times10^{-16}$& -& -$^{a}$& -& -\\[3pt]
PDS~34& 08:49:58.5& -45:53:05.6& $-50.00\pm2.50$& 7& $5.52\pm0.08$& $-55.52\pm2.50$& $7.01\times10^{-17}$& $(3.89\pm0.18)\times10^{-15}$& $-0.26^{+0.07}_{-0.07}$& $1.83^{+0.14}_{-0.14}$& $-6.01^{+0.37}_{-0.38}$\\[3pt]
PDS~290& 09:26:11.0& -52:42:26.9& $-7.00\pm0.35$& 7& $10.26\pm0.06$& $-17.26\pm0.36$& $6.14\times10^{-17}$& $(1.06\pm0.02)\times10^{-15}$& $-1.60^{+0.04}_{-0.04}$& $0.49^{+0.18}_{-0.18}$& $-7.28^{+0.33}_{-0.34}$\\[3pt]
HD~87643& 10:04:30.2& -58:39:52.1& $-145.00\pm14.50$& 8& $3.25\pm0.06$& $-148.25\pm14.50$& $4.22\times10^{-15}$& $(6.26\pm0.61)\times10^{-13}$& -$^{a}$& -& -\\[3pt]
PDS~322& 10:52:08.6& -56:12:06.8& $4.00\pm0.02$& 7& $5.52\pm0.08$& $-1.52\pm0.08$& $1.09\times10^{-16}$& $(1.66\pm0.09)\times10^{-16}$& -$^{a}$& -& -\\[3pt]
PDS~324& 10:57:24.2& -62:53:13.2& $-4.00\pm0.20$& 7& $4.42\pm0.05$& $-8.42\pm0.21$& $8.15\times10^{-17}$& $(6.86\pm0.17)\times10^{-16}$& $-0.75^{+0.10}_{-0.09}$& $1.34^{+0.20}_{-0.19}$& $-6.58^{+0.46}_{-0.40}$\\[3pt]
PDS~138& 11:53:13.2& -62:05:20.9& $-1.35\pm0.15$& 18& $2.18\pm0.04$& $-3.52\pm0.16$& $1.12\times10^{-15}$& $(3.96\pm0.18)\times10^{-15}$& $0.42^{+0.15}_{-0.13}$& $2.51^{+0.24}_{-0.20}$& $-5.27^{+0.44}_{-0.34}$\\[3pt]
GSC~8645-1401& 12:17:47.5& -59:43:59.0& $-9.00\pm0.45$& 7& $9.27\pm0.33$& $-18.27\pm0.56$& $1.58\times10^{-16}$& $(2.89\pm0.09)\times10^{-15}$& $-0.55^{+0.09}_{-0.08}$& $1.54^{+0.18}_{-0.17}$& $-5.64^{+0.23}_{-0.23}$\\[3pt]
Hen~2-80& 12:22:23.1& -63:17:16.8& $-150.00\pm7.50$& 19& $8.07\pm0.07$& $-158.07\pm7.50$& $2.29\times10^{-16}$& $(3.61\pm0.17)\times10^{-14}$& -$^{a}$& -& -\\[3pt]
Hen~3-823& 12:48:42.3& -59:54:34.9& $-25.00\pm1.25$& 19& $5.63\pm0.06$& $-30.63\pm1.25$& $4.05\times10^{-16}$& $(1.24\pm0.05)\times10^{-14}$& -$^{a}$& -& -\\[3pt]
DK~Cha& 12:53:17.2& -77:07:10.7& $-88.00\pm4.40$& 20& $10.64\pm0.33$& $-98.64\pm4.41$& $1.66\times10^{-16}$& $(1.63\pm0.07)\times10^{-14}$& -$^{a}$& -& -\\[3pt]
GSC~8994-3902& 13:19:03.9& -62:34:10.1& $3.39\pm0.07$& 18& $4.21\pm0.06$& $-0.82\pm0.09$& $3.75\times10^{-16}$& $(3.09\pm0.33)\times10^{-16}$& $-1.26^{+0.17}_{-0.15}$& $0.83^{+0.28}_{-0.28}$& $-6.80^{+0.49}_{-0.53}$\\[3pt]
PDS~371& 13:47:31.4& -36:39:49.6& $-40.00\pm2.00$& 12& $2.18\pm0.01$& $-42.18\pm2.00$& $1.26\times10^{-16}$& $(5.30\pm0.25)\times10^{-15}$& -$^{a}$& -& -\\[3pt]
Hen~3-938& 13:52:42.8& -63:32:49.1& $-90.00\pm4.50$& 12& $1.98\pm0.03$& $-91.98\pm4.50$& $9.75\times10^{-16}$& $(8.97\pm0.44)\times10^{-14}$& $1.62^{+0.15}_{-0.13}$& $3.71^{+0.30}_{-0.26}$& $-4.20^{+0.44}_{-0.35}$\\[3pt]
HD~130437& 14:50:50.2& -60:17:10.3& $-53.01\pm0.25$& 19& $3.28\pm0.06$& $-56.29\pm0.26$& $1.89\times10^{-15}$& $(1.06\pm0.01)\times10^{-13}$& $0.95^{+0.09}_{-0.07}$& $3.04^{+0.20}_{-0.18}$& $-4.68^{+0.44}_{-0.38}$\\[3pt]
PDS~389& 15:14:47.0& -62:16:59.7& $-9.00\pm0.45$& 7& $13.64\pm0.09$& $-22.64\pm0.46$& $3.83\times10^{-16}$& $(8.68\pm0.18)\times10^{-15}$& $-0.76^{+0.05}_{-0.04}$& $1.33^{+0.14}_{-0.14}$& $-5.99^{+0.18}_{-0.20}$\\[3pt]
HD~135344& 15:15:48.9& -37:08:55.7& $-6.50\pm0.65$& 8& $8.47\pm0.41$& $-14.97\pm0.77$& $2.26\times10^{-15}$& $(3.38\pm0.17)\times10^{-14}$& $-1.67^{+0.04}_{-0.04}$& $0.42^{+0.18}_{-0.19}$& $-6.89^{+0.20}_{-0.20}$\\[3pt]
Hen~3-1121& 15:58:09.6& -53:51:18.3& $-0.40\pm0.05$& 19& $3.63\pm0.02$& $-4.03\pm0.05$& $5.33\times10^{-16}$& $(2.14\pm0.03)\times10^{-15}$& -$^{a}$& -& -\\[3pt]
CPD-53~6867& 15:58:09.6& -53:51:35.0& $-0.10\pm0.01$& 19& $3.32\pm0.05$& $-3.42\pm0.05$& $1.73\times10^{-15}$& $(5.90\pm0.09)\times10^{-15}$& -$^{a}$& -& -\\[3pt]
HD~143006& 15:58:36.9& -22:57:15.2& $-8.35\pm0.15$& 21& $3.17\pm0.01$& $-11.52\pm0.15$& $3.91\times10^{-16}$& $(4.51\pm0.06)\times10^{-15}$& $-2.41^{+0.04}_{-0.04}$& $-0.32^{+0.22}_{-0.22}$& $-7.73^{+0.25}_{-0.25}$\\[3pt]
WRAY~15-1435& 16:13:06.7& -50:23:20.0& $-20.00\pm1.00$& 19& $4.07\pm0.07$& $-24.07\pm1.00$& $3.48\times10^{-16}$& $(8.37\pm0.35)\times10^{-15}$& $-0.05^{+0.15}_{-0.12}$& $2.04^{+0.21}_{-0.19}$& $-5.80^{+0.48}_{-0.40}$\\[3pt]
Hen~3-1191& 16:27:15.1& -48:39:26.8& $-1093.17\pm23.58$& 22& $4.16\pm0.11$& $-1097.33\pm23.58$& $1.95\times10^{-16}$& $(2.14\pm0.05)\times10^{-13}$& -$^{a}$& -& -\\[3pt]
HD~149914& 16:38:28.6& -18:13:13.7& $9.33\pm0.06$& 23& $11.08\pm0.03$& $-1.75\pm0.07$& $9.56\times10^{-15}$& $(1.68\pm0.07)\times10^{-14}$& $-1.88^{+0.04}_{-0.04}$& $0.21^{+0.19}_{-0.19}$& $-7.21^{+0.26}_{-0.28}$\\[3pt]
V921~Sco& 16:59:06.7& -42:42:08.4& $-194.30\pm19.43$& 24& $2.56\pm0.05$& $-196.86\pm19.43$& $4.16\times10^{-15}$& $(8.20\pm0.81)\times10^{-13}$& $1.79^{+0.17}_{-0.14}$& $3.88^{+0.32}_{-0.28}$& $-3.94^{+0.50}_{-0.44}$\\[3pt]
HD~155448& 17:12:58.7& -32:14:33.5& $4.06\pm0.02$& 25& $8.62\pm0.07$& $-4.55\pm0.07$& $1.09\times10^{-15}$& $(4.95\pm0.08)\times10^{-15}$& $-0.86^{+0.09}_{-0.07}$& $1.23^{+0.19}_{-0.18}$& $-6.11^{+0.29}_{-0.28}$\\[3pt]
PDS~453& 17:20:56.1& -26:03:30.5& $4.00\pm0.20$& 7& $9.41\pm0.35$& $-5.41\pm0.40$& $4.10\times10^{-17}$& $(2.22\pm0.16)\times10^{-16}$& -$^{a}$& -& -\\[3pt]
MWC~878& 17:24:44.7& -38:43:51.4& $-54.00\pm2.70$& 19& $3.28\pm0.06$& $-57.28\pm2.70$& $1.67\times10^{-15}$& $(9.58\pm0.45)\times10^{-14}$& $0.97^{+0.16}_{-0.13}$& $3.06^{+0.27}_{-0.23}$& $-4.66^{+0.51}_{-0.45}$\\[3pt]
HD~319896& 17:31:05.8& -35:08:29.2& $-27.00\pm1.35$& 7& $5.48\pm0.07$& $-32.48\pm1.35$& $6.68\times10^{-16}$& $(2.17\pm0.09)\times10^{-14}$& $0.06^{+0.14}_{-0.11}$& $2.15^{+0.21}_{-0.17}$& $-5.39^{+0.29}_{-0.25}$\\[3pt]
HD~158643& 17:31:24.9& -23:57:45.5& $-3.30\pm0.10$& 3& $11.32\pm0.07$& $-14.62\pm0.12$& $2.37\times10^{-14}$& $(3.47\pm0.03)\times10^{-13}$& $-0.79^{+0.06}_{-0.05}$& $1.30^{+0.16}_{-0.15}$& $-6.07^{+0.22}_{-0.24}$\\[3pt]
\hline
\end{tabular}
\end{adjustbox}
\end{table}
\end{landscape}

\begin{landscape}
\begin{table}
\renewcommand{\thetable}{D1}
\contcaption{.}
\label{tab:D1_Vioque_Mdot4}
\begin{adjustbox}{width=1.33\textwidth}
\begin{tabular}{lccccccccccc}
\hline
Name&		RA&			DEC&	\multicolumn{2}{c}{$EW_{\rm obs}$}&	$EW_{\rm int}$&	$EW_{\rm cor}$&	$F_{\lambda}$&	$F_{\rm line}$&	$\log(L_{\rm line})$&		$\log(L_{\rm acc})$&	$\log(\dot M_{\rm acc})$\\[3pt]	
&				(J2000)&	(J2000)&	(\AA)&		Ref.&		(\AA)&		(\AA)&	(W\,m$^{-2}$\,\AA$^{-1}$)&		(W\,m$^{-2}$)&		[$\rm L_{\sun}$]&		[$\rm L_{\sun}$]&			[$\rm M_{\sun}$\,$\rm yr^{-1}$]\\		
\hline		
HD~323771& 17:34:04.6& -39:23:41.3& $-54.00\pm2.70$& 7& $6.83\pm0.03$& $-60.83\pm2.70$& $2.32\times10^{-16}$& $(1.41\pm0.06)\times10^{-14}$& $-0.30^{+0.11}_{-0.10}$& $1.79^{+0.18}_{-0.18}$& $-5.86^{+0.30}_{-0.28}$\\[3pt]
SAO~185668& 17:43:55.6& -22:05:44.6& -& -& $4.42\pm0.04$& -& $1.87\times10^{-15}$& -& -& -& -\\[3pt]
MWC~593& 17:49:10.1& -24:14:21.2& $-35.00\pm1.75$& 19& $4.81\pm0.06$& $-39.81\pm1.75$& $1.54\times10^{-15}$& $(6.14\pm0.27)\times10^{-14}$& $0.54^{+0.15}_{-0.12}$& $2.63^{+0.25}_{-0.20}$& $-4.85^{+0.32}_{-0.26}$\\[3pt]
PDS~477& 18:00:30.3& -16:47:25.8& $-120.00\pm6.00$& 7& $4.13\pm0.06$& $-124.13\pm6.00$& $1.68\times10^{-16}$& $(2.09\pm0.10)\times10^{-14}$& -$^{a}$& -& -\\[3pt]
HD~313571& 18:01:07.1& -22:15:04.0& $-34.00\pm1.70$& 19& $4.87\pm0.12$& $-38.87\pm1.70$& $9.73\times10^{-16}$& $(3.78\pm0.17)\times10^{-14}$& $0.36^{+0.18}_{-0.13}$& $2.45^{+0.26}_{-0.20}$& $-5.08^{+0.37}_{-0.39}$\\[3pt]
LkHA~260& 18:19:09.3& -13:50:41.2& -& -& $8.22\pm0.05$& -& $7.93\times10^{-17}$& -& -& -& -\\[3pt]
HD~169142& 18:24:29.7& -29:46:49.3& $-13.97\pm0.15$& 21& $13.76\pm0.10$& $-27.73\pm0.18$& $2.82\times10^{-15}$& $(7.83\pm0.05)\times10^{-14}$& $-1.50^{+0.01}_{-0.01}$& $0.59^{+0.15}_{-0.15}$& $-7.09^{+0.26}_{-0.29}$\\[3pt]
MWC~930& 18:26:25.2& -07:13:17.8& -& -& $8.76\pm0.09$& -& $1.50\times10^{-13}$& -& -& -& -\\[3pt]
PDS~520& 18:30:06.1& +00:42:33.5& $-33.00\pm1.65$& 7& $8.32\pm0.28$& $-41.32\pm1.67$& $9.42\times10^{-17}$& $(3.89\pm0.16)\times10^{-15}$& -$^{a}$& -& -\\[3pt]
PDS~530& 18:41:34.3& +08:08:20.7& $-28.00\pm1.40$& 7& $14.71\pm0.22$& $-42.71\pm1.42$& $3.44\times10^{-17}$& $(1.47\pm0.05)\times10^{-15}$& -$^{a}$& -& -\\[3pt]
MWC~953& 18:43:28.4& -03:46:17.0& $-32.00\pm1.60$& 19& $3.66\pm0.05$& $-35.66\pm1.60$& $2.35\times10^{-15}$& $(8.39\pm0.38)\times10^{-14}$& $0.89^{+0.16}_{-0.13}$& $2.98^{+0.28}_{-0.23}$& $-4.57^{+0.48}_{-0.48}$\\[3pt]
HD~174571& 18:50:47.1& +08:42:10.0& $2.00\pm0.20$& 26& $3.64\pm0.04$& $-1.64\pm0.20$& $6.00\times10^{-15}$& $(9.83\pm1.22)\times10^{-15}$& $-0.43^{+0.15}_{-0.13}$& $1.66^{+0.22}_{-0.22}$& $-5.89^{+0.41}_{-0.46}$\\[3pt]
PDS~551& 18:55:22.9& +04:04:35.2& $-50.00\pm2.50$& 7& $2.87\pm0.06$& $-52.87\pm2.50$& $6.02\times10^{-18}$& $(3.18\pm0.15)\times10^{-16}$& -$^{a}$& -& -\\[3pt]
HD~176386& 19:01:38.9& -36:53:26.5& $-0.17\pm0.05$& 4& $12.73\pm0.12$& $-12.89\pm0.13$& $2.68\times10^{-15}$& $(3.46\pm0.03)\times10^{-14}$& $-1.56^{+0.02}_{-0.02}$& $0.53^{+0.16}_{-0.16}$& $-7.08^{+0.27}_{-0.27}$\\[3pt]
TY~CrA& 19:01:40.8& -36:52:33.8& $-5.82\pm0.37$& 15& $13.14\pm0.04$& $-18.96\pm0.37$& $2.48\times10^{-15}$& $(4.71\pm0.09)\times10^{-14}$& -$^{a}$& -& -\\[3pt]
R~CrA& 19:01:53.6& -36:57:08.1& $-85.50\pm4.28$& 27& $14.43\pm0.05$& $-99.93\pm4.28$& $2.87\times10^{-16}$& $(2.87\pm0.12)\times10^{-14}$& -$^{a}$& -& -\\[3pt]
MWC~314& 19:21:33.9& +14:52:56.9& $-125.00\pm15.00$& 28& $3.62\pm0.04$& $-128.62\pm15.00$& $1.43\times10^{-14}$& $(1.84\pm0.21)\times10^{-12}$& $2.70^{+0.20}_{-0.17}$& $4.79^{+0.40}_{-0.36}$& $-2.36^{+0.68}_{-0.66}$\\[3pt]
HD~344261& 19:21:53.5& +21:31:50.5& -& -& $9.23\pm0.23$& -& $1.97\times10^{-16}$& -& -& -& -\\[3pt]
WW~Vul& 19:25:58.7& +21:12:31.3& $-19.10\pm0.57$& 3& $15.06\pm0.10$& $-34.16\pm0.58$& $2.67\times10^{-16}$& $(9.11\pm0.16)\times10^{-15}$& $-1.14^{+0.03}_{-0.03}$& $0.95^{+0.15}_{-0.15}$& $-6.51^{+0.21}_{-0.19}$\\[3pt]
PX~Vul& 19:26:40.2& +23:53:50.7& $-14.40\pm0.43$& 3& $8.37\pm0.32$& $-22.77\pm0.54$& $2.03\times10^{-16}$& $(4.63\pm0.11)\times10^{-15}$& -$^{a}$& -& -\\[3pt]
PDS~581& 19:36:18.8& +29:32:50.8& $-200.00\pm10.00$& 7& $5.16\pm0.07$& $-205.16\pm10.00$& $4.10\times10^{-16}$& $(8.42\pm0.41)\times10^{-14}$& -$^{a}$& -& -\\[3pt]
MWC~623& 19:56:31.5& +31:06:20.1& $-129.00\pm14.00$& 29& $3.85\pm0.04$& $-132.85\pm14.00$& $2.56\times10^{-15}$& $(3.40\pm0.36)\times10^{-13}$& $2.06^{+0.18}_{-0.16}$& $4.15^{+0.35}_{-0.31}$& $-3.19^{+0.45}_{-0.41}$\\[3pt]
V1686~Cyg& 20:20:29.3& +41:21:28.4& $-22.70\pm0.68$& 3& $5.59\pm0.03$& $-28.29\pm0.68$& $1.07\times10^{-16}$& $(3.02\pm0.07)\times10^{-15}$& -$^{a}$& -& -\\[3pt]
MWC~342& 20:23:03.6& +39:29:49.9& $-290.00\pm80.00$& 30& $2.58\pm0.06$& $-292.58\pm80.00$& $5.20\times10^{-15}$& $(1.52\pm0.42)\times10^{-12}$& $2.19^{+0.17}_{-0.19}$& $4.28^{+0.35}_{-0.35}$& $-3.43^{+0.52}_{-0.57}$\\[3pt]
BD+41~3731& 20:24:15.7& +42:18:01.3& -& -& $5.56\pm0.01$& -& $5.45\times10^{-16}$& -& -& -& -\\[3pt]
HBC~694& 20:24:29.5& +42:14:02& -& -& $14.43\pm0.05$& -& $2.10\times10^{-18}$& -& -$^{a}$& -& -\\[3pt]
MWC~1021& 20:29:26.9& +41:40:43.8& -& -& $2.87\pm0.02$& -& $4.74\times10^{-14}$& -& -& -& -\\[3pt]
V1478~Cyg& 20:32:45.6& +40:39:36.1& -& -& $2.87\pm0.02$& -& $3.61\times10^{-14}$& -& -$^{a}$& -& -\\[3pt]
PV~Cep& 20:45:53.9& +67:57:38.6& $-49.90\pm2.50$& 31& $14.43\pm0.05$& $-64.33\pm2.50$& $2.54\times10^{-17}$& $(1.63\pm0.06)\times10^{-15}$& $-2.22^{+0.05}_{-0.05}$& $-0.13^{+0.22}_{-0.22}$& -\\[3pt]
V1977~Cyg& 20:47:37.4& +43:47:24.9& $-32.70\pm0.98$& 3& $8.89\pm0.03$& $-41.59\pm0.98$& $6.92\times10^{-16}$& $(2.88\pm0.07)\times10^{-14}$& $-0.18^{+0.05}_{-0.05}$& $1.91^{+0.12}_{-0.12}$& $-5.49^{+0.15}_{-0.15}$\\[3pt]
HBC~705& 20:51:02.7& +43:49:31.9& $-26.90\pm1.35$& 2& $4.21\pm0.06$& $-31.11\pm1.35$& $5.13\times10^{-16}$& $(1.59\pm0.07)\times10^{-14}$& $0.33^{+0.10}_{-0.09}$& $2.42^{+0.18}_{-0.16}$& $-5.21^{+0.39}_{-0.42}$\\[3pt]
V1493~Cyg& 20:52:04.6& +44:37:30.4& $-9.50\pm0.48$& 2& $12.76\pm0.09$& $-22.26\pm0.48$& $3.03\times10^{-16}$& $(6.75\pm0.15)\times10^{-15}$& -$^{a}$& -& -\\[3pt]
HBC~717& 20:52:06.0& +44:17:16.0& $-19.00\pm0.95$& 2& $6.56\pm0.13$& $-25.56\pm0.96$& $1.41\times10^{-16}$& $(3.61\pm0.14)\times10^{-15}$& -$^{a}$& -& -\\[3pt]
HD~199603& 20:58:41.8& -14:28:59.2& $9.44\pm0.27$& 15& $10.96\pm0.39$& $-1.52\pm0.48$& $1.01\times10^{-14}$& $(1.54\pm0.48)\times10^{-14}$& $-2.42^{+0.13}_{-0.18}$& $-0.33^{+0.31}_{-0.37}$& $-7.65^{+0.32}_{-0.37}$\\[3pt]
\hline
\end{tabular}
\end{adjustbox}
\end{table}
\end{landscape}

\begin{landscape}
\begin{table}
\renewcommand{\thetable}{D1}
\contcaption{.}
\label{tab:D1_Vioque_Mdot5}
\begin{adjustbox}{width=1.33\textwidth}
\begin{tabular}{lccccccccccc}
\hline
Name&		RA&			DEC&	\multicolumn{2}{c}{$EW_{\rm obs}$}&	$EW_{\rm int}$&	$EW_{\rm cor}$&	$F_{\lambda}$&	$F_{\rm line}$&	$\log(L_{\rm line})$&		$\log(L_{\rm acc})$&	$\log(\dot M_{\rm acc})$\\[3pt]	
&				(J2000)&	(J2000)&	(\AA)&		Ref.&		(\AA)&		(\AA)&	(W\,m$^{-2}$\,\AA$^{-1}$)&		(W\,m$^{-2}$)&		[$\rm L_{\sun}$]&		[$\rm L_{\sun}$]&			[$\rm M_{\sun}$\,$\rm yr^{-1}$]\\	
\hline		
V594~Cyg& 21:20:23.3& +43:18:10.2& -& -& $6.39\pm0.08$& -& $3.04\times10^{-15}$& -& -& -& -\\[3pt]
HD~235495& 21:21:27.4& +50:59:47.6& -& -& $14.00\pm0.09$& -& $2.53\times10^{-16}$& -& -& -& -\\[3pt]
AS~470& 21:36:14.2& +57:21:30.8& $-40.00\pm4.00$& 32& $11.44\pm0.04$& $-51.44\pm4.00$& $1.84\times10^{-16}$& $(9.48\pm0.74)\times10^{-15}$& $0.68^{+0.16}_{-0.14}$& $2.77^{+0.26}_{-0.22}$& $-4.37^{+0.44}_{-0.38}$\\[3pt]
GSC~3975-0579& 21:38:08.4& +57:26:47.6& $-6.90\pm0.35$& 13& $14.83\pm0.21$& $-21.73\pm0.40$& $9.95\times10^{-17}$& $(2.16\pm0.04)\times10^{-15}$& $-1.22^{+0.05}_{-0.04}$& $0.87^{+0.17}_{-0.17}$& $-6.55^{+0.25}_{-0.26}$\\[3pt]
BH~Cep& 22:01:42.8& +69:44:36.4& $-6.20\pm0.19$& 3& $7.70\pm0.44$& $-13.90\pm0.48$& $1.80\times10^{-16}$& $(2.50\pm0.09)\times10^{-15}$& $-2.06^{+0.02}_{-0.02}$& $0.03^{+0.19}_{-0.19}$& $-7.34^{+0.23}_{-0.25}$\\[3pt]
BO~Cep& 22:16:54.0& +70:03:44.9& $-7.50\pm0.23$& 3& $7.71\pm0.41$& $-15.21\pm0.47$& $7.33\times10^{-17}$& $(1.11\pm0.03)\times10^{-15}$& $-2.31^{+0.02}_{-0.02}$& $-0.22^{+0.20}_{-0.20}$& $-7.69^{+0.23}_{-0.22}$\\[3pt]
V669~Cep& 22:26:38.7& +61:13:31.5& $-127.00\pm60.00$& 33& $6.60\pm0.02$& $-133.60\pm60.00$& $3.41\times10^{-16}$& $(4.56\pm2.05)\times10^{-14}$& -$^{a}$& -& -\\[3pt]
MWC~655& 22:38:31.8& +55:50:05.3& $-14.30\pm0.72$& 26& $3.47\pm0.07$& $-17.77\pm0.72$& $7.05\times10^{-16}$& $(1.25\pm0.05)\times10^{-14}$& $0.26^{+0.11}_{-0.09}$& $2.35^{+0.19}_{-0.16}$& $-5.40^{+0.43}_{-0.38}$\\[3pt]
MWC~657& 22:42:41.8& +60:24:00.6& $-185.35\pm13.00$& 34& $3.20\pm0.03$& $-188.55\pm13.00$& $1.73\times10^{-15}$& $(3.26\pm0.22)\times10^{-13}$& $2.00^{+0.13}_{-0.11}$& $4.09^{+0.30}_{-0.27}$& $-3.41^{+0.94}_{-0.71}$\\[3pt]
BP~Psc& 23:22:24.6& -02:13:41.3& $-13.17\pm0.35$& 35& $3.15\pm0.45$& $-16.32\pm0.58$& $1.64\times10^{-16}$& $(2.68\pm0.09)\times10^{-15}$& -$^{a}$& -& -\\[3pt]
LkHA~259& 23:58:41.6& +66:26:12.7& $-24.00\pm1.20$& 2& $12.03\pm0.18$& $-36.03\pm1.21$& $8.27\times10^{-17}$& $(2.98\pm0.10)\times10^{-15}$& $-1.28^{+0.05}_{-0.05}$& $0.81^{+0.17}_{-0.18}$& $-6.54^{+0.21}_{-0.23}$\\[3pt]
\hline
\end{tabular}
\end{adjustbox}
\\
\begin{flushleft}
{\bf Notes.}
$^{(a)}$ Stars which have low quality parallaxes in the {\em Gaia} DR2 Catalogue (see the text for discussion).
References: (1) \citet{Herbig1988}; (2) \citet{Hernandez2004}; (3) \citet{Mendigutia2011a}; (4) \citet{Pogodin2012}; (5) ESO Programme 082.A-9011(A); (6) ESO Programme 076.B-0055(A); (7) \citet{Sartori2010}; (8) \citet{Baines2006}; (9) \citet{Miroshnichenko1999}; (10) \citet{Boehm1995}; (11) \citet{Wheelwright2010}; (12) \citet{Vieira2011}; (13) \citet{Hernandez2005}; (14) \citet{Oudmaijer1999}; (15) ESO Programme 085.A-9027(B); (16) ESO Programme 082.D-0061(A); (17) ESO Programme 084.A-9016(A); (18) ESO Programme 083.A-9013(A); (19) \citet{Carmona2010}; (20) \citet{Spezzi2008}; (21) \citet{Dunkin1997}; (22) ESO Programme 075.D-0177(A); (23) ESO Programme 60.A-9022(C); (24) \citet{BorgesFernandes2007}; (25) ESO Programme 073.D-0609(A); (26) \citet{Ababakr2017}; (27) \citet{Manoj2006}; (28) \citet{Frasca2016}; (29) \citet{Polster2012}; (30) \citet{Kucerova2013}; (31) \citet{Acke2005}; (32) \citet{Nakano2012}; (33) \citet{Miroshnichenko2002}; (34) \citet{Miroshnichenko2000}; (35) \citet{Zuckerman2008}.
\end{flushleft}
\end{table}
\end{landscape}

%%%%%%%%%%%%%%%%%%%%%%%%%%%%%%%%%%%%%%%%%%%%%%%%%%

% Don't change these lines
%\bsp	% typesetting comment
%\label{lastpage}
\end{document}